\documentclass[a4paper,12pt]{article}

\usepackage[dvipdfmx]{graphicx}
\usepackage{tikz}
\usetikzlibrary{decorations.pathmorphing}

\usepackage{amsmath}

\usepackage[english]{babel}

\usepackage{amssymb,amsmath,wrapfig,subfig}
\usepackage{amsfonts}
\usepackage{latexsym}
\usepackage{graphicx}
\usepackage{graphicx,color}
\usepackage{epsfig}
\usepackage{epsf}
\usepackage{bbold}
\usepackage{verbatim}
\usepackage{amsthm}
\usepackage{amsbsy}
\usepackage{hyperref}
\usepackage{multirow,array,float}
\usepackage{cite}
\definecolor{link}{rgb}{.8,.15,.1}
\hypersetup{colorlinks=true,linkcolor=link,citecolor=link,urlcolor=link,linktocpage}
\usepackage{mathrsfs}

\usepackage{bm}
\usepackage{color}
\usepackage{ulem}
\usepackage{amssymb}

\DeclareMathOperator{\Tr}{Tr}

\usepackage{geometry}
\numberwithin{equation}{section}
\allowdisplaybreaks

\newcommand{\ie}{{\it i.e.}, }
\newcommand{\eg}{{\it e.g.}, }

\newcommand{\etc}{{\it etc.}}

\begin{document}

\newcommand{\hiduke}[1]{\hspace{\fill}{\small [{#1}]}}
\newcommand{\aff}[1]{${}^{#1}$}
\renewcommand{\thefootnote}{\fnsymbol{footnote}}

\begin{titlepage}
\begin{flushright}
{\footnotesize KIAS-P18037}
\end{flushright}
\begin{center}
{\Large\bf
The Thouless time for mass-deformed SYK
}\\
\bigskip\bigskip
{\large Tomoki Nosaka,\footnote{\tt nosaka@yukawa.kyoto-u.ac.jp}}\aff{1}
{\large Dario Rosa\footnote{\tt Dario85@kias.re.kr}}\aff{1}
{\large and Junggi Yoon\footnote{\tt junggi.yoon@icts.res.in}}\aff{2}\\
\bigskip\bigskip
\aff{1}: {\small
\it School of Physics, Korea Institute for Advanced Study\\
85 Hoegiro Dongdaemun-gu, Seoul 02455, Republic of Korea
}\\
\bigskip
\aff{2}: {\small
\it International Centre for Theoretical Sciences (ICTS-TIFR), \\
Shivakote, Hesaraghatta Hobli, Bengaluru 560089, India.
}
\end{center}

\begin{abstract}
We study the onset of RMT dynamics in the mass-deformed SYK model (\ie an SYK model deformed by a quadratic random interaction) in terms of the strength of the quadratic deformation.
We use as chaos probes both the connected unfolded Spectral Form Factor (SFF) as well as the Gaussian-filtered SFF, which has been recently introduced in the literature.
We show that they detect the chaotic/integrable transition of the mass-deformed SYK model at different values of the mass deformation: the Gaussian-filtered SFF sees the transition for large values of the mass deformation; the connected unfolded SFF sees the transition at small values. 
The latter is in qualitative agreement with the transition as seen by the OTOCs.
We argue that the chaotic/integrable deformation affect the energy levels inhomogeneously: for small values of the mass deformation only the low-lying states are modified while for large values of the mass deformation also the states in the bulk of the spectrum move to the integrable behavior.

\end{abstract}

\bigskip\bigskip\bigskip

\end{titlepage}

\renewcommand{\thefootnote}{\arabic{footnote}}
\setcounter{footnote}{0}

\tableofcontents

\section{Introduction and Summary}
\label{sec: introduction}

The common feature of classical chaotic systems is their high sensitivity to the initial conditions. 
Concretely, they show an exponential dependence of the dynamical evolution to the initial conditions (for example the initial position) of the form
\begin{align}
\label{eq:classical_lyapunov}
\frac{\partial \, x(t)}{\partial \, x(0)}\, \propto \, e^{\lambda_L \, t} \ ,
\end{align}
where $\lambda_L$ is the Lyapunov exponent which measures the strength of chaos in the system under investigation. 

The extension of the concept of classical chaos to the quantum mechanical framework is not straightforward: from equation \eqref{eq:classical_lyapunov} it is clear that the classical notion of chaos relies on the notion of trajectory and so its extension to the quantum world is problematic.
Traditionally, quantum chaos has been associated to the well-known Bohigas-Giannoni-Schmit (BGS) conjecture \cite{Bohigas:1983er}, which can be stated as follows: {\it spectra of quantum systems, whose classical analogues are chaotic, show the same fluctuation properties as predicted by random matrix theory (RMT)}. 
From the BGS conjecture we see that the quantum chaotic properties of a system are encoded in the {\it statistical} properties of its spectrum, and so they can be analyzed using statistical methods. 
In particular, quantum chaotic systems should show the typical features of RMT, i.e. {\it level repulsion} and {\it spectral rigidity}.\footnote{For an introduction to the BGS conjecture, the field of quantum chaos and RMT techniques, the reader can look at, for example, \cite{Mehta:2004,GUHR1998189,Bohigas:1984}. 
}
The BGS conjecture have received in the years a large number of numerical and analytical confirmations and it is nowadays accepted as true, even though a proof is still lacking.

On the other hand,  another diagnostics of quantum chaos has become very popular in recent days for the study of many-body\footnote{Similar studies have been performed also for single-body quantum systems \cite{Hashimoto:2017oit} but the results are not conclusive yet.} quantum systems: starting from an old work by Larkin and Ovchinnikov \cite{1969JETP...28.1200L} and treating equation \eqref{eq:classical_lyapunov} semi-classically, Kitaev has proposed the out-of-time-ordering correlators (OTOCs) 
\begin{align}
    \label{eq:OTOC}
  C_{W V} (\beta , t) \, \propto \,  \langle W(t) V(0) W(t) V(0) \rangle_\beta \ ,
\end{align}
where $W$ and $V$ are $2$ {\it generic} operators and $\beta$ is the inverse temperature, as an alternative way to characterize quantum chaos.
In particular, if $C_{W V} (\beta , t)$ displays a behavior of the form 
\begin{align}
    \label{eq:decay_OTOC}
    C_{W V} (\beta , t) \, =\,  1 - \frac 1N e^{\lambda_L t} \ , \qquad \beta \ll t \ll \beta \log N \ ,
\end{align}
where $N$ is the number of degrees of freedom, the system is considered to be chaotic.
The definition of quantum chaos based on \eqref{eq:OTOC} has some very nice features, that have been explored heavily in recent years: it provides a {\it quantum} definition of the Lyapunov exponent, which has been later shown \cite{Maldacena:2015waa} to satisfy the universal bound
\begin{align}
    \label{eq:OTOC_bound}
    \lambda_L \, \leq \, \frac{2 \pi}{\beta} \ ,
\end{align}
saturated by systems which are dual to Einstein gravity --- the so-called maximally chaotic systems.\footnote{However some maximally chaotic systems not directly dual to Einstein gravity have been recently discussed~\cite{Murata:2017rbp,deBoer:2017xdk}.}
Moreover, \eqref{eq:OTOC} provides an interesting time scale --- the {\it scrambling time} --- which can be viewed as the time at which the exponential behavior in \eqref{eq:decay_OTOC} becomes dominant and quantum effects of chaos appear.

Given the two definitions of chaos we just reviewed, it is natural to ask how (and if) they are related to each other.
Of course, finding such a connection would be very important, in order to construct a unique theory of quantum chaos and to find a unique definition of quantum chaotic systems.\footnote{Indeed, a priori there could be systems which are chaotic according to one definition and not chaotic according to the other.}
However, from the way in which they are defined, it is clear that finding a connection between the two definitions is very hard.
From a technical point of view, they use completely different tools and languages: the definition of chaos based on the BGS conjecture uses a statistical approach and statistical tools, while the definition of chaos in terms of the OTOCs is formulated in field theoretical (or quantum mechanical) terms. 
From a conceptual point of view, the two definitions should apply at time scales which are in principle very different from each other: the definition based on the OTOCs is semi-classical and so it is valid at earlier times than the definition based on the BGS conjecture which applies at very late times, when the system has lost completely the memory of the initial dynamics and it can be described by some random hamiltonians which just preserve the symmetry of the original dynamics.

To address this kind of questions, the SYK model~\cite{Sachdev:1992fk,KitaevTalks,Polchinski:2016xgd,Jevicki:2016bwu,Maldacena:2016hyu,Jevicki:2016ito} and its generalizations and variants\footnote{ For example, see the $U(1)$ symmetry and flavor generalizations~\cite{Sachdev:2015efa,Gross:2016kjj,Davison:2016ngz,Bulycheva:2017uqj,Yoon:2017nig,Bhattacharya:2017vaz}, supersymmetric generalizations~\cite{Fu:2016vas,Murugan:2017eto,Yoon:2017gut,Peng:2017spg,Narayan:2017hvh,Bulycheva:2018qcp}, SYK(-like) models without disorder~\cite{Nishinaka:2016nxg,Witten:2016iux,Gurau:2016lzk,Carrozza:2015adg,Klebanov:2016xxf,Peng:2016mxj,Krishnan:2016bvg,Ferrari:2017ryl,Narayan:2017qtw,Chaudhuri:2017vrv,deMelloKoch:2017bvv,Azeyanagi:2017drg,Choudhury:2017tax,Krishnan:2017lra,Prakash:2017hwq,Benedetti:2018goh}, gravity duals~\cite{Maldacena:2016upp,Mandal:2017thl,Das:2017pif,Das:2017hrt,Das:2017wae,Gaikwad:2018dfc,Nayak:2018qej,Forste:2017apw} and the Schwarzian theory~\cite{Stanford:2017thb,Kitaev:2017awl,Qi:2018rqm}.} can be useful toy models: Indeed, one of the main attractive features of SYK is that the OTOC can be computed (exactly in the large $N$ and large $q$ limit and numerically in the large $N$ limit with finite $q$) and shows maximal chaos. 
Moreover, starting from the works \cite{Cotler:2016fpe,Garcia-Garcia:2016mno,You:2016aa}, it has been shown that its spectrum clearly shows RMT features. In particular, \cite{Cotler:2016fpe} has studied the so-called Spectral Form Factor (SFF) for the SYK model, which is nothing but the analytically-continued thermal partition function (we denote with $\beta$ the inverse temperature)
\begin{align}
    \label{eq:SFF_intro}
    g (t, \, \beta)\, \propto\, |\Tr(e^{- (\beta + i t) H_{SYK}})|^2 \ , 
\end{align}
and has shown that for large enough values of the time variable $t$ it shows the same behavior of RMT.
One of the appealing feature of the SFF is that it provides a quantum mechanical flavor to the usual RMT quantities (like the level repulsion and the spectral rigidity) and so it could be in principle closer to the language used in the OTOCs description.

Indeed, some studies addressing the connection between the RMT properties of SYK and the OTOCs appeared in recent days: very recently,  \cite{Gharibyan:2018jrp} has studied the onset of RMT behavior in the (Gaussian filtered) SYK SFF\footnote{
As we will review, the Gaussian filtered SFF is simply the SFF in which the contribution coming from the states in the bulk of the spectrum are {\it exponentially} magnified.
In this way, it has been shown in \cite{Gharibyan:2018jrp} that the onset of RMT behavior can be detected to earlier times than in the standard SFF.} 
and found that, in agreement with the results of \cite{Roberts:2018mnp}, the time-scale at which the RMT behavior appears --- which we will call {\it Thouless time} --- is {\it not} the scrambling time.
Moreover, in \cite{Cotler:2017jue} it has been argued that, while the RMT dynamics is able to capture the relevant physics at very late times, it fails to reproduce the early time features of chaos and that, for example, RMT would scramble almost immediately --- an unwanted feature which would violate the chaos bound \eqref{eq:OTOC_bound}.
All together, these results suggest that the connection between the early-time definition of chaos, based on the OTOCs diagnostics, and the late-time definition of chaos, based on RMT analysis, could be very hard (or impossible) to find --- even in a very particular model like SYK.

In this paper, motivated by the negative results just mentioned, we will try to test a weaker possibility: even though it is by now clear that the scrambling time and the Thouless time are  not the same time-scale --- and even though it is clear that the RMT description is unable to capture aspects of the early-time manifestations of chaos --- it is still possible that the Thouless time and the scrambling time are functionally related, in a perhaps highly non-trivial way, and that systems which are more chaotic according to the OTOCs criterion develop the RMT behavior at earlier times --- i.e. their Thouless time is smaller than the Thouless time of systems which have smaller Lyapunov exponents.

To this purpose, we will focus our attention on a specific variant of the SYK model --- which we will call ``mass-deformed" SYK model --- introduced and studied in \cite{Garcia-Garcia:2017bkg}. 
It is characterized by an hamiltonian that, beyond the quartic interaction term of the standard SYK model, contains an additional quadratic piece, a random mass term
\begin{align}
    \label{eq:hamiltonian_intro}
    H = \sum_{i < j < k < l }^N J_{ijkl} \, \chi^i \, \chi^j \, \chi^k \, \chi^l + i \, \kappa \, \sum_{i < j}^N K_{ij} \, \chi^i \, \chi^ j \ . 
\end{align}
As we will review following \cite{Garcia-Garcia:2017bkg}, this model shows a transition from a chaotic behavior, when the constant $\kappa$ is small enough and so the dynamics is dominated by the quartic chaotic piece, to an integrable behavior, when the constant $\kappa$ is large and the dynamics is dominated by the quadratic, integrable, term.

The transition from chaotic to integrable behavior has been observed, in \cite{Garcia-Garcia:2017bkg}, both by a RMT analysis and by studying numerically (and analytically in the usual large $q$ limit) the large $N$ behavior of the OTOCs.
However, the two diagnostics of chaos do {\it not} see the transition in the same way: the RMT analysis shows the transition from chaos to integrable behavior for values of $\kappa \sim 15$ or larger, while the OTOCs see the transition for values of $\kappa$ which are much lower, around $\kappa \sim 1$.
We will argue that this discrepancy is not just due to the fact that the OTOCs are computed in the large $N$ limit and the RMT analysis is performed at finite $N$.
Instead we will provide arguments that this discrepancy is a consequence of the way in which the chaotic/integrable transition affects the spectrum of the model: indeed, we will see that the integrable transition starts from the low-lying states, i.e. the states closer to the ground state and only after, i.e. for larger values of $\kappa$, it affects the states in the bulk of the spectrum.
In agreement with this pattern, we will conclude that the OTOCs are mostly affected by the tail of the spectrum, i.e. by the states which are closer to the ground state, while they are pretty insensitive to the states in the bulk of the spectrum, and for this reason they see the chaotic/integrable transition for lower values of $\kappa$. 
On the other hand, the RMT analysis is mostly affected by the states residing in the bulk of the spectrum and so it sees the transition at much higher values of $\kappa$ when also the states in the bulk show the transition to the integrable behavior.
We will provide confirmations of this picture both by means of RMT observables depending only on the energy spectrum (level spacing distribution and $\tilde r$ statistics) and by RMT observables depending also on the eigenvectors, like the {\it Inverse Participation Ratios} (IPR).
All these results point towards a clear separation of the chaotic/integrable transition between the low-lying modes and the states in the bulk, as can be  seen in Figure~\ref{fig: ipr average level} which shows how the IPR displays a different behavior between the low-lying states and the states in the bulk of the spectrum.

With the aim of finding a connection between the behavior of the OTOCs and the RMT analysis, and to find a quantity which can provide an estimation of the Thouless time in agreement with the scrambling time, we will look for RMT observables which could be able to capture the same physics of the OTOCs and that, in particular, are able to see the chaotic/integrable transition for lower values of $\kappa$.
To this end, we will test both the Gaussian filtered SFF, as introduced by \cite{Gharibyan:2018jrp}, and the {\it connected unfolded SFF}, which has been introduced in \cite{Garcia-Garcia:2018ruf} and that is simply the SFF computed using the {\it unfolded} spectrum and removing the disconnected piece. 
The Gaussian filtered SFF detects the chaotic/integrable transition at large values of $\kappa$, in agreement with the RMT analysis of \cite{Garcia-Garcia:2017bkg}.
This result is not surprising: by construction the Gaussian filtered SFF magnifies the bulk of the spectrum with respect to the tail and hence it sees the transition at large values of $\kappa$.\footnote{This result can be seen as an additional confirmation that the chaotic/integrable transition affects the energy levels inhomogeneously, starting from the tail and then moving to the bulk.}

On the other hand, we will see that the Thouless time computed using the connected unfolded SFF, especially for low enough values of the temperature, shows a pattern which is qualitatively similar to the pattern seen by the OTOCs. 
In particular, we will see that the chaotic/integrable transition arises for low values of $\kappa$ and that, at $\kappa \sim 1$ the chaotic features mostly disappear.
We will then use the connected unfolded SFF to estimate the Thouless time for different values of $\kappa$ and $N$ and we will see that the chaotic/integrable transition arises around $k \sim 1$ and that it becomes sharper and sharper when $N$ increases.
These features, and the qualitative agreement between the OTOCs and the connected unfolded SFF, suggests that the connected unfolded SFF could be an important observable to study the relation between the early-time chaos diagnostics and the BGS conjecture, in the mass-deformed SYK model as well as in other models. Of course, our study has several technical and conceptual limitations, that we plan to address elsewhere and that we will discuss in the final section of this paper.

The paper is organized as follow. In \textbf{Section~\ref{sec:model}}, we will review the main features of the mass-deformed SYK model, as introduced in \cite{Garcia-Garcia:2017bkg}. 

In \textbf{Section~\ref{sec:thouless}}, we will review the notion of the Thouless time as the time-scale at which the RMT behavior appears. 
We will mention about the differences between this time-scale and the scrambling time and we will introduce our main diagnostic tools: the connected unfolded SFF, as introduced in \cite{Garcia-Garcia:2018ruf}, and the Gaussian-filtered SFF, introduced very recently in \cite{Gharibyan:2018jrp}.

In \textbf{Section~\ref{sec:numerics_preliminary}}, we will present our numerical results for the computation of the connected unfolded SFF. 
We will discuss that it seems to capture a chaotic/integrable transition at {\it small} values of $\kappa \sim 1$ (roughly compatible with the transition measured by the OTOCs) as well as that it suggests that the transition could be {\it temperature-dependent}. 
This second observation leads us to conclude that the chaotic/integrable transition does not affect the spectrum homogeneously, but it mostly affects the states in the tail of the spectrum.

In \textbf{Section~\ref{sec:chaos_energy}}, we will test this possibility by studying three diagnostics of chaos: the level spacing distribution, the Inverse Partecipation Ratios (IPR) and the IR diversity.
The latter two are especially interesting since they do not require any unfolding procedure and, more importantly, because they involve also the energy eigenvectors, and not just the energy spectrum. 
We will see that all these observables point to a clear separation of the chaotic/integrable transitions for the states in the tail of the spectrum and the states in the bulk, with the latter requiring a higher value of $\kappa$ to move to the integrable dynamics.

In \textbf{Section~\ref{sec:thouless_computation}}, we will compute the Thouless time as measured by the connected unfolded SFF. 
We will confirm that the Thouless time exhibits a clear transition for small values of $\kappa$. 
Transition that seems to be sharper and sharper when $N$ increases.

In \textbf{Section~\ref{sec:gaussian_filter}}, we will discuss the behavior of the Gaussian filtered SFF for different values of $N$ and $\kappa$.
We will see, in agreement with the results of \cite{Gharibyan:2018jrp}, that its behavior is, at the level of precision we can achieve, almost independent on $N$.
Moreover, we will see that it detects clearly a transition to the integrable dynamics for {\it large} values of $\kappa$, approximately $\kappa \sim 15$.

In \textbf{the discussion}, we will present our conclusions and prospects of future work. In \textbf{Appendix~\ref{app:r-statistics}}, we will support the evidences of Section \ref{sec:chaos_energy} by computing the $\tilde r$-parameter statistics.
We will also discuss some puzzles related with the unfolding procedure that we hope to clarify in the future.

\section{The model}
\label{sec:model}

We will analyze the following model, introduced and studied in \cite{Garcia-Garcia:2017bkg}, which generalizes the SYK hamiltonian adding a quadratic piece (a so-called random mass term) 
\begin{align}
\label{eq:hamiltonian}
H \,=\, \sum_{i < j < k < l }^N J_{ijkl} \, \chi^i \, \chi^j \, \chi^k \, \chi^l \,+\, i \, \kappa \, \sum_{i < j}^N K_{ij} \, \chi^i \, \chi^ j \ . 
\end{align}
As usual, the $\chi^i$ are $N$ Majorana fermions in $1$ dimension  satisfying $\left\{ \chi^i , \, \chi^j \right\}  = \delta^{ij}$ and  the coupling constants $J_{ijkl}$ are random variables extracted from a Gaussian distribution of mean value $0$ and variance $\frac{6}{N^{3}}$. 
The random mass term contains the antisymmetric tensor $K_{ij}$, whose entries are randomly extracted from a Gaussian of mean value $0$ and variance $\frac 1N$. 
Finally, we have the parameter $\kappa$ which controls the chaotic properties of the model: for $\kappa = 0$ the model is equivalent to the standard SYK model and hence it is maximally chaotic. 
As we mentioned in the Introduction, turning on $\kappa$ the model undergoes a chaotic/integrable transition which, however, is detected differently by the OTOCs and the RMT analysis: it appears around $\kappa \sim 1$ when we consider the OTOCs and around $\kappa \sim 15$ when we consider RMT tools, \cite{Garcia-Garcia:2017bkg}.

Since our goal is to find RMT diagnostics which can be, in principle, sensitive to the scrambling physics --- and hence to the type of chaos that the OTOCs diagnose --- we will look for RMT quantities which are mostly affected by the tail of the spectrum.
To this purpose, we will propose the connected unfolded Spectral Form Factor (SFF) as an interesting RMT quantity to study. 
We will show that it matches more closely the behavior of the OTOCs, even if we stress that a very precise comparison between the OTOCs results and the results we will obtain in this paper would require to extrapolate our analysis to the large $N$ limit, a problem that we would like to address in the future.
Instead, our focus in this paper will be mostly on the behavior of the connected unfolded SFF as a function of $\kappa$.
We will focus most of our computations on the case of $N = 30$, but we will also comment some features and caveats of the $N = 28$ and $N = 32$ cases.\footnote{As we will see, the case of $N = 30$ has a better behavior at small values of $\kappa$ with respect to the cases of $N = 28$ and $N = 32$. 
The reason for this, which will be discussed extensively in Appendix \ref{app:r-statistics}, is that the quadratic deformation in \eqref{eq:hamiltonian} forces the RMT ensemble to fall in the GUE class.
When $N = 30$ also the pure SYK model, with $\kappa = 0$, belongs to GUE, while the cases of $N = 28$ and $N =32$ at $\kappa = 0$ belong to GSE and GOE, respectively. 
In this way, by turning on $\kappa$ we do not break any symmetry and we do not change the RMT ensemble, obtaining a more homogeneous behavior of the Thouless time for small $\kappa$.}

\section{The Thouless time $t_{\text{Th}}$}
\label{sec:thouless}

In the characterization of the chaotic properties of a system via the OTOCs,  the strength of chaos is usually quantified by the Lyapunov exponents: systems with higher values of the Lyapunov exponents are more chaotic than systems with lower (or vanishing) values of the Lyapunov exponents.
Moreover, there is a notion of maximally chaotic systems: they are systems with Lyapunov exponents saturating the chaos bound
\begin{align}
& \lambda_L \, \leq \, \frac{2 \, \pi}{\beta} \ ,
\end{align}
with $\beta$ the inverse temperature.
These quantitative features of chaos are not usually discussed in the RMT approach: in this framework one is interested in studying how much the spectrum of the model under investigation resembles the spectra occuring in RMT. 
Hence, quantities that are usually studied in this approach are the level repulsion and the spectral rigidity, for example; whereas the concept of maximal chaos and Lyapunov exponent (or the associated scrambling time)  are not considered in this case. 
Given the implications that the scrambling time and maximal chaos have in holography, it would be very interesting to introduce these concepts also in the RMT approach to chaos.

Given these motivations, we conjecture that some remnants of the early-time chaos (as diagnosed by OTOCs and Lyapunov exponents) could be analyzed in the RMT regime by a precise characterization of the so-called {\it Thouless time}. 
The Thouless time is defined as the time after which the SFF of the system under study behaves like the SFF of RMT and we conjecture that systems with higher Lyapunov exponents (which are more chaotic and scramble earlier) should show RMT behaviors at earlier times with respect to systems which are less chaotic. 
Hence, our goal is to compute numerically the Thouless time for the model (\ref{eq:hamiltonian}) at different values of the variable $\kappa$, with the aim of testing whether the chaotic/integrable transition as seen by this observable is compatible with the transition seen by the OTOC.

Let us emphasize that, very recently, \cite{Gharibyan:2018jrp} has argued that the time in which the RMT behavior appears is {\it not} the scrambling time, in agreement with the argument of \cite{Roberts:2018mnp}. 
However, we are suggesting the weaker possibility that, even though the time we will measure is not the scrambling time, it could be functionally related to the scrambling time -- in a perhaps highly non-trivial way --- and so it could be a useful indirect probe of the scrambling time and of the scrambling physics.

A way to evaluate the Thouless time has been discussed, for a different model, in \cite{Garcia-Garcia:2018ruf}. 
The computation goes as follows: one computes the {\it connected} piece of the SFF using the {\it unfolded} spectrum. 
Concretely, by connected unfolded SFF we will mean the following quantity:
\begin{align}
    \label{eq:conn_unf_SFF_def}
    g_c(t, \, \beta) \, =\,  \frac{\langle|\sum_i e^{ - (\beta - i \, t) \, \tilde E_i}|^2\rangle}{\langle|\sum_i e^{ - \beta  \, \tilde E_i}|^2\rangle} \, - \, \biggl|\frac{\langle\sum_i e^{ - (\beta - i \, t) \, \tilde E_i}\rangle}{\langle\sum_i e^{ - \beta  \, \tilde E_i}\rangle} \biggr|^2 \ ,
\end{align}
where we denoted by $\{\tilde E_1 , \, \tilde E_2 , \, \tilde E_{2^{N/2}} \}$ the {\it unfolded} energy levels (we will denote the original energy levels with $\{E_1 , \, E_2 , \, E_{2^{N/2}} \}$) and the symbol $\langle \dots \rangle$ denotes the average over many ensemble realizations.\footnote{We caution the reader that in this paper we will use {\it annealed} quantities instead of quenched ones. 
While we know that this procedure is in principle not correct, see for example \cite{Nishinaka:2016nxg}, it has been noticed already in \cite{Cotler:2016fpe} that for numerical evaluations the two procedures are almost equivalent at the level of precision we can achieve.}
The time at which the connected unfolded SFF develops the {\it ramp} typical of RMT is the Thouless time. 

\begin{figure}[t!]
\centering
\includegraphics[width=8cm]{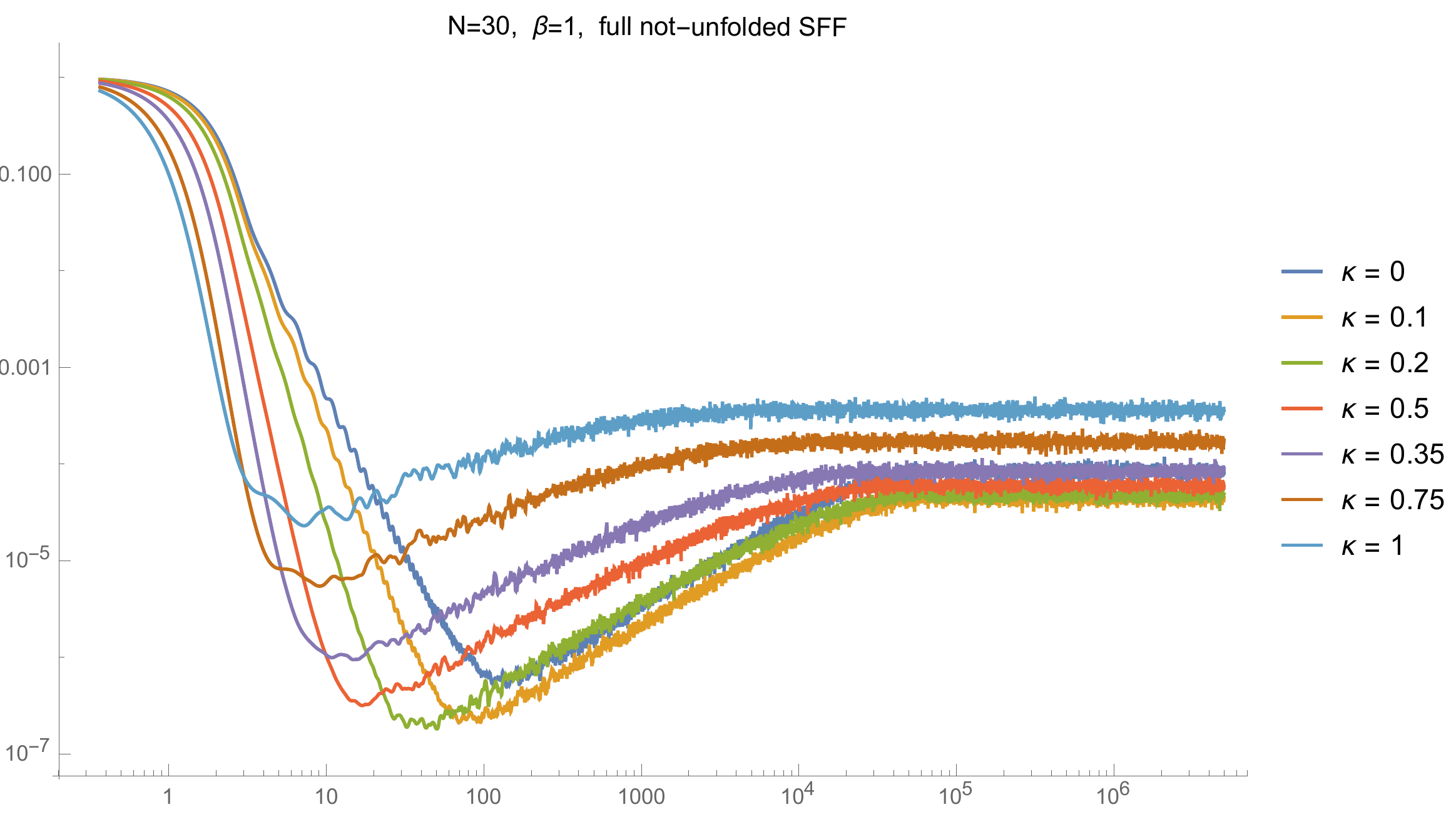}\quad
\includegraphics[width=8cm]{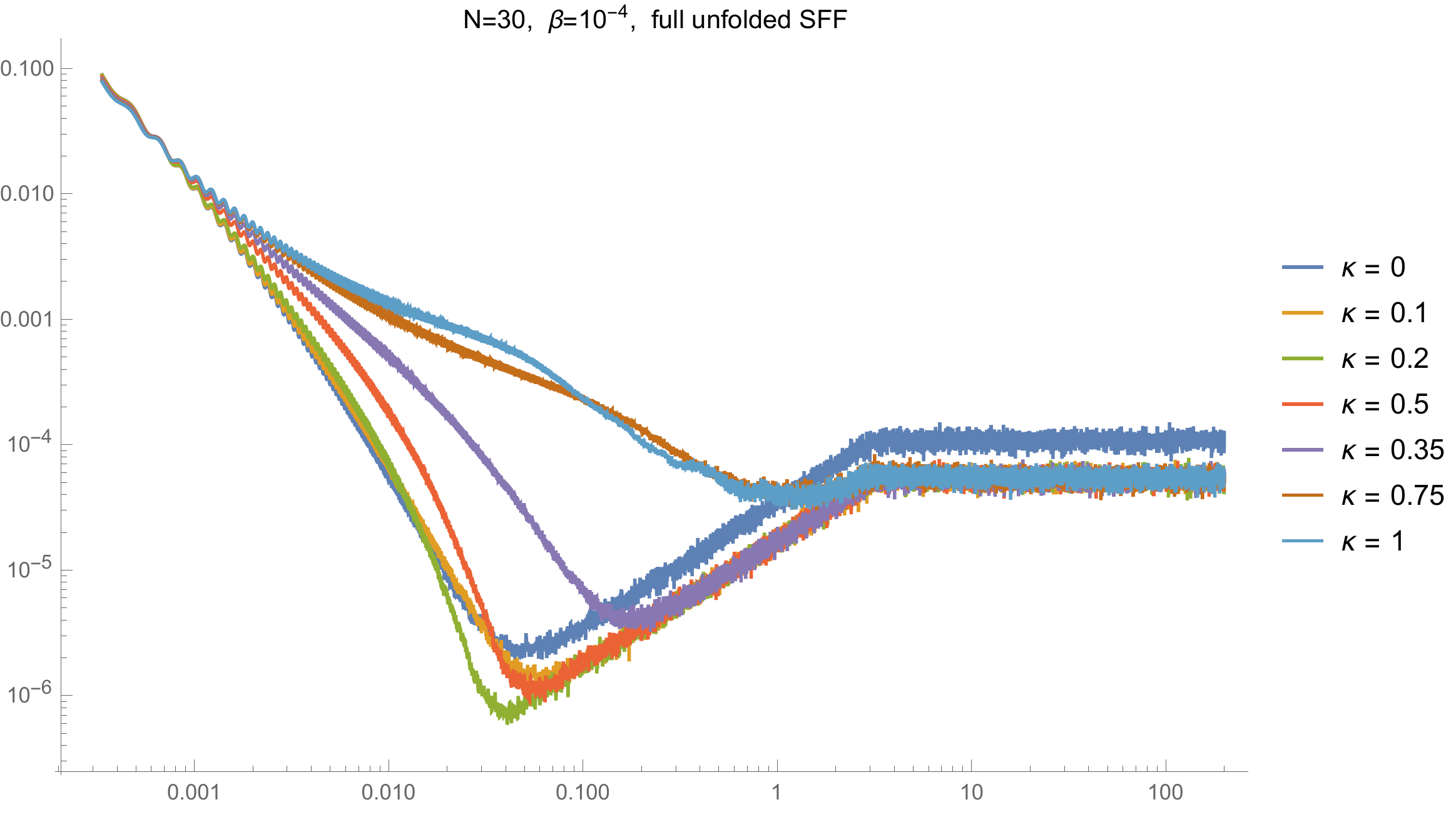}
\caption{Left: the full, not-unfolded, SFF for the case of $N=30$ and for increasing values of $\kappa$.
Right:  the full, unfolded, SFF for the case of $N=30$ and for the same values of $\kappa$.
In both cases, the average is performed over $M = 100$ ensembles.}
\label{fig:fullSFF}
\end{figure}

Let us discuss, via an example, the importance of considering the connected unfolded SFF. 
In Figure \ref{fig:fullSFF} we report the {\it full} SFFs, computed for the case of $N = 30$ and with increasing values of $\kappa$, before and after the unfolding procedure.
We recognize that, without doing the unfolding, the time at which the ramp starts to appear is {\it decreasing} as a function of $\kappa$ and this is not the behavior we expect. 
The models with higher values of $\kappa$ are less chaotic (as expected) because the ramp are more horizontal, but it would be hard to relate this feature to the Thouless time.
After performing the unfolding, we see that the time at which the ramp appears is closer to be an increasing function of $\kappa$ (as expected) and also the ramps show the same slope but, still, at low values of $\kappa$ it is not possible to find a monotonically increasing trend.

The necessity of considering the connected piece of the SFF is easy to understand: as already discussed in \cite{Cotler:2016fpe}, the full SFF is dominated by the disconnected piece at early times, which has a slow fall-off (it goes like $1 / t^3$ in SYK) and ``hides'' the ramp behavior of the connected SFF.
The typical way to cure this problem simply consists in removing the disconnected piece of the SFF and consider the connected piece and this is the strategy we will follow in this paper.
 
On the other hand, it has been suggested in \cite{Gharibyan:2018jrp} that, in many-body quantum systems like SYK, removing the disconnected piece could be not enough: for this kind of systems, also the connected SFF shows a fall-off which goes like $1 / t^3$ and, hence, it still hides part of the ramp behavior.
For this reasons, a different observable has been proposed \cite[eq. (19)]{Gharibyan:2018jrp}, which can be viewed as a ``Gaussian filtered'' SFF
\begin{align}
\label{eq:gaussianfilter}
|Y(\alpha, \, t)|^2 \, =\,  \frac{\langle|\sum_i e^{- \alpha \, E_i^2 - it  E_i}|^2\rangle}{\langle\bigl(\sum_i e^{- \alpha \, E_i^2 }\bigr)^2\rangle} \ ,
\end{align}
with $\alpha$ a free parameter and without unfolding the spectrum.
We can intuitively think about this observable as the SFF computed for a spectrum in which the density of the states $\rho (E)$ has been rescaled to $e^{- \alpha \, E^2}  \rho (E)$ and it is immediate to see that such a rescaling increases the importance of the states of the bulk of the spectrum (we are assuming that the spectrum is centered at $E  = 0$) with respect to the states in the tail. 
They show that such an observable is more suitable for many-body chaotic systems like SYK, in the sense that it is able to detect the onset of RMT behavior at earlier times than the connected SFF.
Of course, this observable by construction is almost insensitive to changes of the spectrum in the tail but it has the nice feature that it is not necessary to unfold the spectrum.\footnote{As we will discuss in Section \ref{sec: discussion} the unfolding is a very tricky procedure: many RMT observables are dependent on the unfolding and it is not always clear how to perform an ``honest" unfolding, see for example \cite{Gharibyan:2018jrp} and \cite{Flores:2000ew}.}

In Section \ref{sec:gaussian_filter}, we will compare the results, for the values of $\kappa$ we will probe, for both the connected unfolded SFF and for the Gaussian filtered SFF and we will give an explanation of their different behaviors as functions of $\kappa$.

\section{Numerical results on the connected unfolded SFF}
\label{sec:numerics_preliminary}

Following the suggestion outlined in \cite[App.~A]{Gharibyan:2018jrp} we used a fifth order polynomial fitting function to unfold each ensemble realization separately.

\begin{figure}[t!]
\begin{center}
\fbox{\includegraphics[width=8cm]{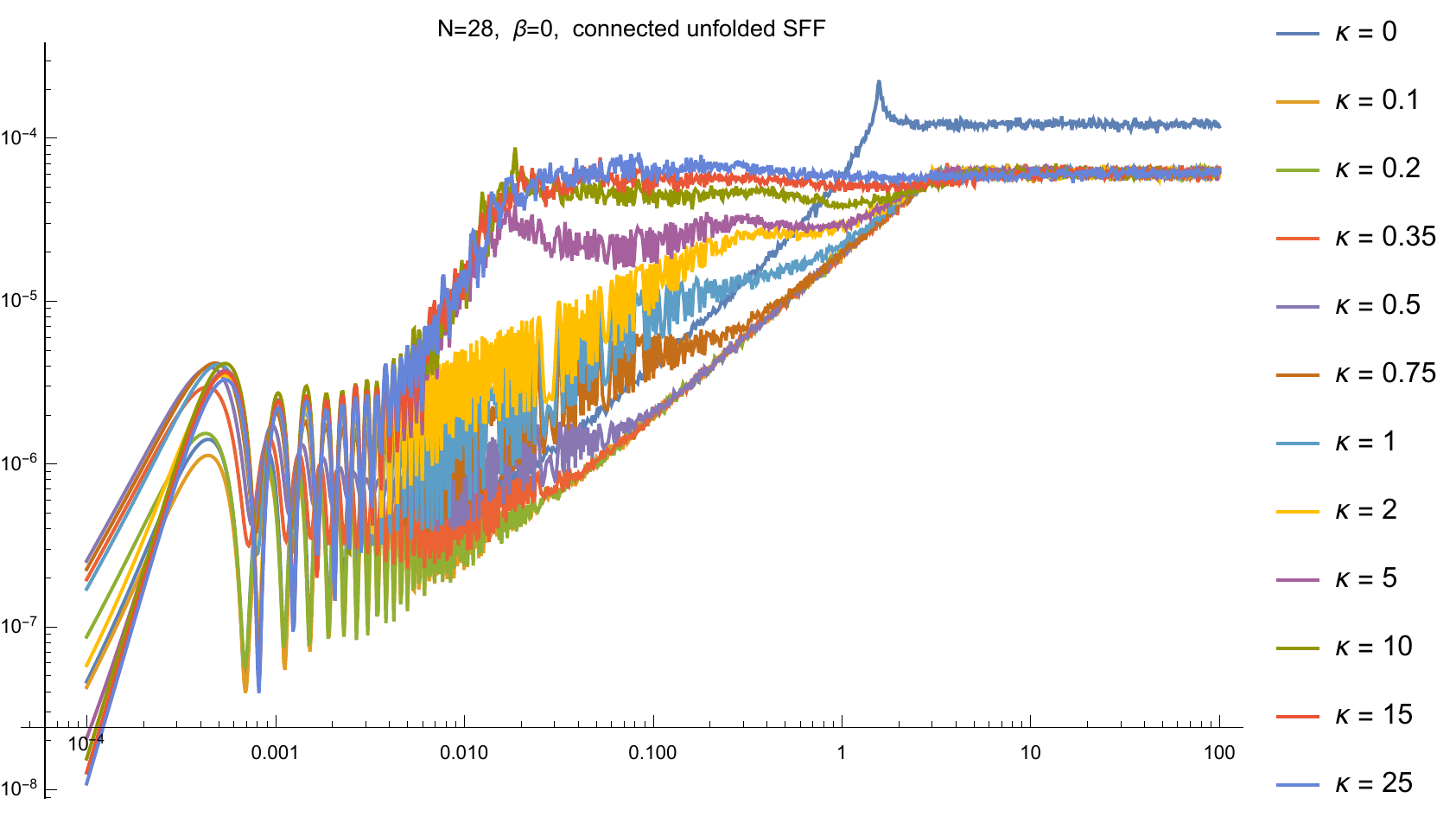}}\quad
\fbox{\includegraphics[width=8cm]{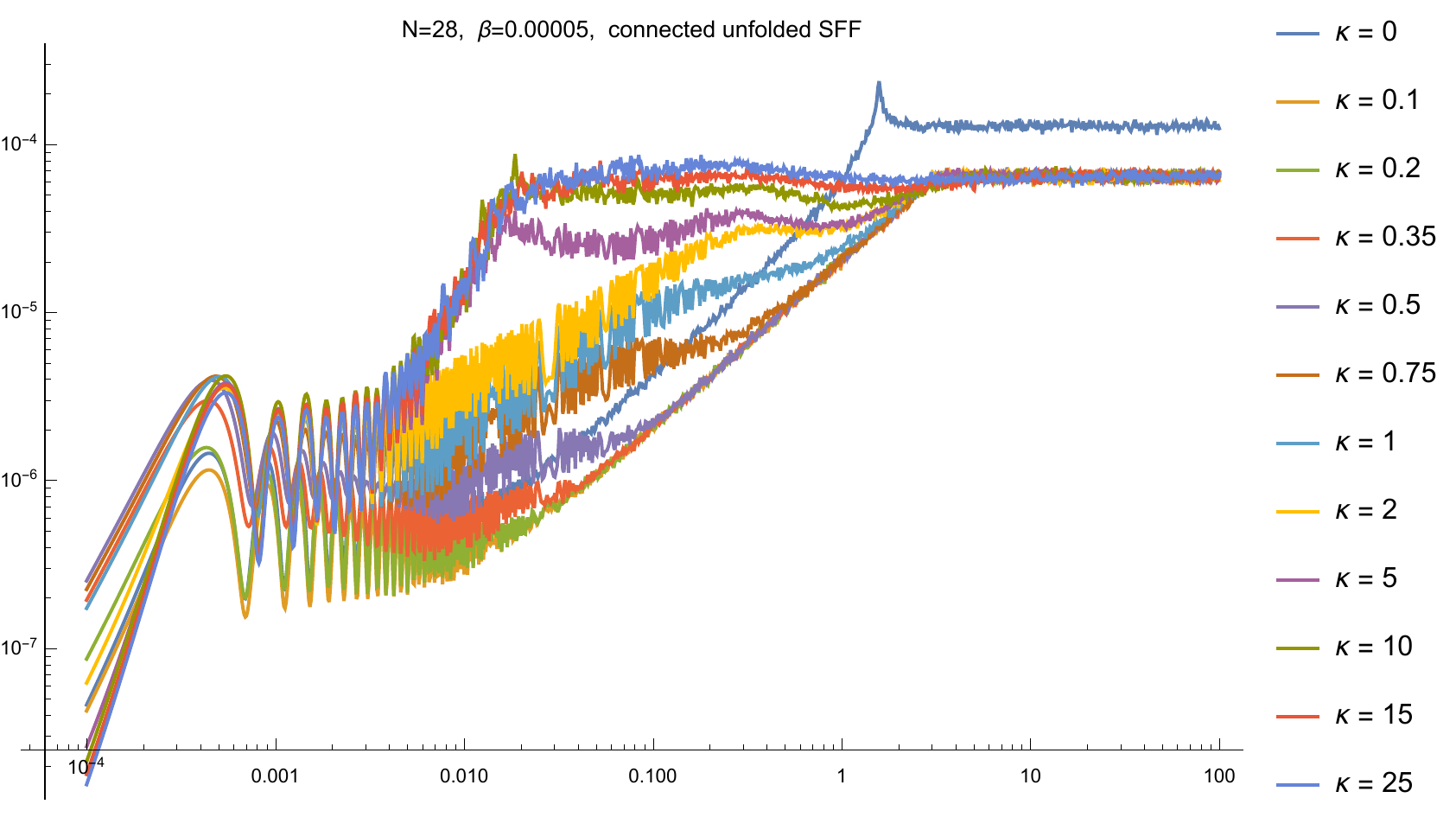}}\\
\fbox{\includegraphics[width=8cm]{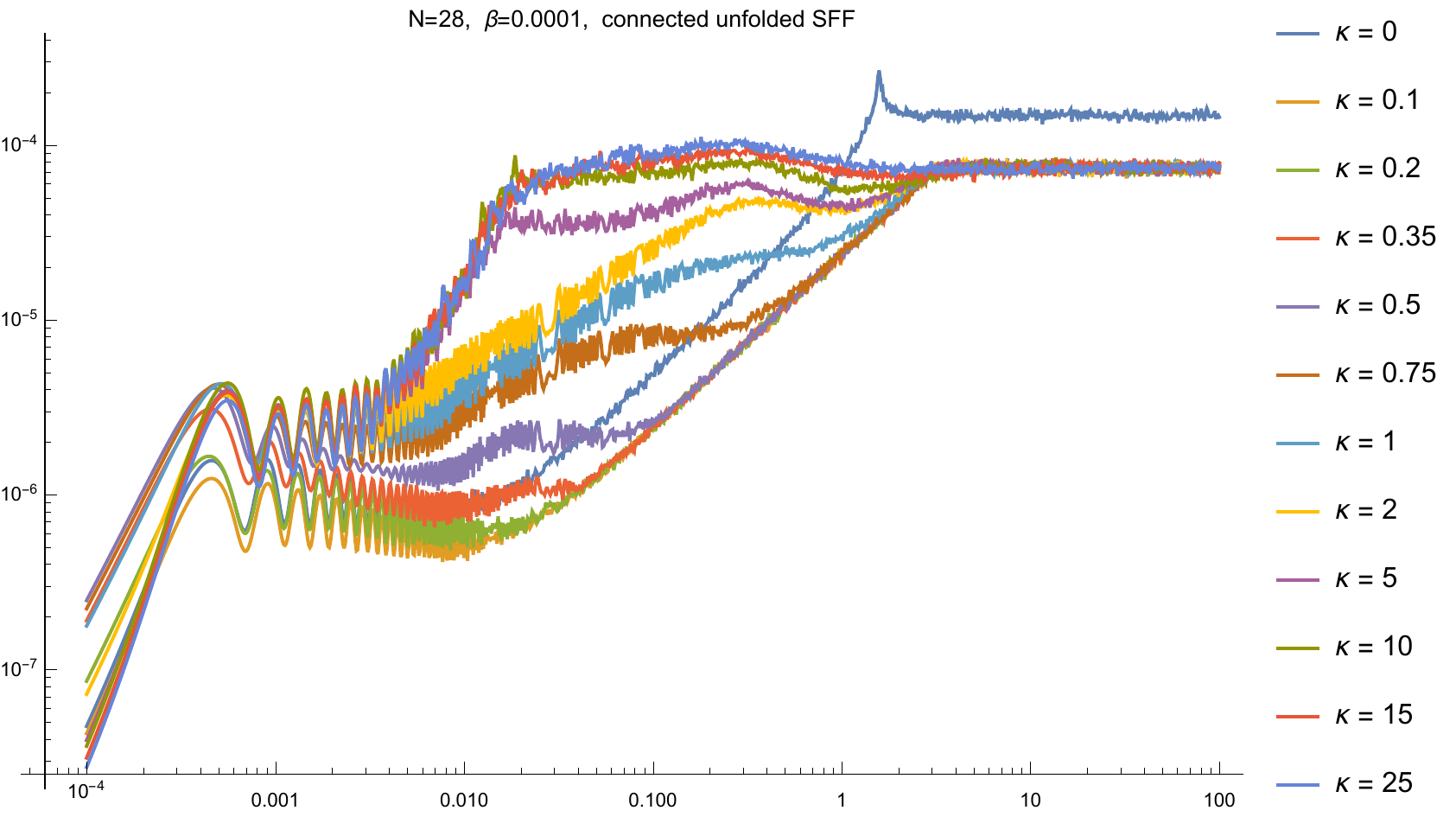}}\quad
\fbox{\includegraphics[width=8cm]{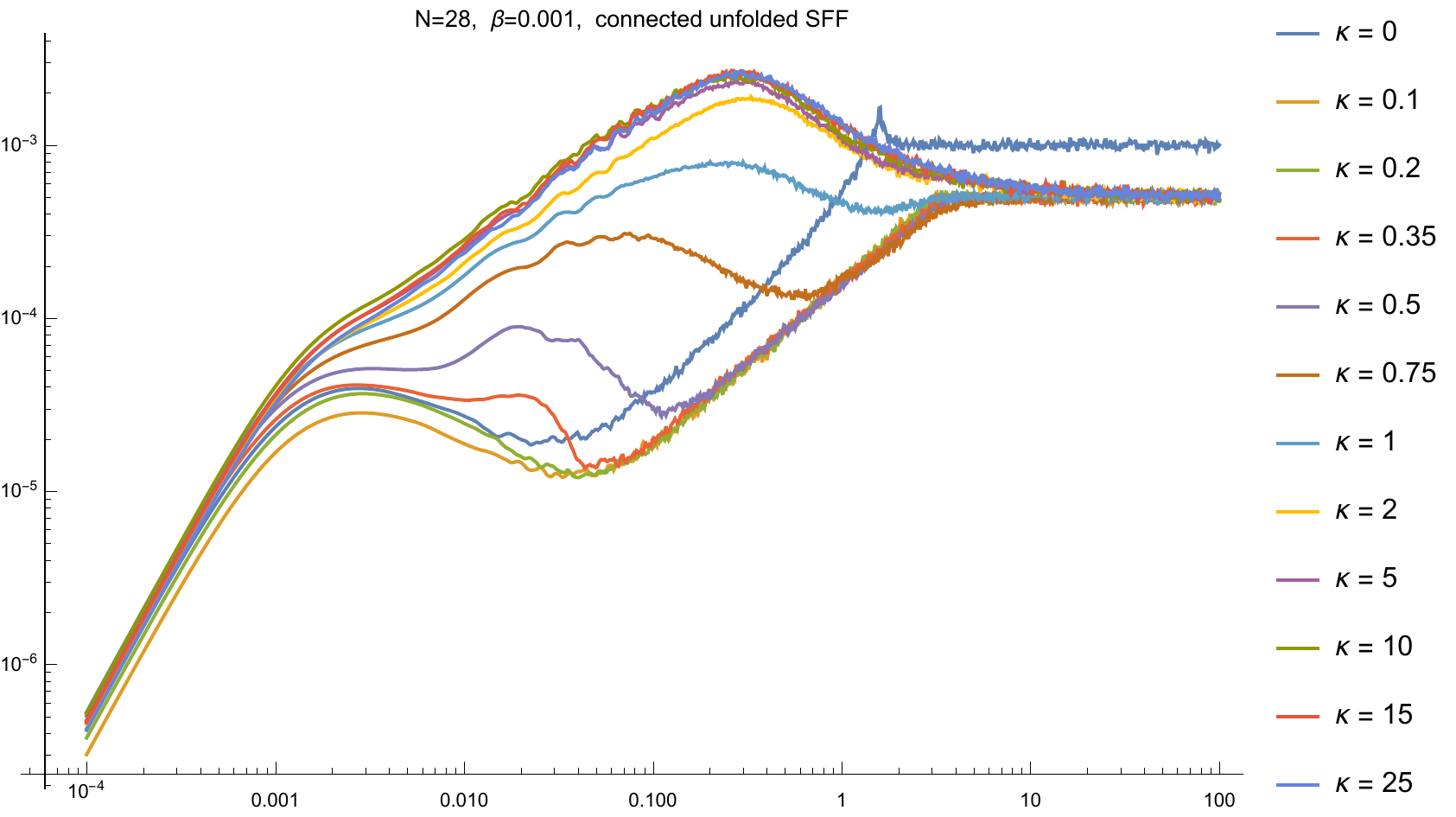}}\\
\fbox{\includegraphics[width=8cm]{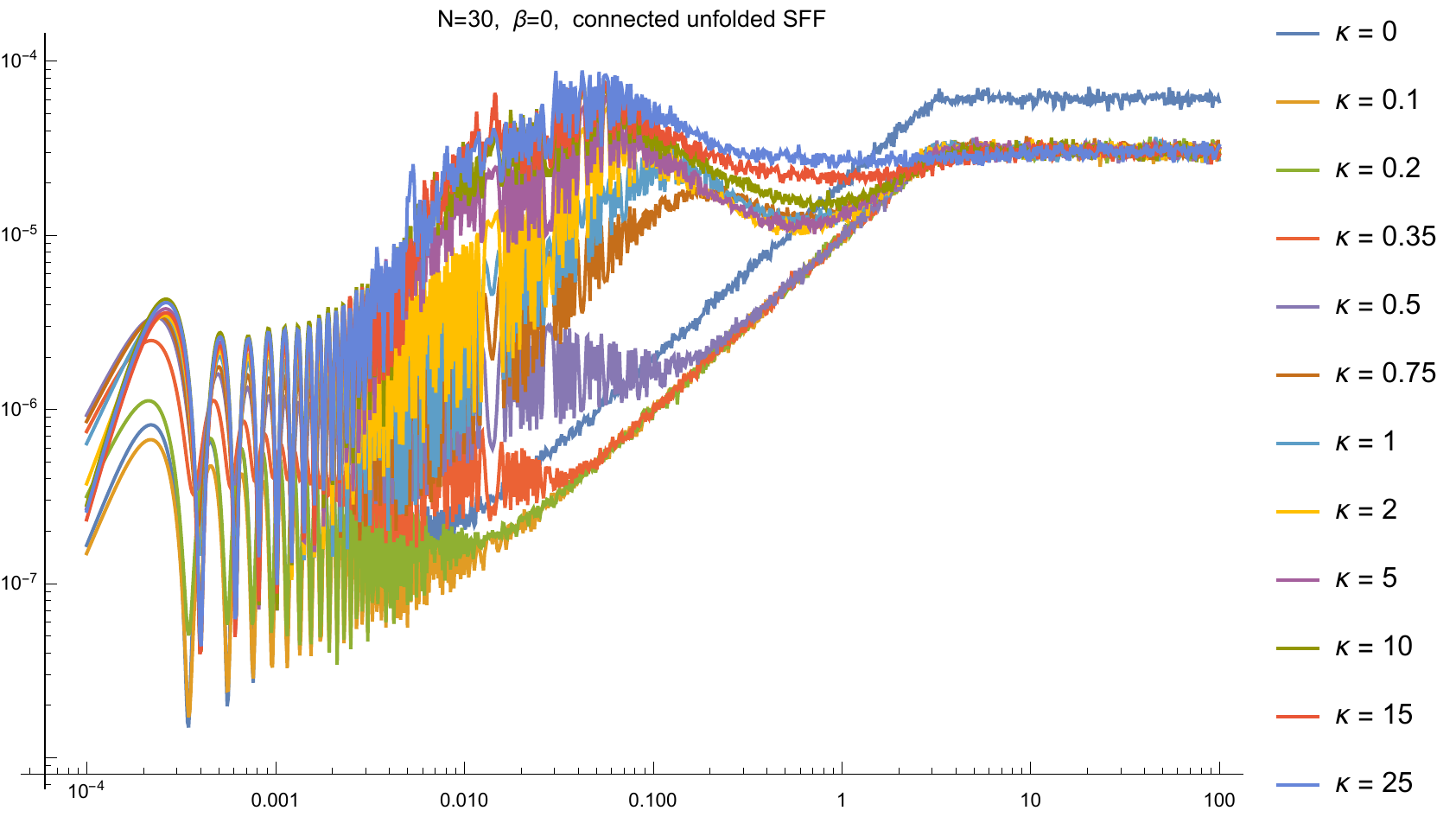}}\quad
\fbox{\includegraphics[width=8cm]{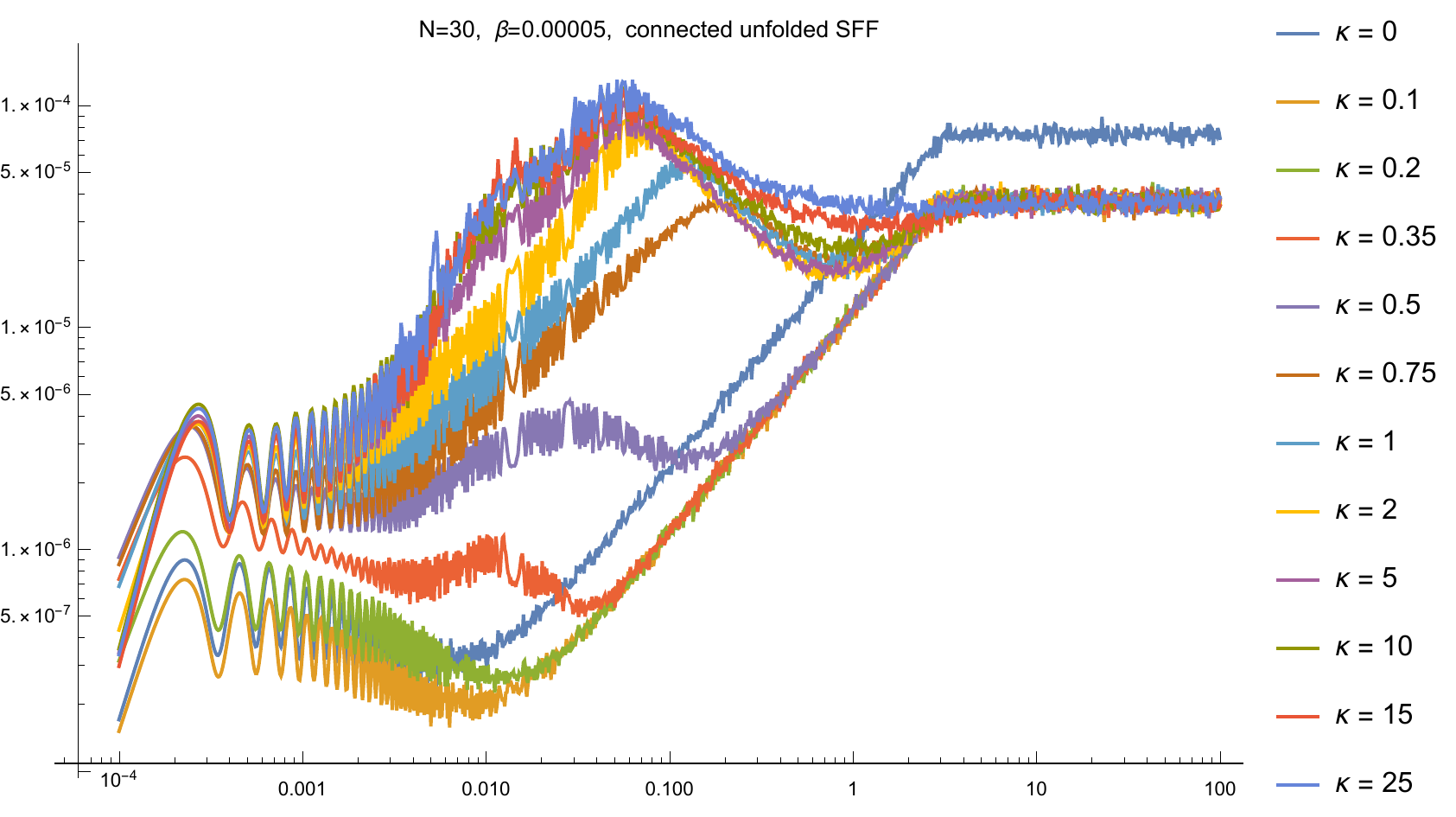}}\\
\fbox{\includegraphics[width=8cm]{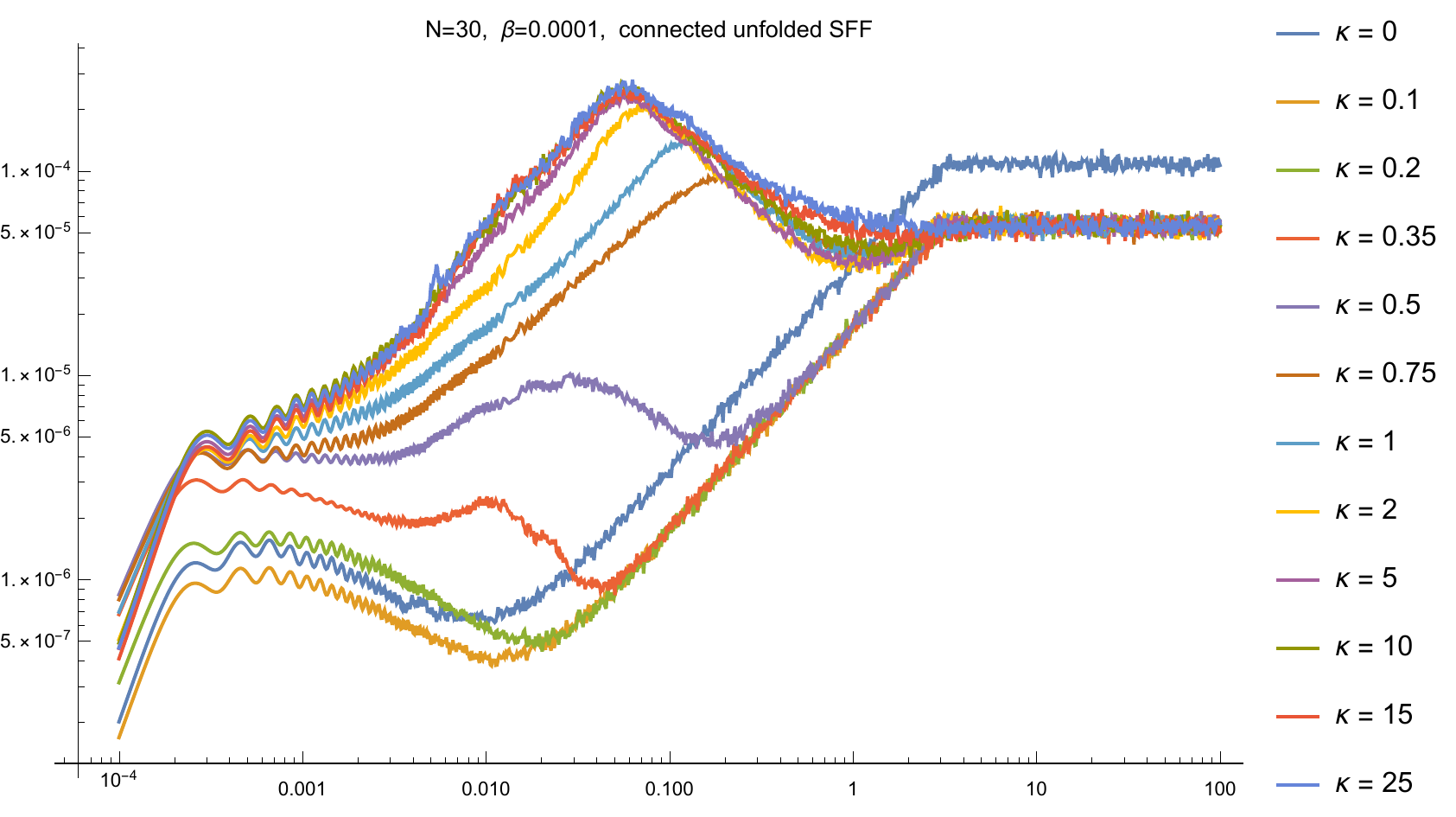}}\quad
\fbox{\includegraphics[width=8cm]{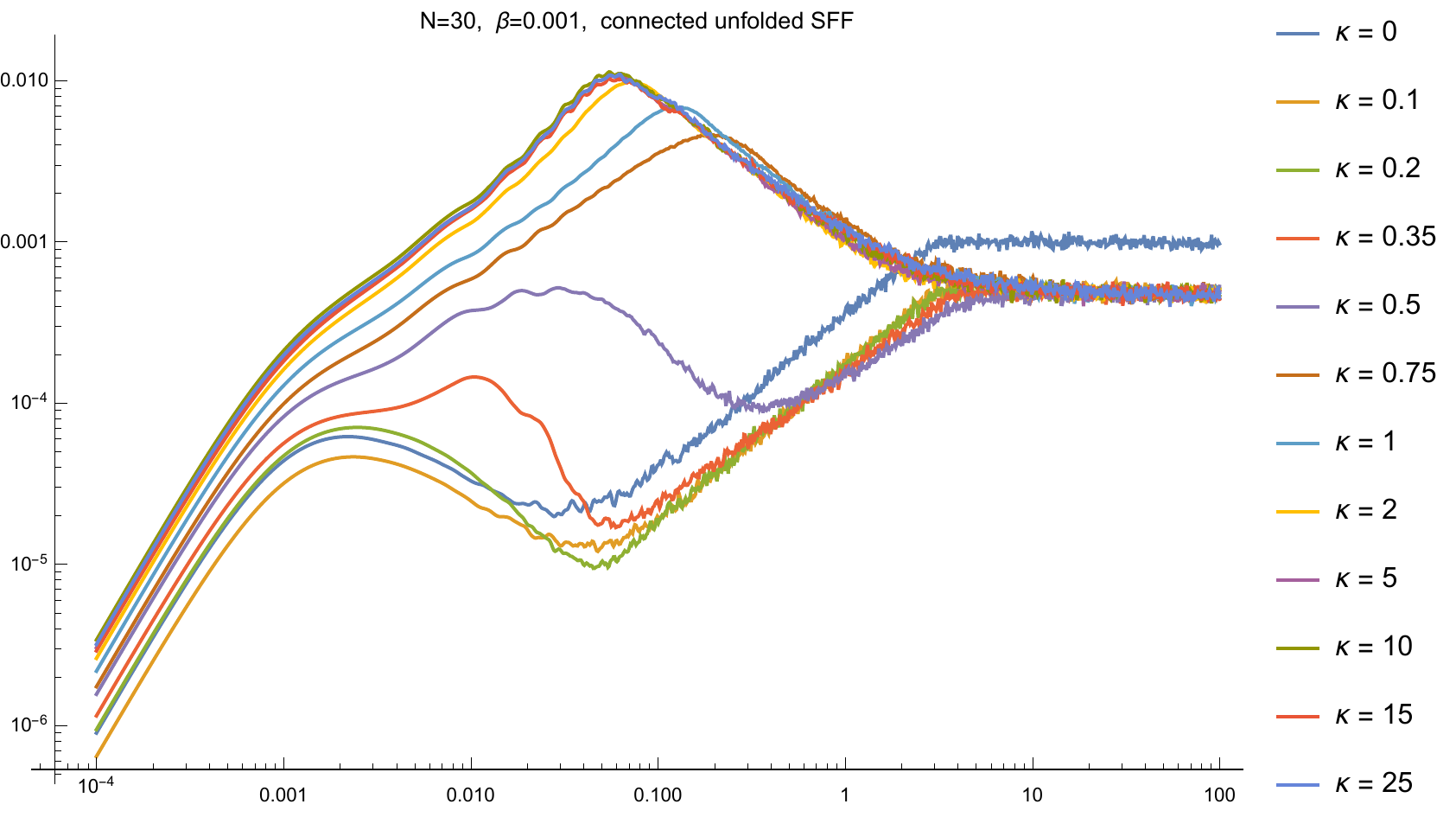}}
\end{center}
\caption{
The connected unfolded SFF for $N=28, \, 30$ and $\beta=0, \, 5\times 10^{-5}, \, 10^{-4}, \, 10^{-3}$.
The average is performed over $M=603$ realizations for $N=28$ and $M=312$ realizations for $N=30$ respectively.
}
\label{fig:beta0}
\end{figure}

\begin{figure}[t!]
\begin{center}
\fbox{\includegraphics[width=8cm]{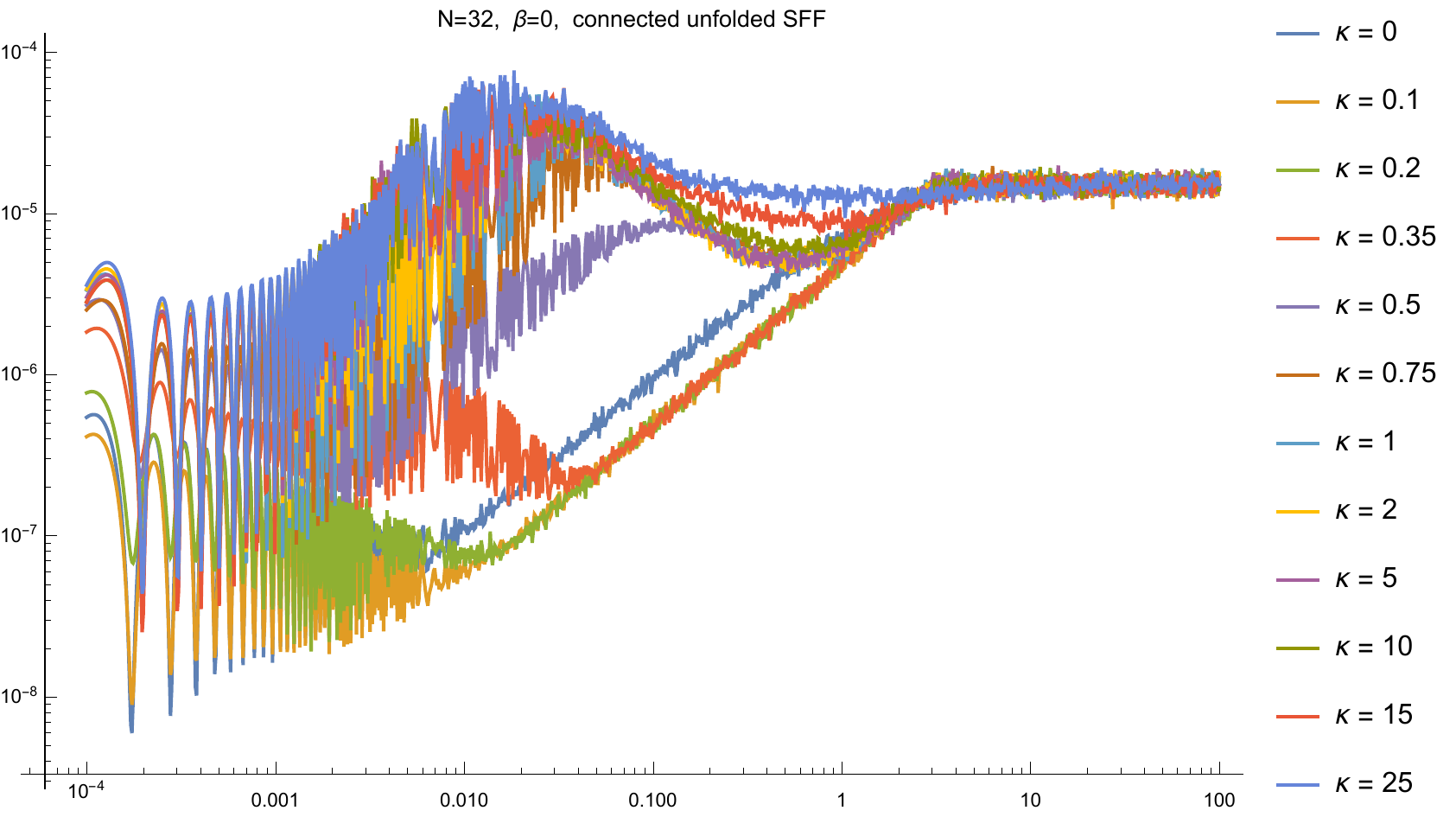}}\quad
\fbox{\includegraphics[width=8cm]{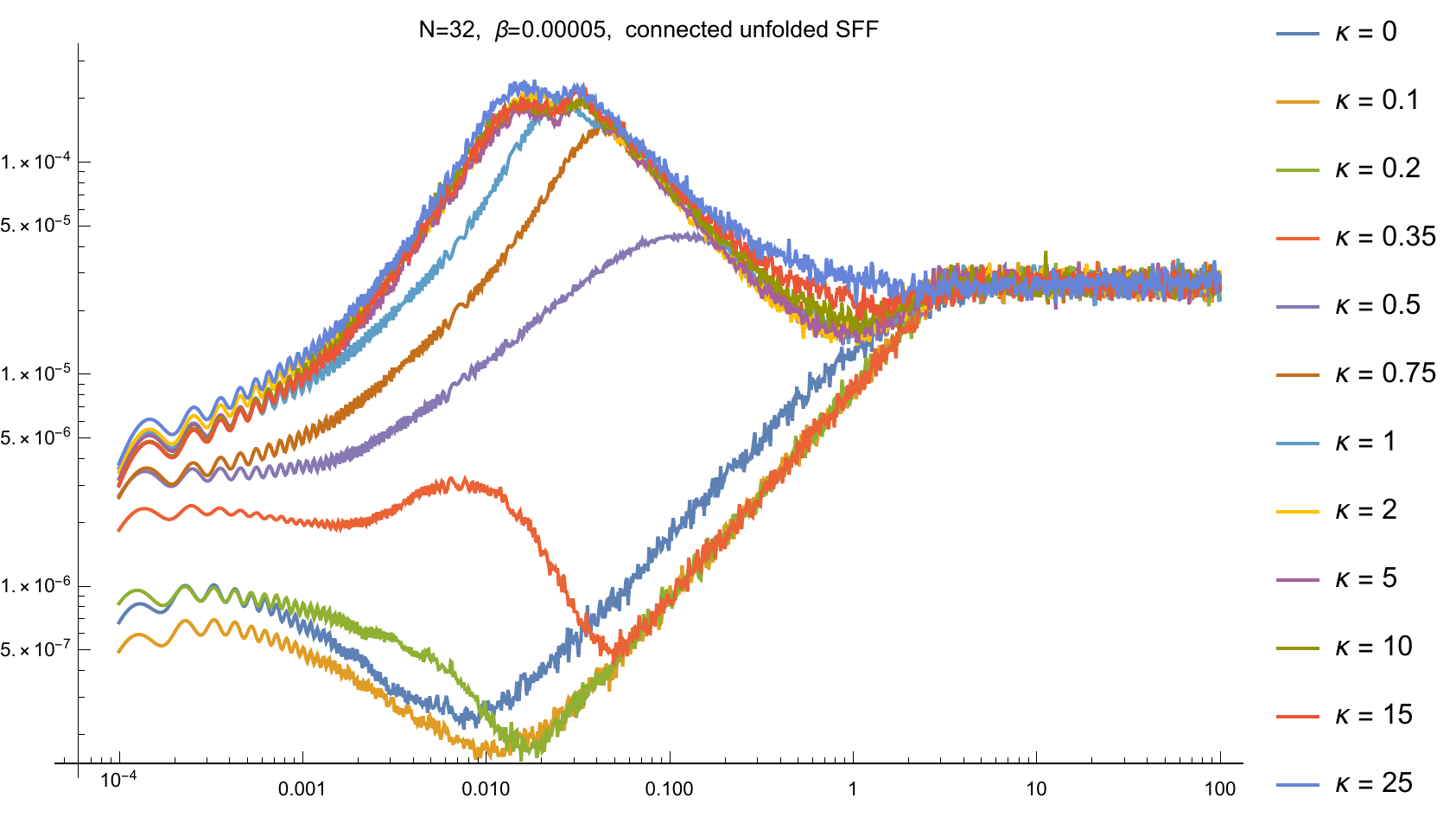}}\\
\fbox{\includegraphics[width=8cm]{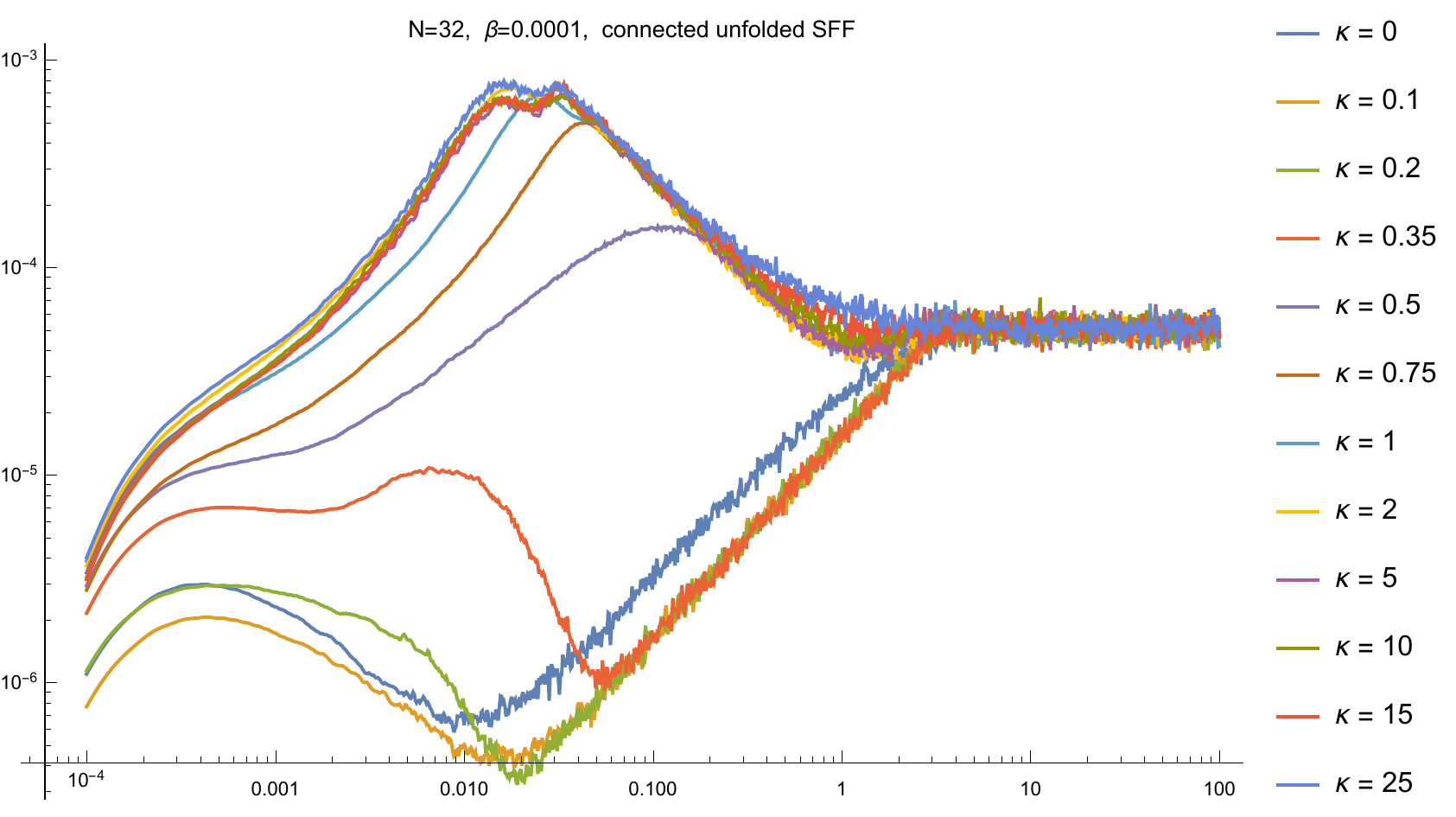}}\quad
\fbox{\includegraphics[width=8cm]{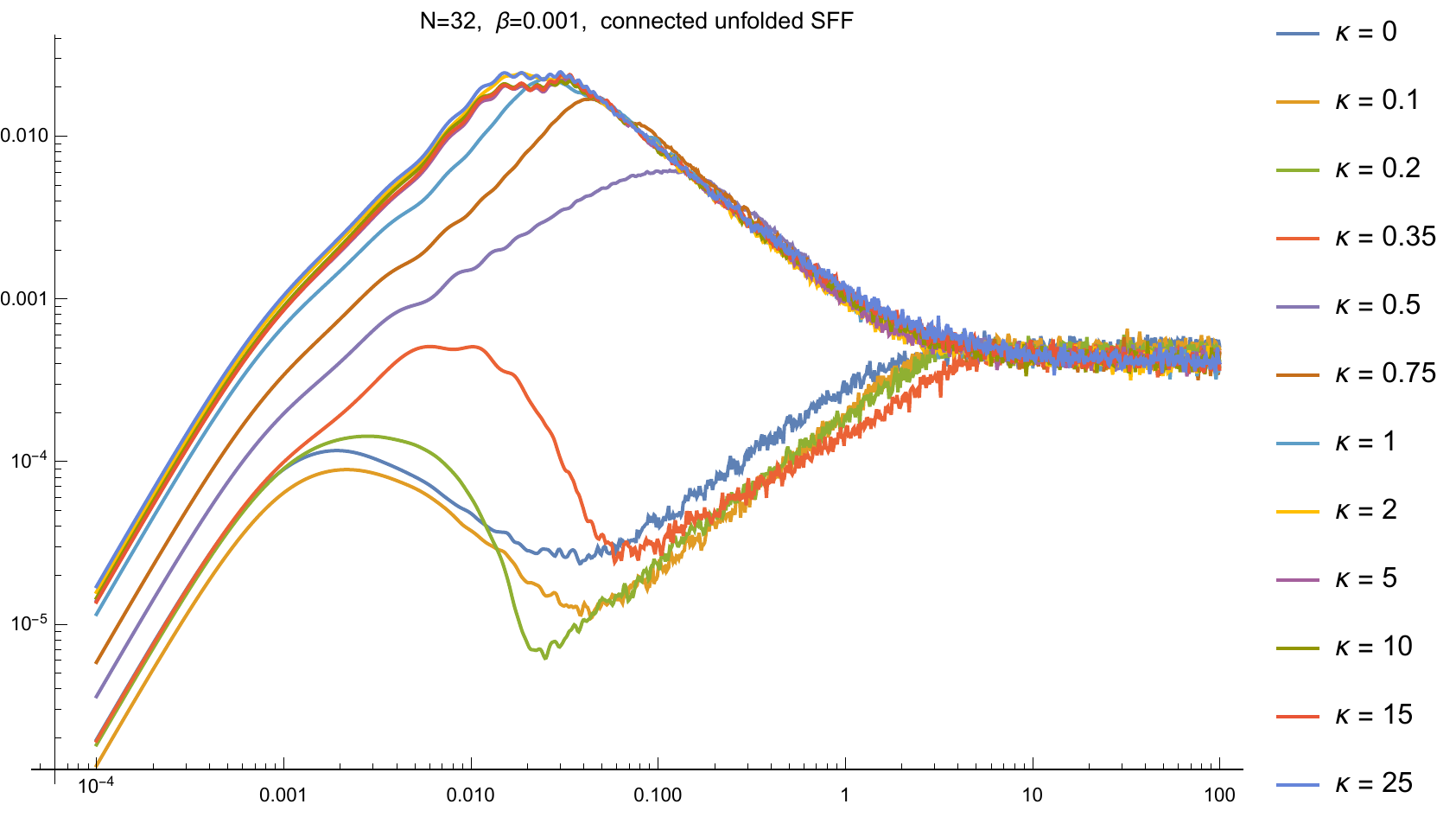}}
\end{center}
\caption{
The connected unfolded SFF for $N=32$ and $\beta=0, \, 5\times 10^{-5}, \, 10^{-4}, \, 10^{-3}$.
The average is performed over $M=150$ realizations.
}
\label{fig:counSFFN32}
\end{figure}

\begin{figure}[t!]
\fbox{\includegraphics[width=8cm]{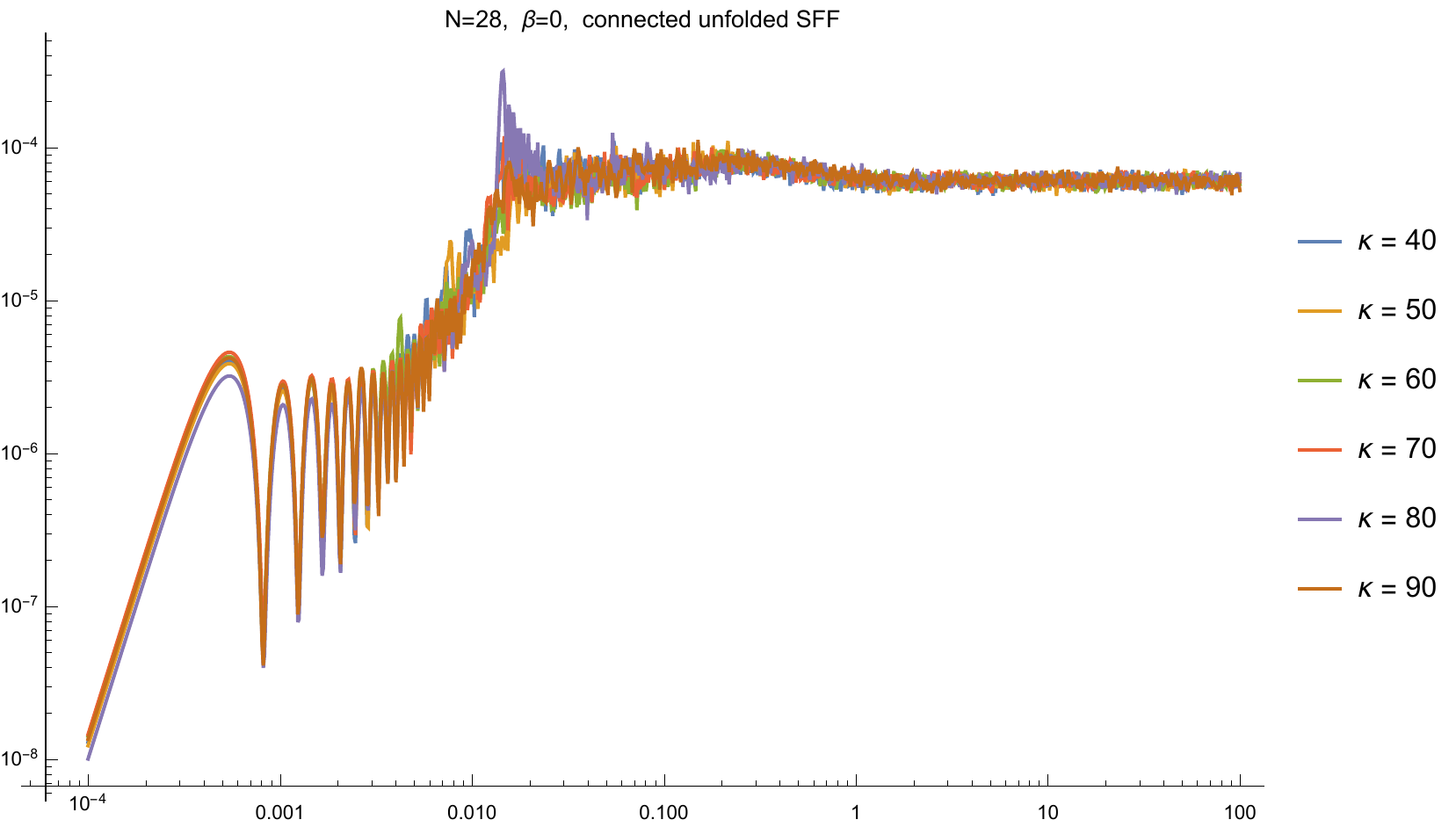}}\quad
\fbox{\includegraphics[width=8cm]{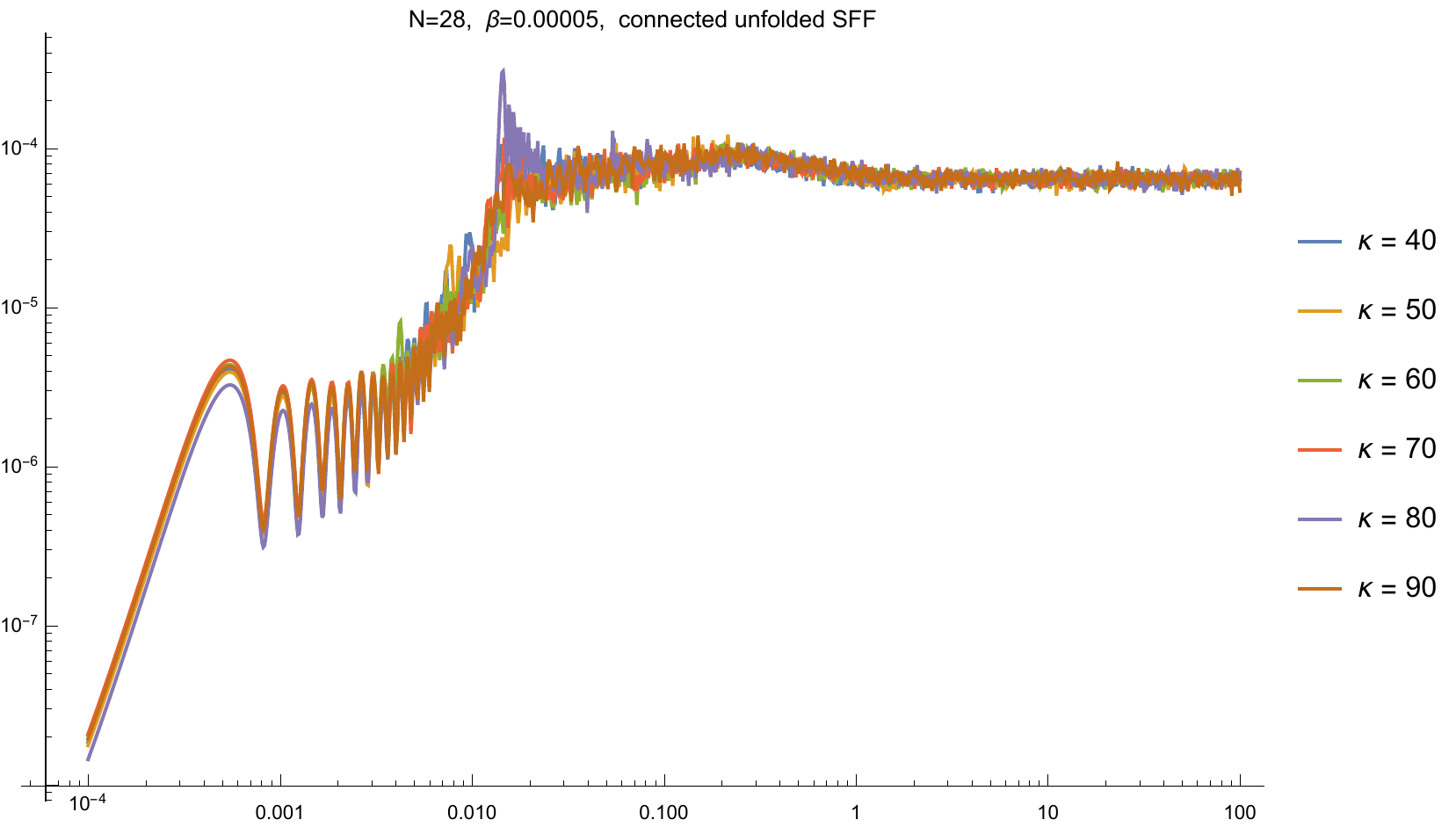}}\\
\fbox{\includegraphics[width=8cm]{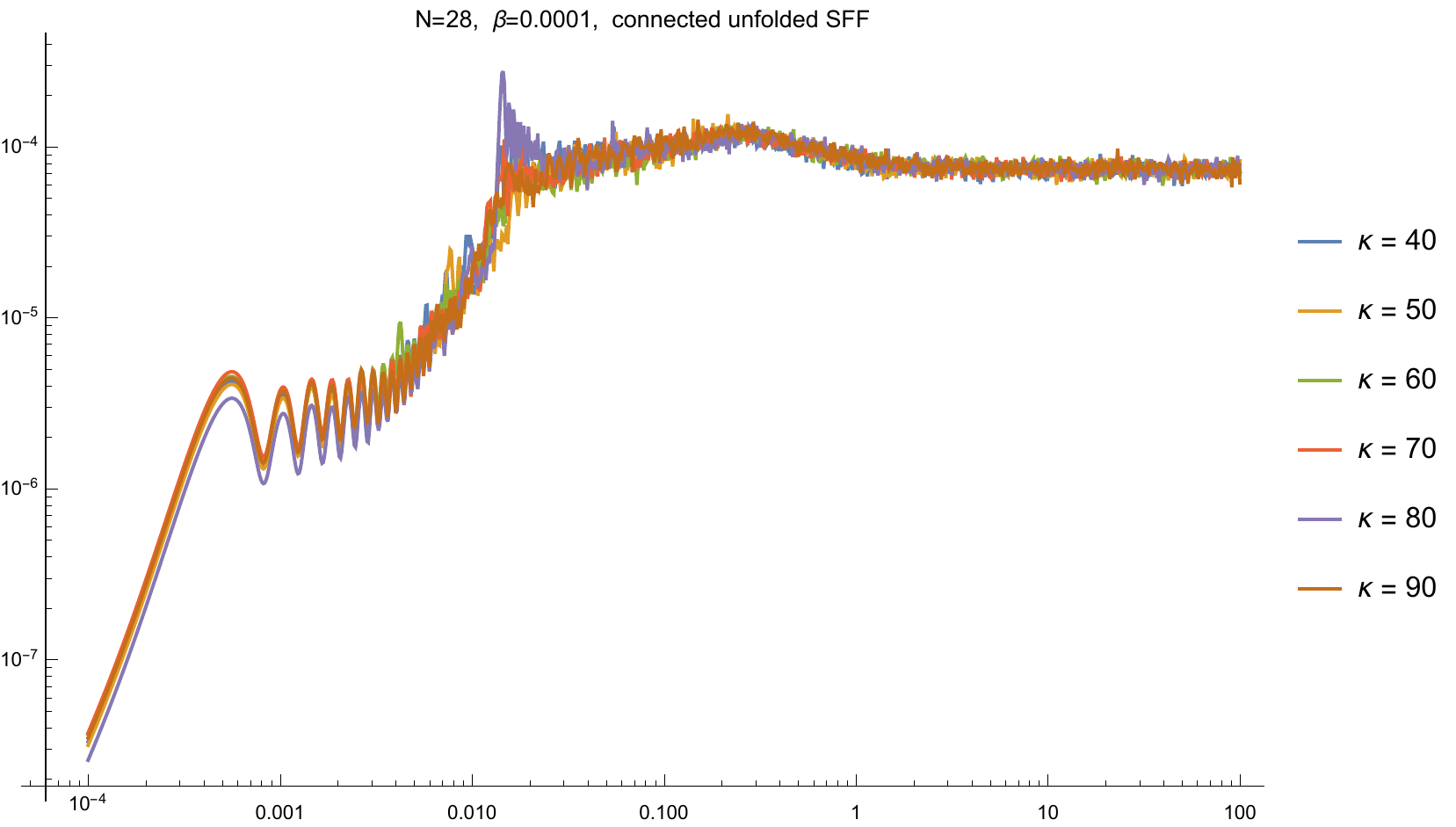}}\quad
\fbox{\includegraphics[width=8cm]{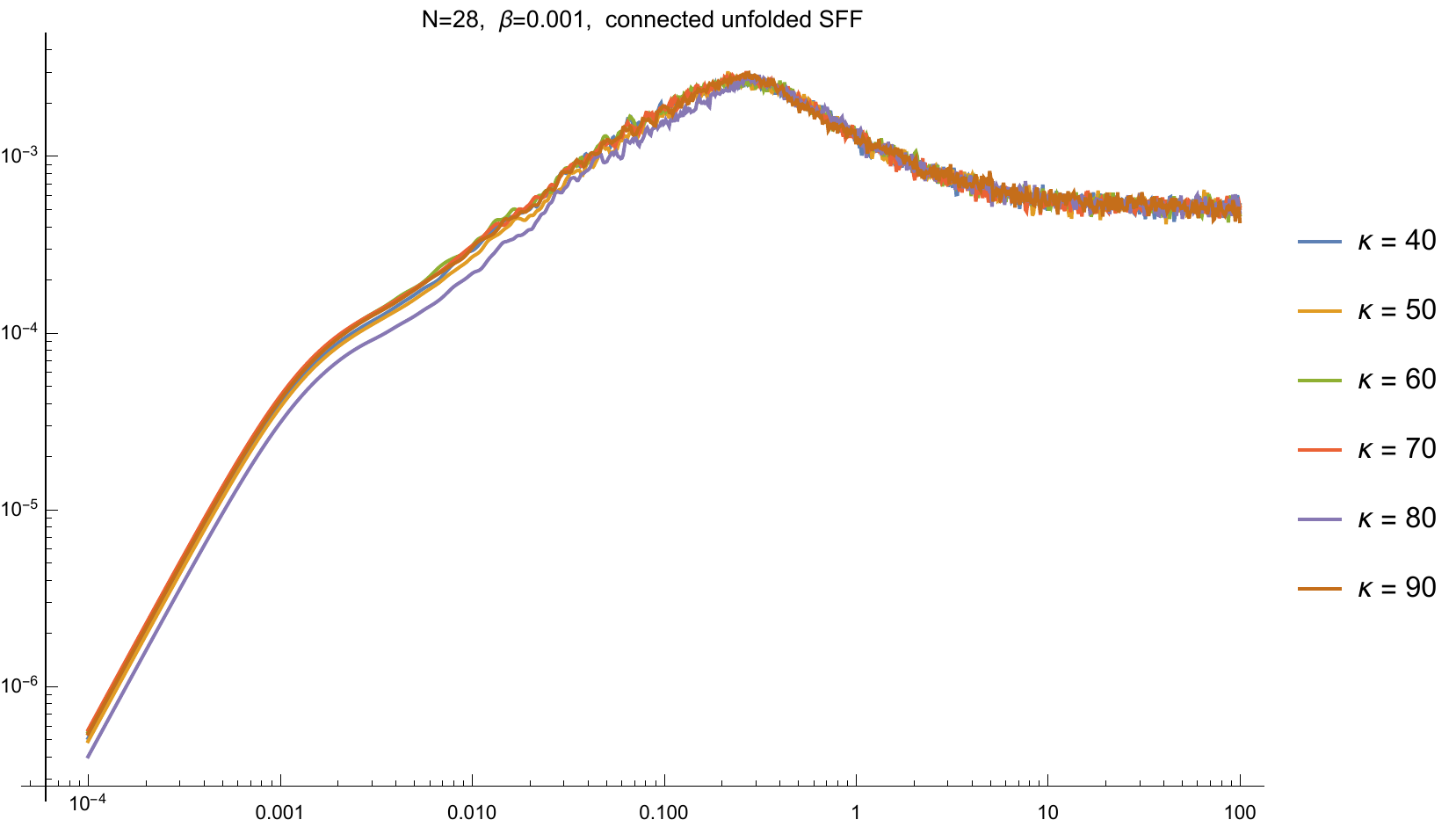}}\\
\fbox{\includegraphics[width=8cm]{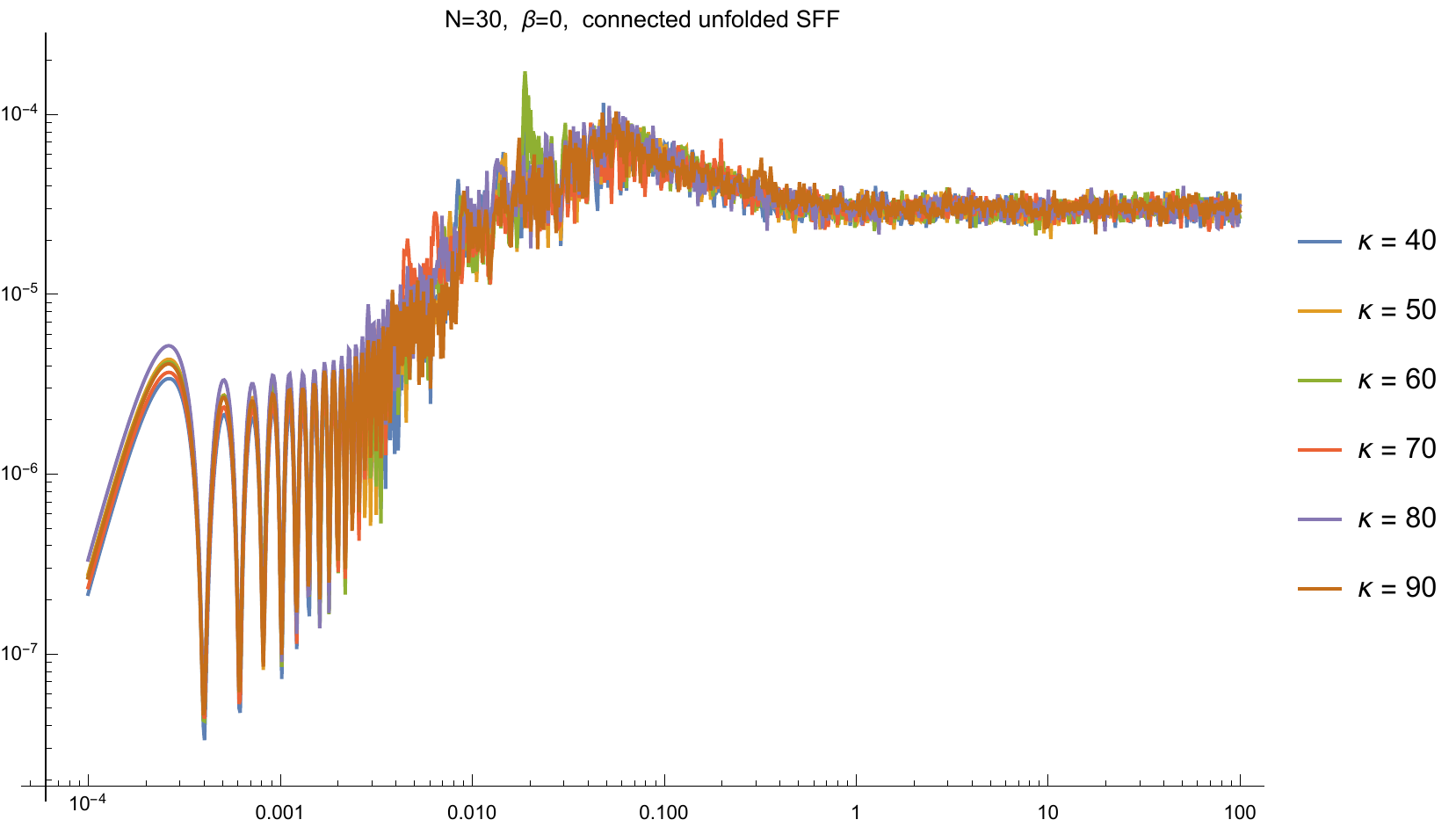}}\quad
\fbox{\includegraphics[width=8cm]{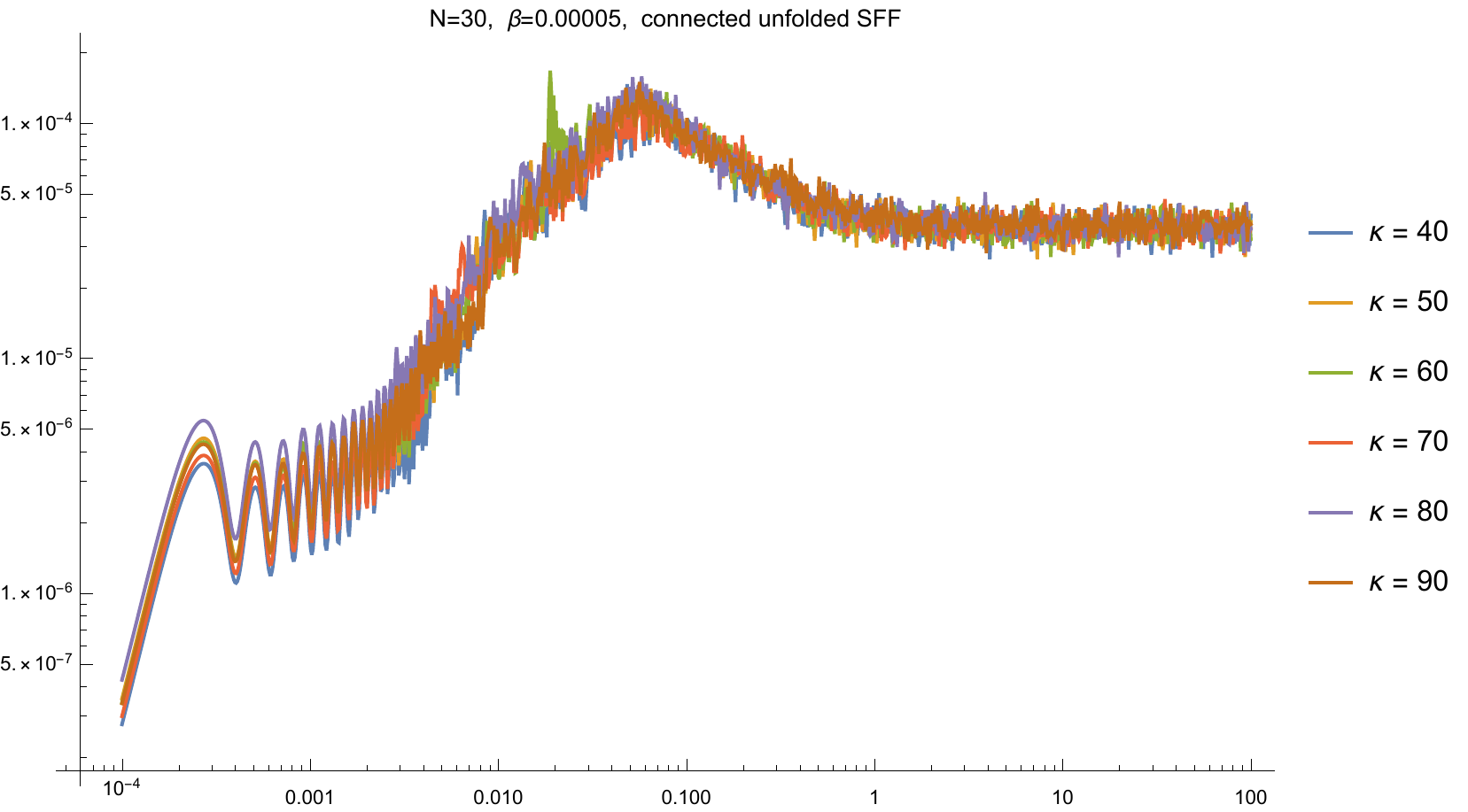}}\\
\fbox{\includegraphics[width=8cm]{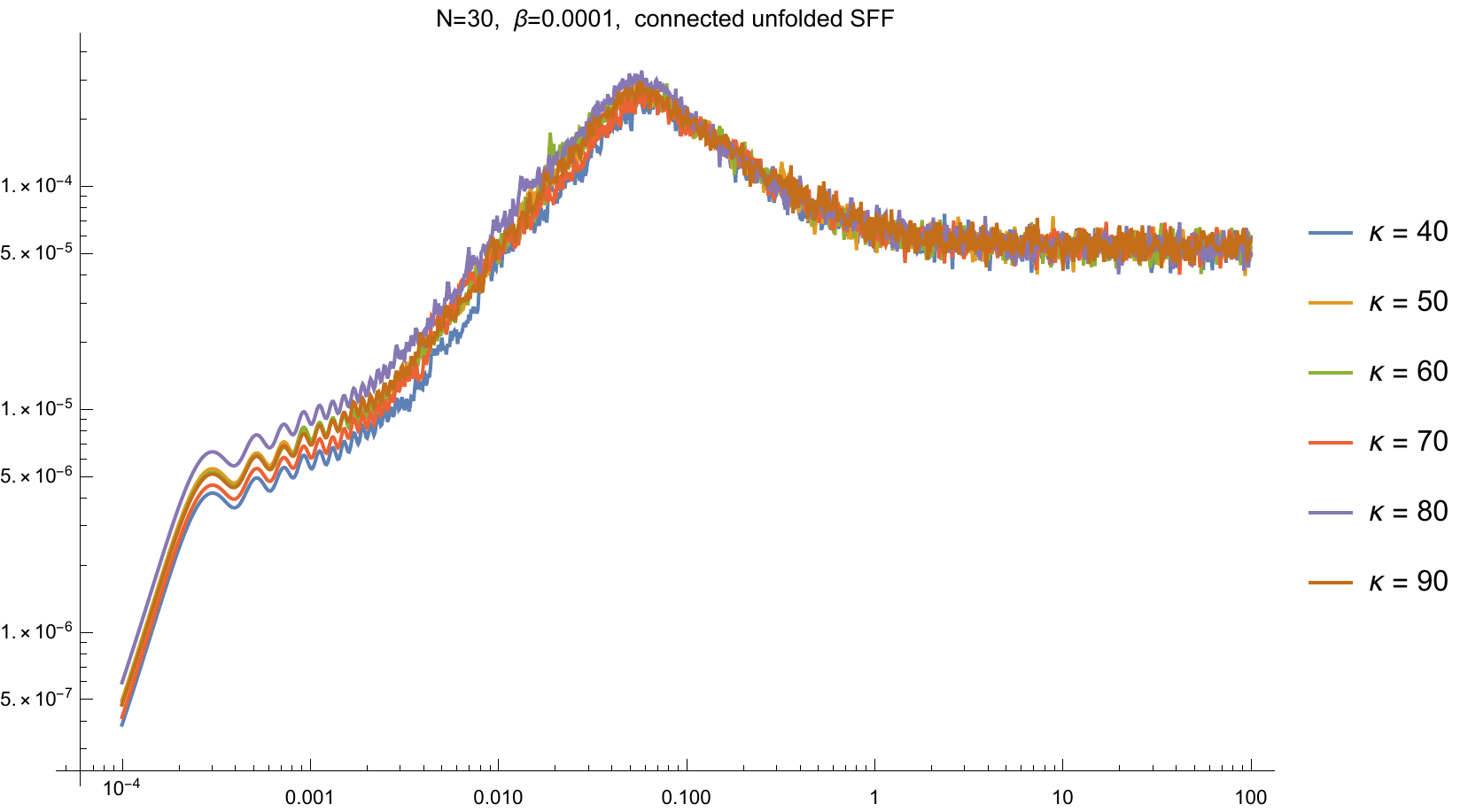}}\quad
\fbox{\includegraphics[width=8cm]{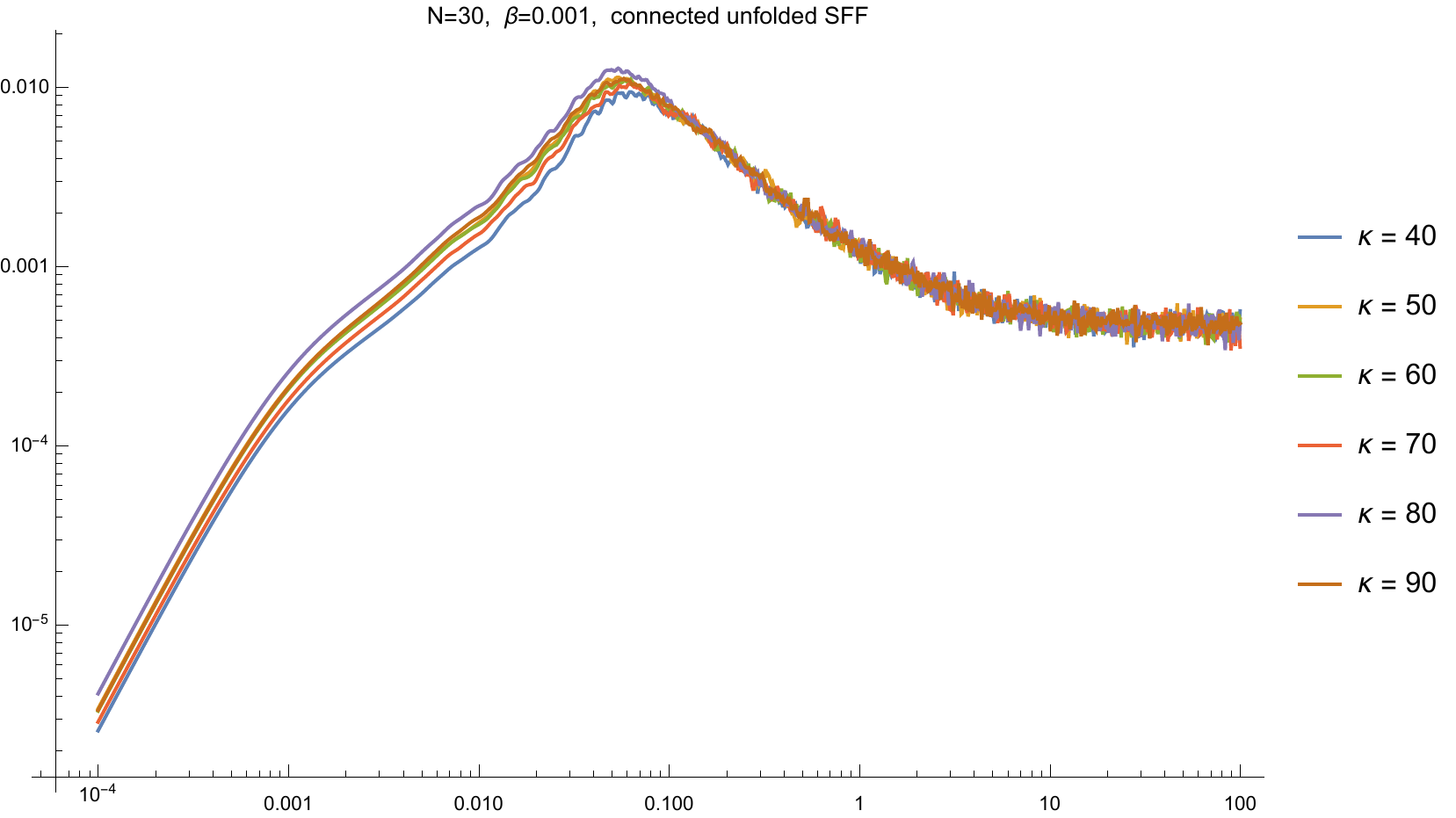}}
\caption{
The connected unfolded SFF for large $\kappa$.
The average is performed over $M=201$ realizations for $N=28$ and $M=101$ realizations for $N=30$.
}
\label{fig:beta_10-4_high_k}
\end{figure}

We computed the connected unfolded SFFs for $N = 28,\,30,\,32$ and for the cases with  $\kappa = 0 , \, 0.1 , 0.2,\,0.35,\, 0.5 ,\,0.75, \, 1 ,\,2,\,5 , \, 10,\,15,\,25$ as well as the cases with higher values of $\kappa$ ($40\le \kappa\le 90$).
We have worked at $\beta = 0 , \, 5\times 10^{-5},\,10^{-4} , \, 10^{-3}$.
The results for $\kappa\le 25$ are summarized in Figure \ref{fig:beta0} and \ref{fig:counSFFN32}, while those for $\kappa\ge 40$ are displayed in Figure~\ref{fig:beta_10-4_high_k}.
For $\beta=0$ we immediately see that the oscillatory behaviors, typical of the SFFs at very high temperatures, make very hard to identify the Thouless times.
For this reason, in the following we will mostly focus our attention on the higher values of $\beta$.

Moving to the case $\beta = 5\times 10^{-5},\,10^{-4}$, some more interesting behaviors appear.
We recognize that the expected behavior of the Thouless time clearly shows up. 
For $\kappa$ sufficiently small (say, $\kappa \leq 1$) the time at which the RMT behavior appears is clearly an increasing function of $\kappa$: in this sense, we can say that the recipe based on the connected unfolded SFF is working as expected and that the time at which the RMT behavior becomes manifest in the connected unfolded SFF is sensitive to the value of $\kappa$.
From this point of view, we see that the connected unfolded SFF ``cures" the problems we pointed out in Figure \ref{fig:fullSFF}.

Another feature which is worth to remark is that, especially for $N = 30,\,32$, the connected unfolded SFF is almost unchanged when passing from $\kappa = 1$ to $\kappa = 10$ and that, for these values of $\kappa$, the hints of chaos in the graph are very low: this suggests that the transition from chaos to integrability happens for $\kappa \sim 1$ and that, after the transition, the connected unfolded SFF is insensitive to the value of $\kappa$.
This intuition is confirmed by Figure \ref{fig:beta_10-4_high_k}, in which we plot the connected unfolded SFF for large values of $\kappa$.
We see clearly that, for these values of $\kappa$, we cannot find any remaining hints of chaos in the graph and that everything is unchanged when $\kappa$ gets increased.

From the results for $\beta=10^{-3}$,
an interesting point which is important to observe is that the chaotic features at $\kappa \geq 1$, that at $\beta = 10^{-4}$ were still vaguely visible, are now completely washed out: at these values of $\beta$ and $\kappa$ the unfolded connected SFF is already showing the integrable behavior.
On the other hand, we notice that, for $N = 30$ and $N = 32$ the ramps of the connected unfolded SFF do not have exactly the same slope.
Indeed the various slopes are slightly different from each other, a behavior that we already noticed in a much more prominent way in the left Figure \ref{fig:fullSFF} before the unfolding procedure: we believe that this behavior is signaling that our unfolding procedure is not entirely correct in the tail of the spectrum and that as a result the slopes show some differences.

All together, these preliminary observations suggest that the connected unfolded SFF is able to see the transition from the chaotic to integrable regime in the mass-deformed SYK model: when the relevant parameter $\kappa$ gets increased the Thouless time also increases accordingly.
Moreover, the transition from chaos to integrability seems to be temperature-dependent and at $\beta = 5\times 10^{-5},\,10^{-4}$ happens around $\kappa \sim 1$, while it happens for smaller values of $\kappa$ at $\beta = 10^{-3}$. 
Let us emphasize that the value at which the chaotic/integrable transition appears, $\kappa \sim 1$, is  compatible with the transition which in  \cite{Garcia-Garcia:2017bkg} was observed by studying the OTOCs.
This is different from the transition observed by the RMT analysis which instead appeared at higher values of $\kappa$, say $\kappa \sim 15$ or even bigger.
However, we stress that our results have several limitations, that we already mentioned in the Introduction and that we will discuss more in Section \ref{sec: discussion}.

In the next Section, we will confirm and will provide an explanation to these features.

%
\section{Chaoticity depends on energy levels}
\label{sec:chaos_energy}
%

In this Section we will argue that the chaotic/integrable transition is not affecting the energy levels homogeneously but it start to affect the energy levels close to the ground state --- the tail of the spectrum --- and then, i.e. for larger values of $\kappa$, it moves to the bulk of the spectrum.

This pattern is suggested by the results of Section~\ref{sec:numerics_preliminary}, in which we have described that the connected unfolded SFF is seeing a transition from chaotic to integrable regime for values of $\kappa \sim 1$ and that the precise value of $\kappa$ at which this transition appears is {\it temperature-dependent}.
Let us recall how the value of $\beta$ affects the SFF (we will consider, for simplicity, the expression for the full, not-unfolded, SFF).
Given a certain energy spectrum $\left\{ E_1 , \, E_2, \, \dots , \, E_N \right\}$, the SFF can be written as
\begin{align}
\label{eq:SFF_expanded}
g (\beta , \, t) \propto \sum_{i , j} \, e^{- \beta \, (E_i + E_j)}  e^{- i (E_i - E_j) \, t} \ .
\end{align}
From \eqref{eq:SFF_expanded} it is clear that the inverse temperature $\beta$ plays the role of a cut-off, telling us how many energy levels effectively contribute to the SFF: indeed, when the products $\beta \, (E_i + E_j)$ become large, the contributions to the SFF are exponentially suppressed.
Combining this observation with the results of Section~\ref{sec:numerics_preliminary} --- which showed that the chaotic hints in the connected SFF for $\kappa \sim 1$ disappear when passing from $\beta = 10^{-4}$ to $\beta = 10^{-3}$ --- it becomes natural to guess that the transition from chaos to integrability starts to affect the energy levels close to the ground state and that later (\ie for higher values of $\kappa$) move to the bulk of the spectrum.
This would explain the behavior, as a function of the temperature, of the connected unfolded SFF for $\kappa \sim 1$: at $\beta = 10^{-4}$ the SFF is still probing a certain amount of the states in the bulk of the spectrum, which are in the chaotic regime for $\kappa \sim 1$; on the other hand, for $\beta = 10^{- 3}$ the SFF is dominated by states which are closer to the ground state where, at $\kappa \sim 1$ the transition to the integrable regime would have already taken place.

To confirm the behavior just described we will compute three different diagnostics of quantum chaos in RMT: the level spacing distribution, the Inverse Participation Ratios (IPR) and the IR diversity (and the structural entropy).
We will compute them, at various values of $\kappa$, for different portions of the spectrum, obtained by keeping only a finite number of levels, $L$, ordered from the ground state.
In this way, we will see that the chaotic/integrable transition arises at lower values of $\kappa$ for small values of $L$, while appears at larger values of $\kappa$ when $L$ is large.

We want to stress that, contrary to the level spacing distribution, IPR and IR do not require any unfolding procedure. 
The fact that they also show a separation in the chaotic/integrable transition between the tail and the bulk suggests that this separation is not an artifact due to the unfolding.
Moreover  they could be  more easily related to the physics of the OTOCs, since their definition involves also the energy eigenvectors and not just the eigenvalues.

\subsection{Level spacing distribution}
\label{sec:level_repulsion}

\begin{figure}[t!]
\centering
\fbox{\includegraphics[width=8cm]{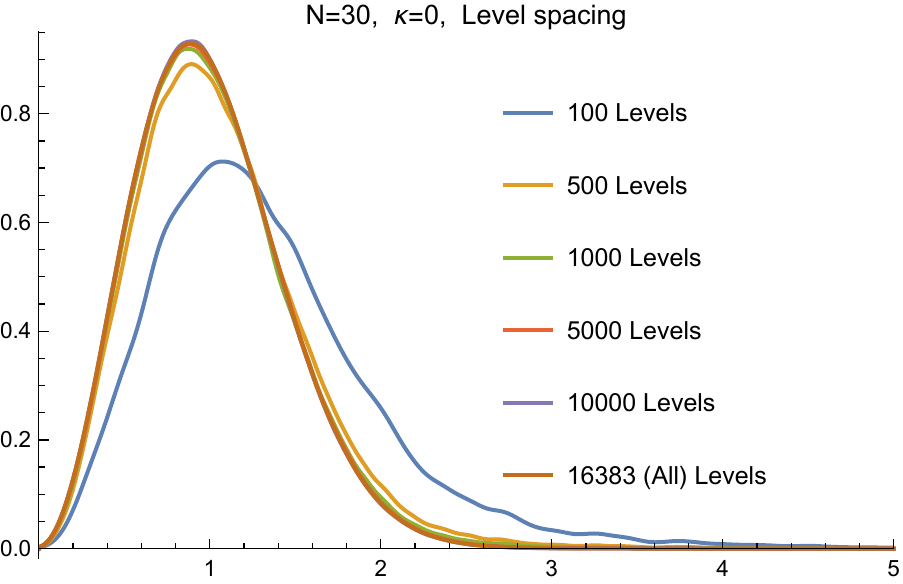}}\quad
\fbox{\includegraphics[width=8cm]{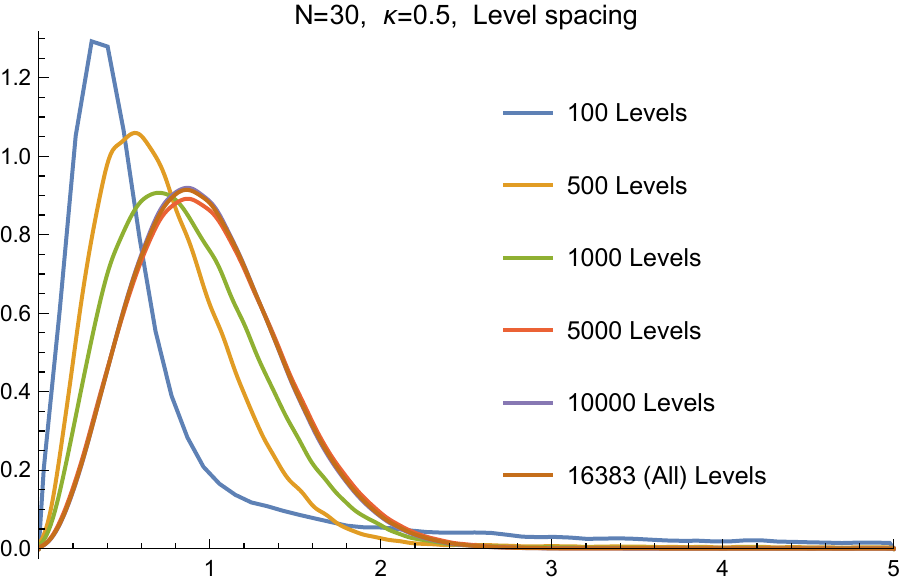}}\\
\fbox{\includegraphics[width=8cm]{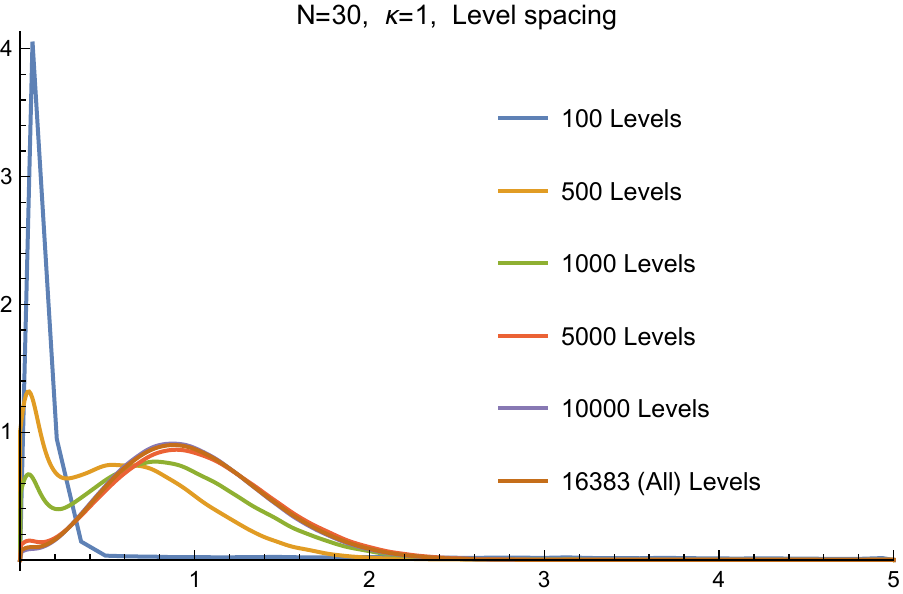}}\quad
\fbox{\includegraphics[width=8cm]{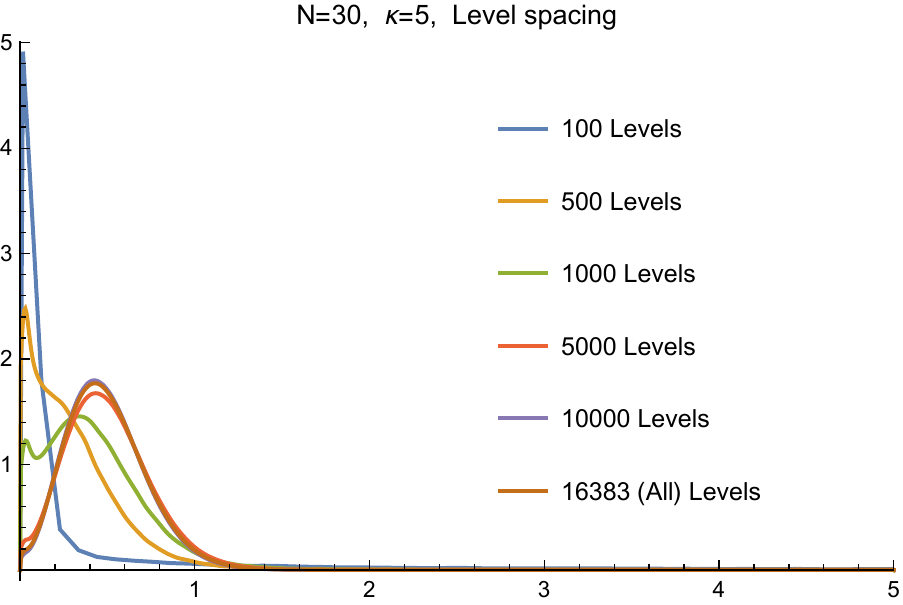}}
\caption{The level spacing distribution, averaged over $M=312$ realizations for the $N = 30$ Hamiltonian with $\kappa=0, \, 0.5, \, 5$.
The distributions are computed on subsectors of the spectra including $L$ levels starting from the ground state.
}
\label{fig:spacing_N30}
\end{figure}

We studied the level spacing distribution\footnote{We caution the reader that the SFF and the level spacing distribution are actually probing two, a priori, {\it different} manifestations of quantum chaos: the level spacing distribution is a short-range observable, while the SFF is mostly affected by the global structure of the spectrum (it is a long-range observable).
However, in typical systems both the long-range and the short-range observables show a similar behavior, chaotic or integrable.} for the unfolded upper block of the $N=30$ hamiltonians averaged over $312$ ensembles, including only the first $L = 100, \, 500, \, 1000, \, 5000, \, 10000$ ordered energy levels,  for the values of $\kappa = 0 , \, 0.5, \, 1, \, 5$.
Note that here we unfold the whole spectrum first, and then keep the first $L$ eigenvalues in the unfolded spectrum.
This is what the factor $e^{-\beta\,(\tilde E_i+\tilde E_j)}$ in the connected unfolded SFF \eqref{eq:SFF_expanded} effectively does.
 The results are presented in Figure~\ref{fig:spacing_N30}.

We clearly see that, for $\kappa = 0$, the spectrum shows level repulsion even for energy levels very close to the ground state ($L =100$); for $\kappa = 0.5$ we  see that the level repulsion for the levels closer to the ground state is reduced but it is still clearly present for intermediate ranges ($L = 500$ and $L = 1000$) and for very long ranges ($L = 5000$ and $L = 10000$). 
This feature is lost for $\kappa = 1$ and $\kappa = 5$: for these values of $\kappa$ even at the intermediate ranges of $L = 500$ and $L = 1000$ the level repulsion is reduced, while it is still clearly present for the levels in the bulk ($L = 5000$ and $L = 10000$). 

This result  confirms our hypothesis that the non-chaotic features start to appear close to the ground state and that, in passing from $\beta = 10^{-4}$ to $\beta = 10^{-3}$ the SFF probes regions of the spectrum where at $k = 1 , 5$ a transition between chaotic and integrable behaviors is happening. 
Indeed, after the unfolding, we remind that the SFF effectively probes around $1000$ energy levels at $\beta=10^{-3}$, and around $10000$ energy levels at $\beta = 10^{-4}$, and we see from Figure~\ref{fig:spacing_N30} that these are the values of $L$ at which the transition is taking place for $\kappa \sim 1$.

Another conclusion we reach is that, given the qualitative agreement between the transition from chaos to integrable behavior as seen by the OTOCs and by the connected unfolded SFF, also the behavior of the OTOCs is mostly controlled by the tail of the spectrum, while it is pretty insensitive to the features of the spectrum in the bulk.

In Appendix~\ref{app:r-statistics}, we also study a dimensionless quantity, the $\tilde{r}$-parameter, which characterizes the level spacing statistics. The averaged $\tilde{r}$-parameter also supports the distinct transition at $\kappa\sim 1$ and $10$ for the tails and the bulk of the spectrum, respectively.
However, we have to say that such separation is less evident (but still visible) for the $\tilde{r}$-parameter computed without performing the unfolding (Figure \ref{fig: all rpara middle 30 level}).

\subsection{Inverse Participation Ratios (IPR)}
\label{sec: ipr}

To provide further evidences to the picture we described in Section~\ref{sec:level_repulsion}, we compute other chaos diagnostics, {\it inverse participation ratio}~(IPR) and {\it IR entropy}, as defined in \cite{IZRAILEV1990299,ZELEVINSKY199685,Santos:2010aa,Radicevic:2016kpf,Ho:2017nyc}. In particular, \cite{Ho:2017nyc} analyzed the IR diversity, which is the normalized exponential of the IR entropy, of the SYK model. These quantities measure the randomness, or the delocalization, of the energy eigenvectors.

Let us consider a normalized energy eigenstate $|E_n\rangle$ ($n=1,2,\cdots, D$):
\begin{equation}
    |E_n\rangle\, =\, \sum_{m=1}^D \psi_{mn} |e_m \rangle\ ,
\end{equation}
where $\{|e_m\rangle\}$ is a set of given basis vectors and $D$ is the dimension of Hilbert space. Then, the {\it inverse participation ratios}~(IPR) $\xi_n$ is defined by
\begin{equation}
    \xi_n\,\equiv\,  {1\over \;\;\displaystyle{\sum_{m=1}^D |\psi_{mn}|^4}\;\;}\ .\label{def: ipr}
\end{equation}
In addition, the {\it IR entropy} $S^{\text{\tiny IR} }_n$ is given by
\begin{equation}
    S^{\text{\tiny IR} }_n\,\equiv\, -\sum_{m=1}^D |\psi_{mn}|^2 \log |\psi_{mn}|^2 \ ,
\end{equation}
and it is convenient to evaluate the {\it IR diversity}~\cite{Ho:2017nyc}:
\begin{equation}
    \Omega_{n}\,\equiv\, {\exp S^{\text{\tiny IR} }_n\over D}\quad,\hspace{5mm}(\; 0\leqq \Omega_{n}\leqq 1 \;)\ .\label{def: ir diversity}
\end{equation}
Note that the IR diversity $\Omega_n$ and the IPR $\xi_n$ correspond to the first R\'enyi entropy $S^{(1)}_n$ and (the exponential of) the second R\'enyi entropy $S^{(2)}_n$, respectively, defined by
\begin{equation}
    S^{(\alpha)}_n\,\equiv\,-{1\over \alpha-1} \log\left[ \sum_{m=1}^D |\psi_{mn}|^{2\alpha}\right] \ .
\end{equation}
These observables measure how widely an energy eigenstate spreads over a given basis. Therefore, they depend on the basis but nevertheless they have played a successful role in characterizing quantum chaos. Of course, these quantities are useful if a ``natural" basis is at disposal in the system under investigation. For example, if one choose the energy eigenstates as a basis, the energy eigenstate would be completely localized with respect to that basis, a result which is not very satisfactory. On the other hand, by picking up the site basis in a lattice system, one can study how the system is spatially localized. 


\begin{figure}[t!]
\centering
\subfloat[]{\includegraphics[width=8cm]{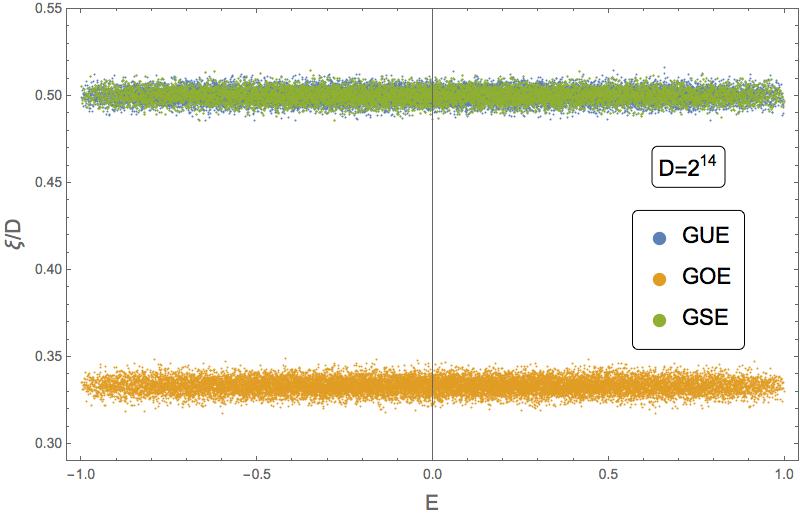}\label{fig: random matrix ipr dist}}\quad
\subfloat[]{\includegraphics[width=8cm]{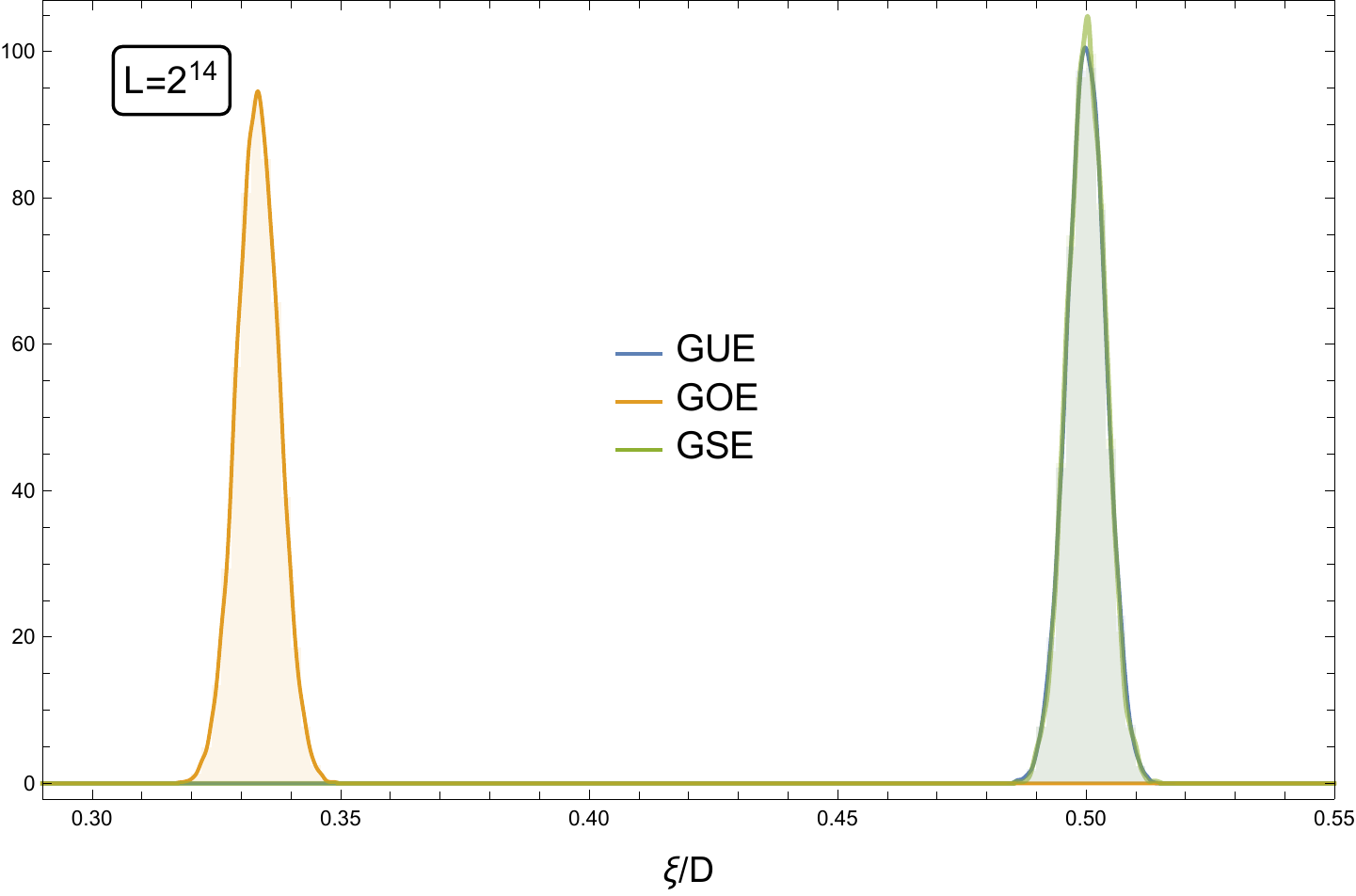}\label{fig: random matrix ipr dist2}}
\caption{Left: The distribution of the normalized IPR of the GUE, GOE and GSE of dimension $D=2^{14}$ versus the normalized energy $E$. The distributions of the GUE and GSE overlap. Right: The distribution of the normalized IPR of the GUE, GOE and GSE of dimension $D=2^{14}$. The normalized IPR of GUE, GOE and GSE follows a Gaussian distribution with mean $\approx 0.500, 0.333$ and $0.500$ and deviation $\approx 0.00394,0.00423$ and $0.00385$, respectively.}
\label{fig: random matrix ipr dist all}
\end{figure}
In this Section, we focus on the IPR of the mass-deformed SYK model, and we will present a qualitatively similar result for the IR diversity  in Section~\ref{sec: ir diversity}. The IPRs of eigenvectors of a random matrix drawn from GUE, GOE and GSE form a Gaussian distribution around the mean value $0.500, 0.333$ and $0.500$ with deviation $\approx 0.00394,0.00423$ and $0.00385$, respectively (See Figure~\ref{fig: random matrix ipr dist all}). Here, we evaluated  a normalized IPR, where the normalization factor is given by the dimension of the eigenvector $D=2^{14}$.
Of particular significance are the values of the standard deviations: they are very small, which means that all the IPRs are very close to the average values.

\begin{figure}[t!]
\centering
\subfloat[]{\includegraphics[width=8cm]{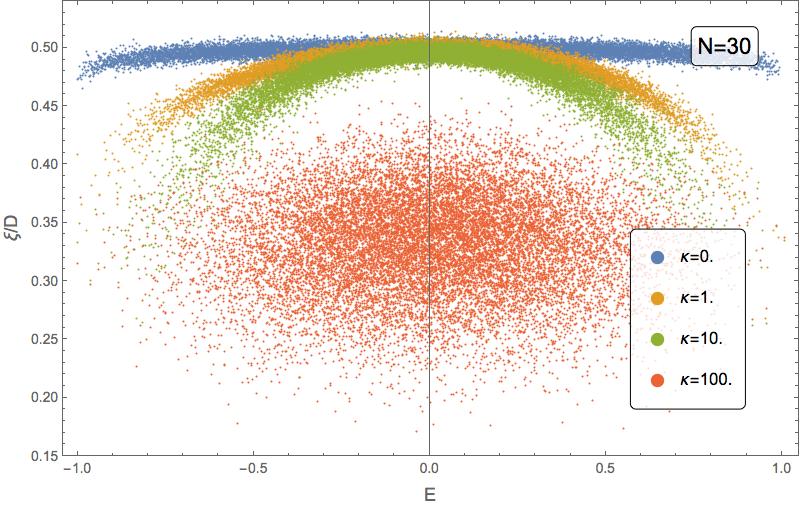}\label{fig: ipr dist}}\quad
\subfloat[]{\includegraphics[width=8cm]{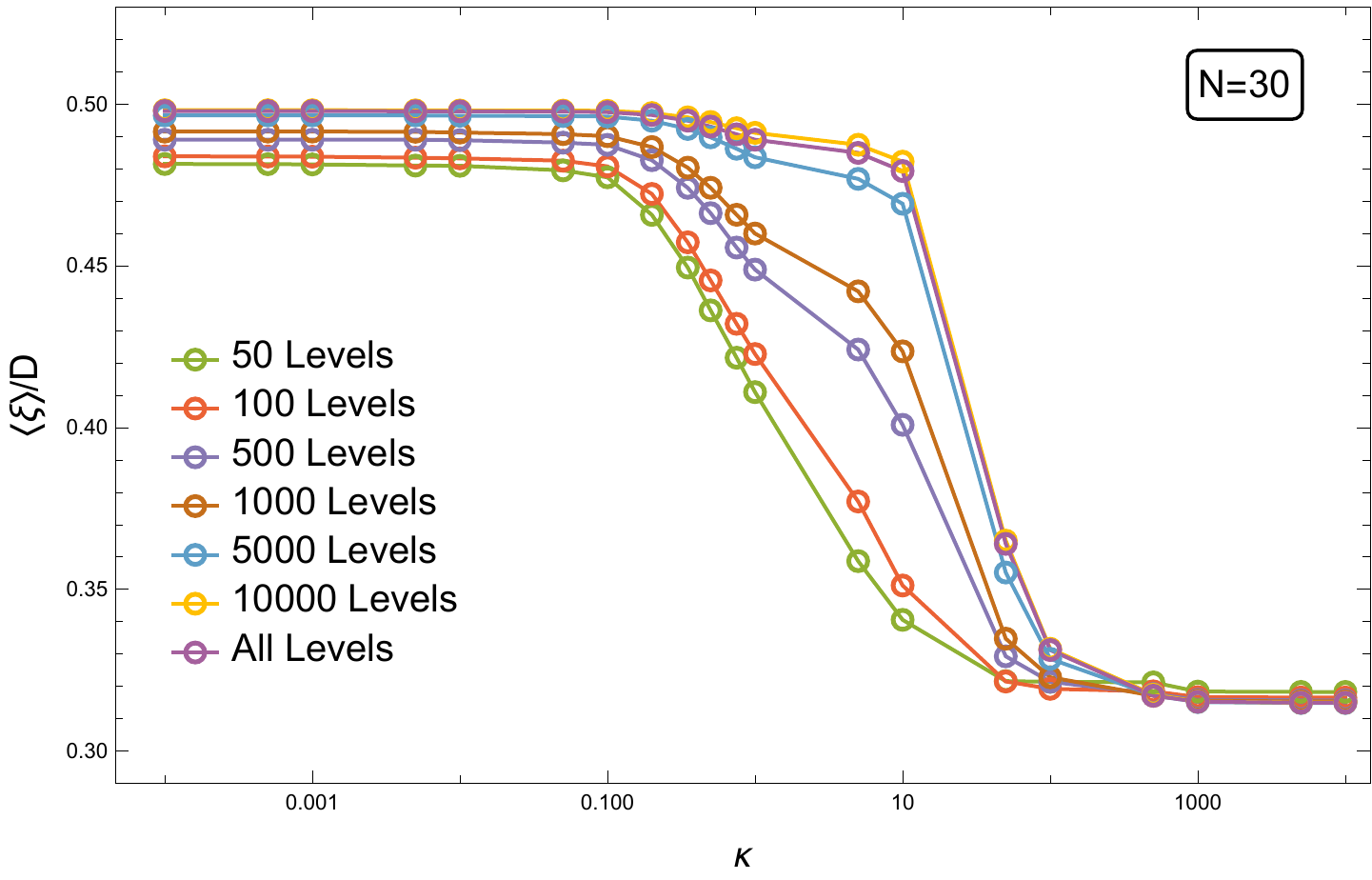}\label{fig: ipr average level}}
\caption{Left: The distributions of the normalized IPR of the even parity sector~($D=2^{14}$) of $N=30$ SYK model with $\kappa=0, 1, 10$ and $100$ versus normalized energy $E$, respectively. Right: The mean value of the normalized IPR of the subsectors of the spectra including $L$ levels starting from the ground state of the even parity sector~($D=2^{14}$) of $N=30$ SYK model.}
\label{fig: all ipr dist}
\end{figure}
For the IPR of the mass-deformed SYK model, one can naturally choose the ``spin-chain'' basis where our hamiltonian corresponds to all-to-all random interactions among $N/2$ sites. Then, the IPR (and, the IR diversity) will measure the ``localization'' with respect to this basis. We focus on the $N=30$  mass-deformed SYK,  which clearly exhibits the distinct chaotic/integrable transition of the low-lying energy subsectors as we have seen in the level spacing distribution of Section~\ref{sec:numerics_preliminary} (also, see the $\tilde{r}$-parameter analysis of Appendix~\ref{app:r-statistics}). Moreover, we analyze the even parity subsector, and therefore we normalize the IPRs of the energy eigenvectors by $D=2^{14}$.

\begin{figure}[t!]
\centering
\subfloat[$L=50$]{\includegraphics[width=8cm]{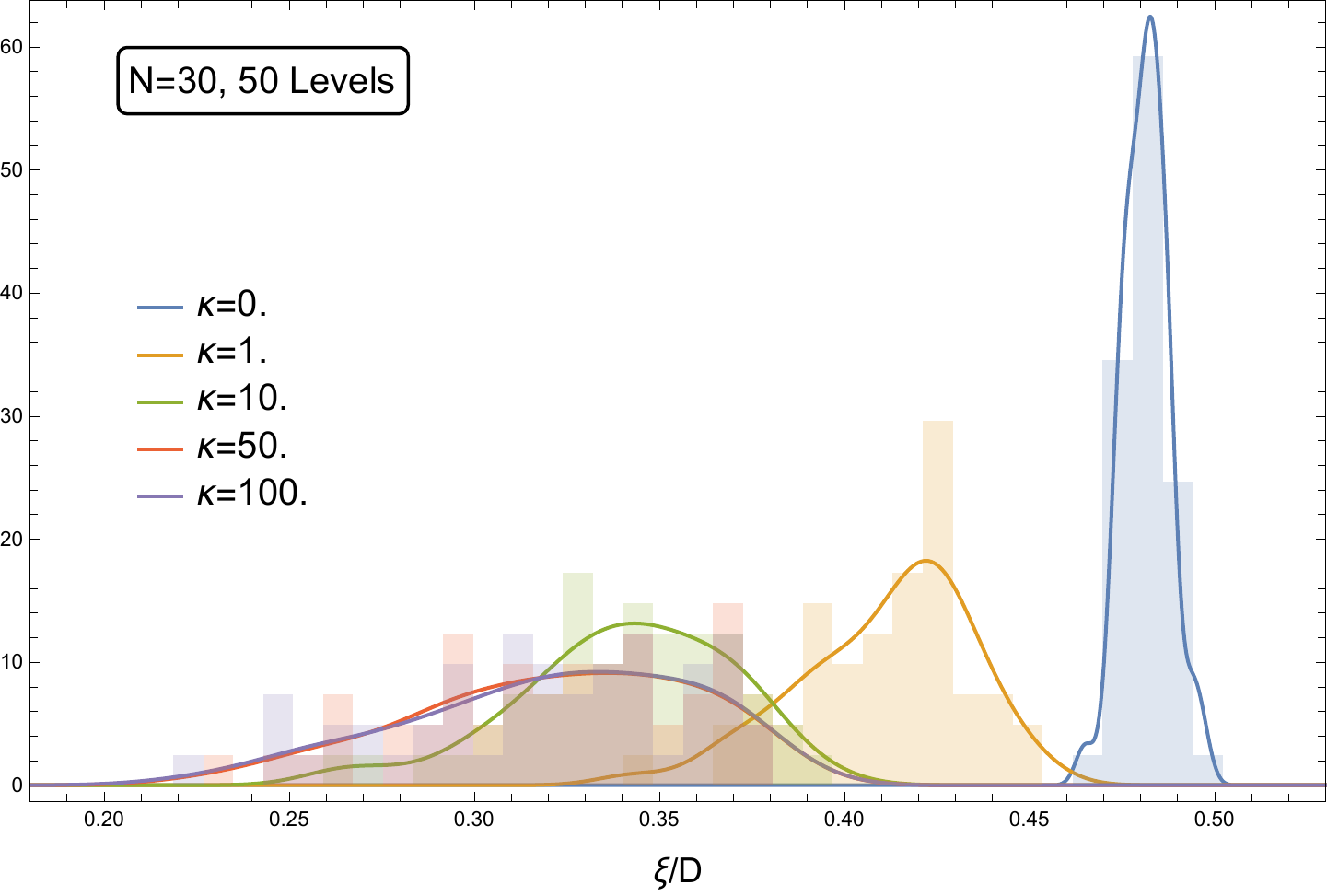}\label{fig: ipr dist 50}}\quad
\subfloat[$L=100$]{\includegraphics[width=8cm]{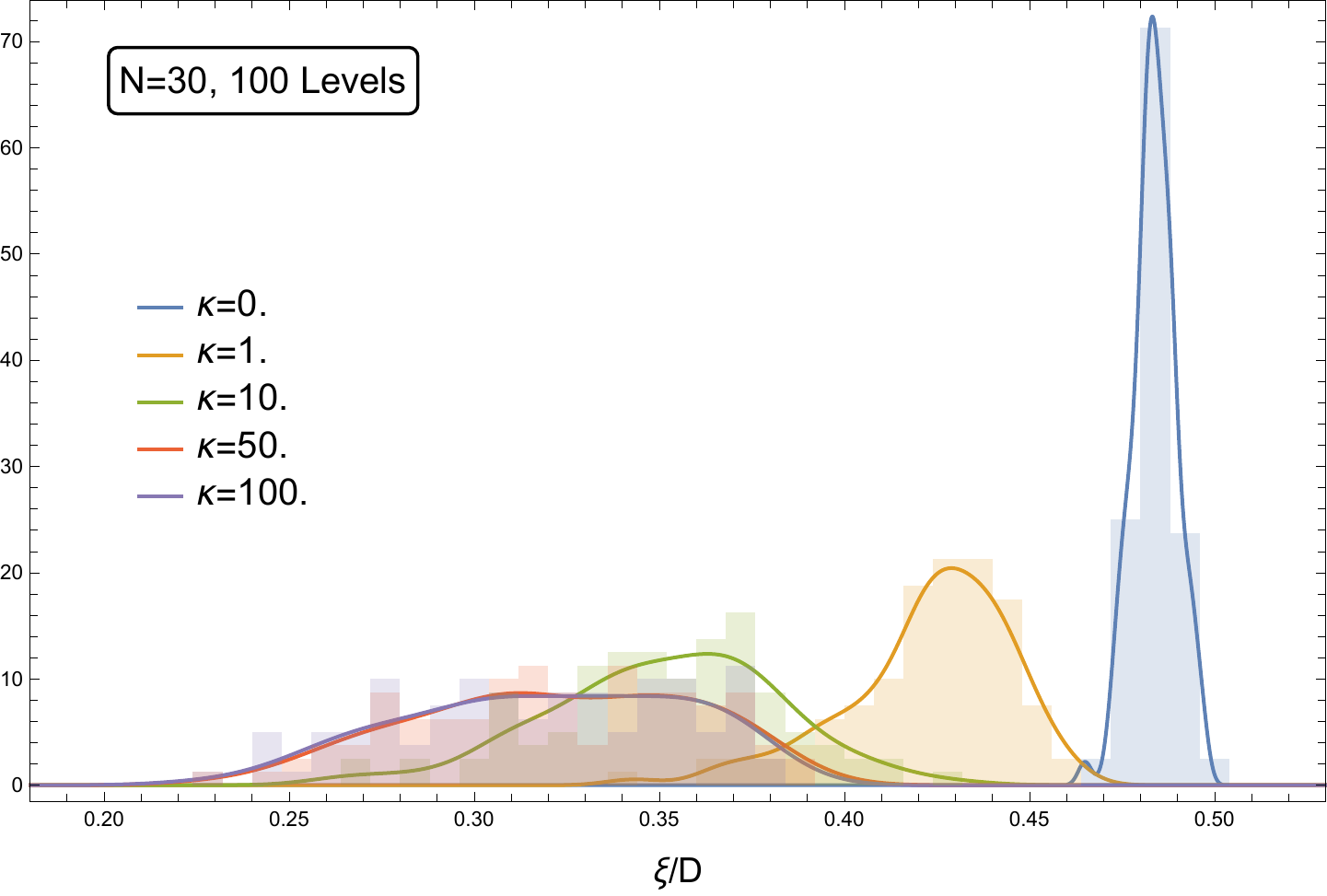}\label{fig: ipr dist 100}}\\
\subfloat[$L=5000$]{\includegraphics[width=8cm]{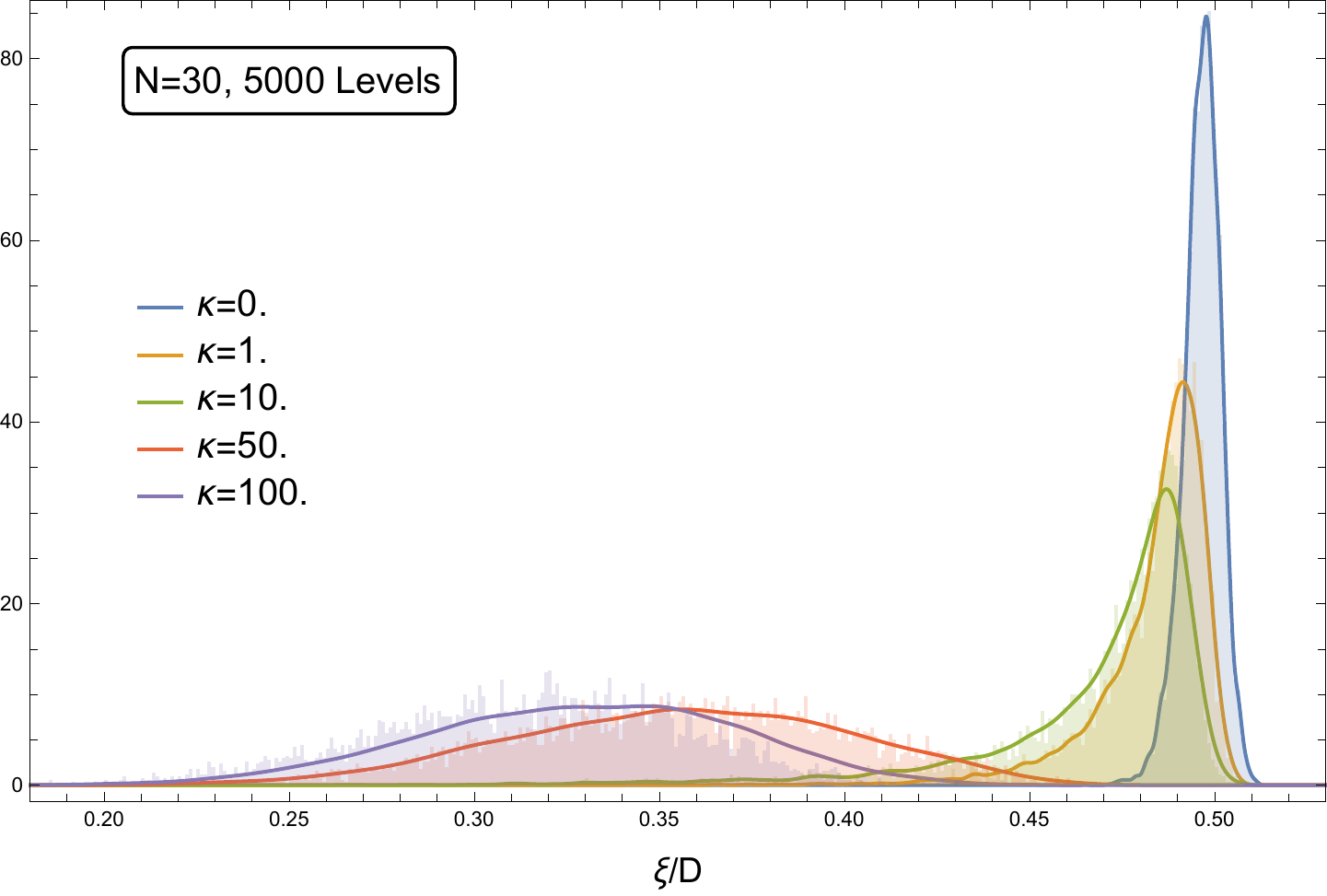}\label{fig: ipr dist 5000}}\quad
\subfloat[$L=10000$]{\includegraphics[width=8cm]{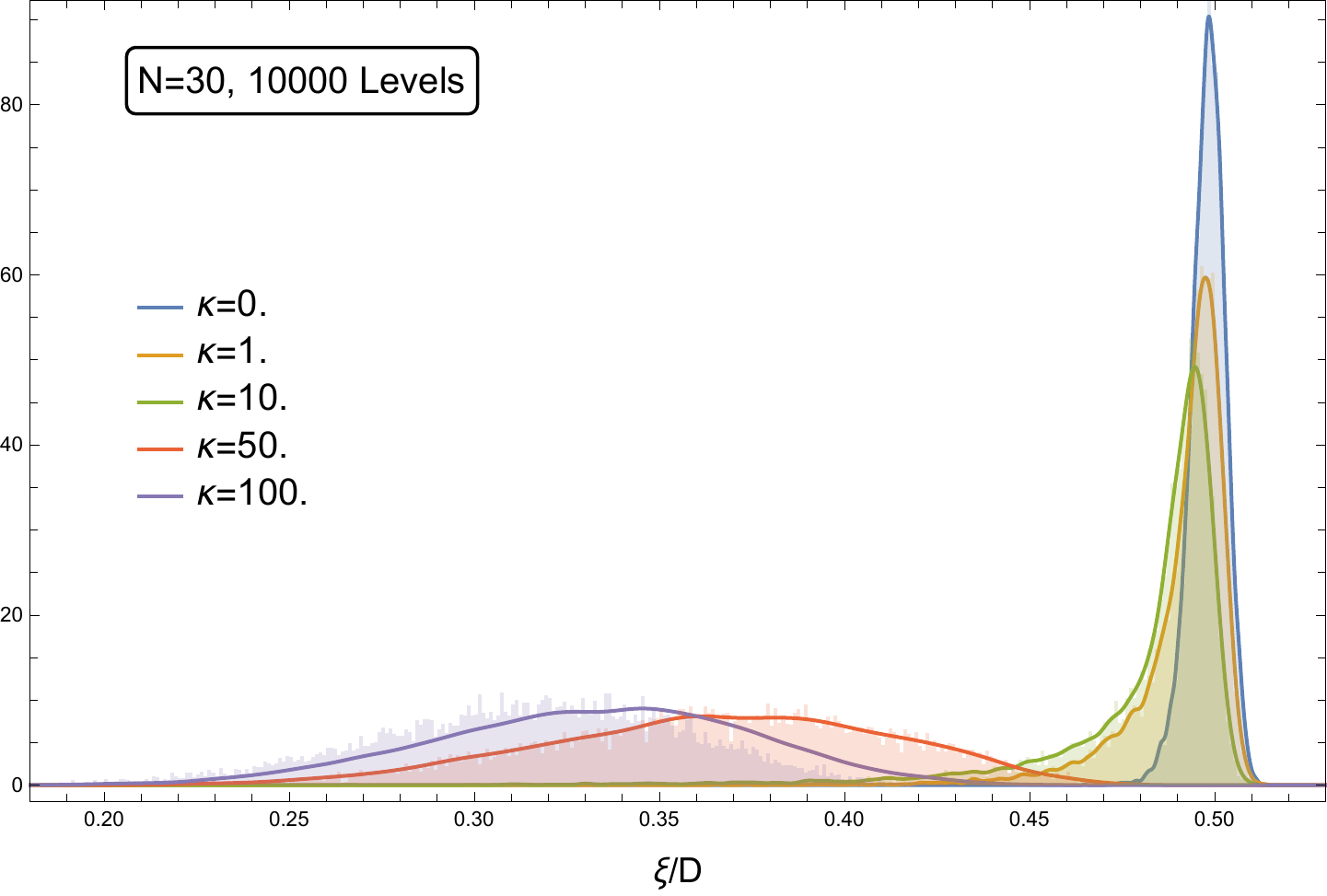}\label{fig: ipr dist 10000}}\\
\caption{The distribution of the IPR of the subsectors of the spectra including $L$ levels starting from the ground state of the even parity sector~($D=2^{14}$) of $N=30$ SYK model with $\kappa=0,1,10, 50, 100$ ($L=50,100,5000,10000$).}
\label{fig: ipr dist level}
\end{figure}
We compute the IPR of $N=30$ mass-deformed SYK model with various $\kappa$'s. In particular, we present the distribution of the IPR for $\kappa=0, 1, 10, 100$ with respect to the normalized energy in Figure~\ref{fig: ipr dist}. For small $\kappa\ll 1$, the IPRs are sharply concentrated around $\xi\approx0.5$ over the whole spectrum. As expected for $N=30$ SYK model, this distribution is close to the that of the GUE. As $\kappa$ is increased (\textit{e.g.,} $0<\kappa<10$), the IPRs of the low-lying spectrum are decreasing and dispersing while the bulk of spectrum keeps staying around $\xi\approx 0.5$ (See Figure~\ref{fig: ipr dist level}). 
Once again, of particular importance are the values of the variances more than the mean values. 
Indeed, it is from the values of the variances that we can really measure a departure from the RMT behavior.

This implies that the energy eigenstates of the low energy spectrum become ``more localized'' with respect to the basis at $\kappa\sim 1$ and move away from the delocalization regime in $1<\kappa<10$. On the other hand, the bulk of the spectrum still belongs to delocalization regime of the $GUE$ even at $\kappa=10$. For $\kappa>10$, the bulk of spectrum begins to deviate from the delocalization regime. 
%

%
The distinct transitions of the IPRs of the tails and the bulk of spectrum can also be seen in the mean value of the IPR versus $\kappa$ in Figure~\ref{fig: ipr average level}. For full spectrum, the transition of the IPR takes place at $\kappa\sim 10$. On the other hand, the low-lying spectrum ($L=50, 100$) cross the transition at $\kappa\sim 1$.
However, we stress that from the mean value of the IPR one cannot immediately conclude that a chaotic/integrable transition is taking place, since it could be also a transition from GUE to GOE: what really discriminates between these two cases are the variances of the distributions, plotted in Figure \ref{fig: ipr dist level}.\footnote{In Section \ref{sec: ir diversity} we will see that also the structural entropy allows to distinguish between these two extreme cases.}

The behavior we described is consistent with the fact that the energy eigenvectors of the tails of the spectrum are typically more sensitive to a perturbation than the bulk of the spectrum.\footnote{We would like to thank Subhro Bhattacharjee for pointing out this.}

This difference in the delocalization/localization transitions of the low-lying and the bulk of spectrum is also consistent with the previous results based on the spectral observable, the level spacing distribution, of Section~\ref{sec:level_repulsion} (as well as the $\tilde{r}$-parameter in Appendix~\ref{app:r-statistics}). The IPR analysis of the distinct transitions is based on the eigenvectors, which provides independent confirmation of our hypothesis on the chaotic/integrable transition. Note that correlation functions, in particular the OTOCs, also depend not only on the spectrum but also on the eigenvectors. In particular, since the low-lying spectrum have a significant influence on the OTOCs, the Lyapunov exponent should be greatly affected by the chaotic/integrable transition of the IPRs of the low-lying spectrum rather than the bulk one. 
Hence, we believe that this result provides a further supporting evidence that the connected unfolded SFF, for sufficiently large values of $\beta$, indeed sees the same physics of the OTOCs.

\subsection{IR diversity and structural entropy}
\label{sec: ir diversity}

%
\begin{figure}[t!]
\centering
\subfloat[]{\includegraphics[width=8cm]{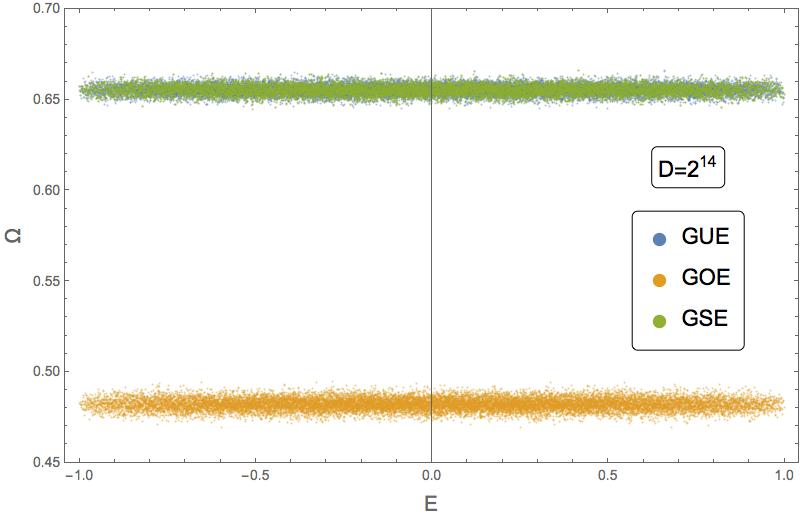}\label{fig: random matrix ir diversity 2}}\quad
\subfloat[]{\includegraphics[width=8cm]{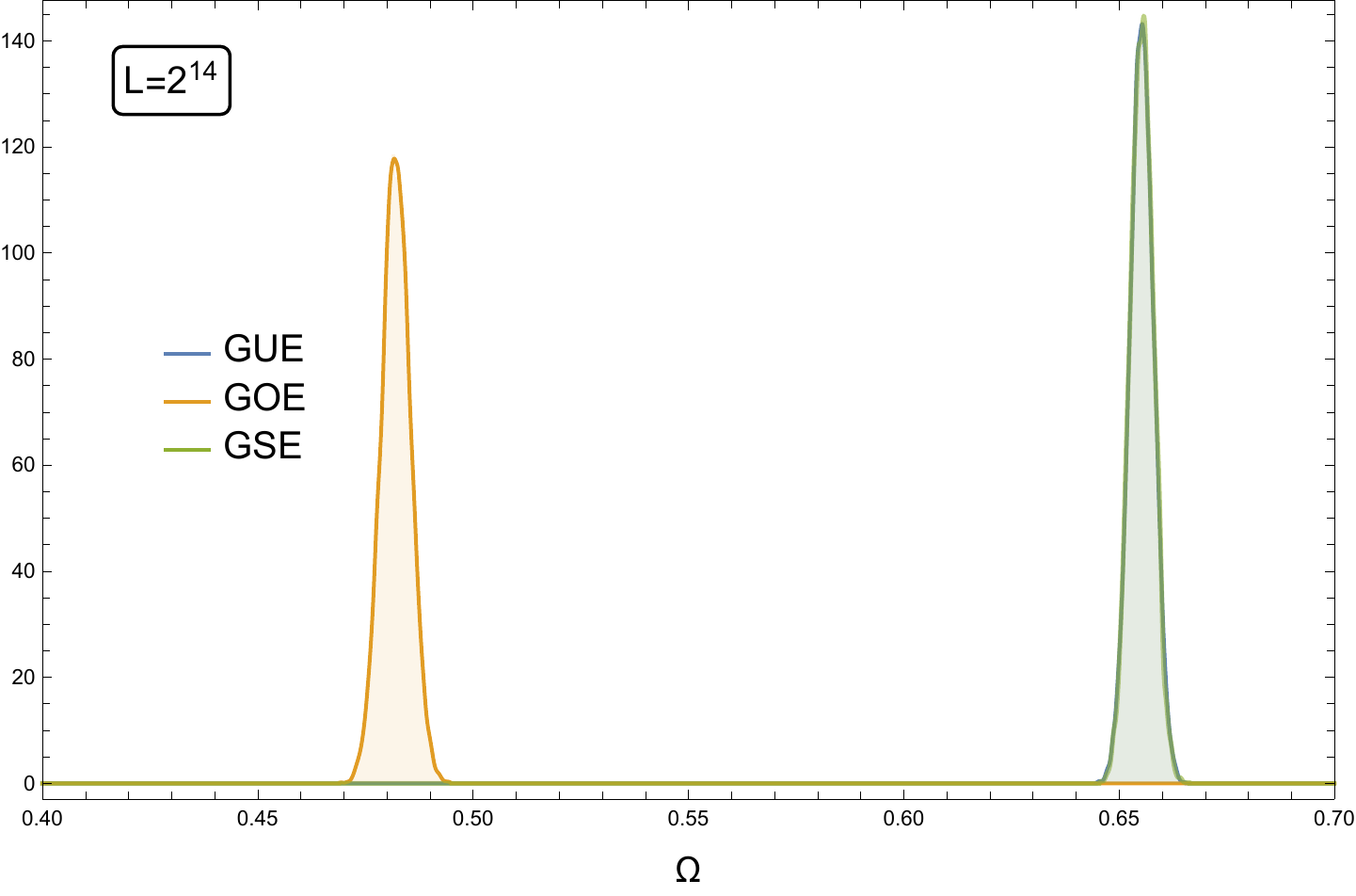}\label{fig: random matrix ir diversity}}
\caption{Left: The distribution of the IR diversity of the GUE, GOE and GSE of dimension $D=2^{14}$ versus the normalized energy $E$. The distributions of the GUE and GSE overlap. Right: The distribution of the IR diversity of the GUE, GOE and GSE of dimension $D=2^{14}$. The IR diversity of GUE, GOE and GSE follows the Gaussian distribution with mean $\approx 0.655, 0.482$ and $0.655$ and deviation $\approx 0.00278,0.00336$ and $0.00273$, respectively.}
\label{fig: random matrix ir diversity all}
\end{figure}
\begin{figure}[t!]
\centering
\subfloat[]{\includegraphics[width=8cm]{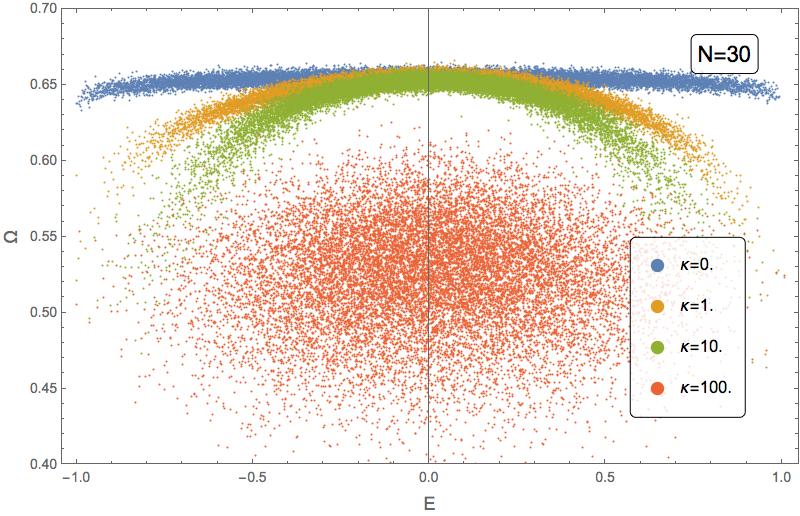}\label{fig: ir diversity dist all}}\quad
\subfloat[]{\includegraphics[width=8cm]{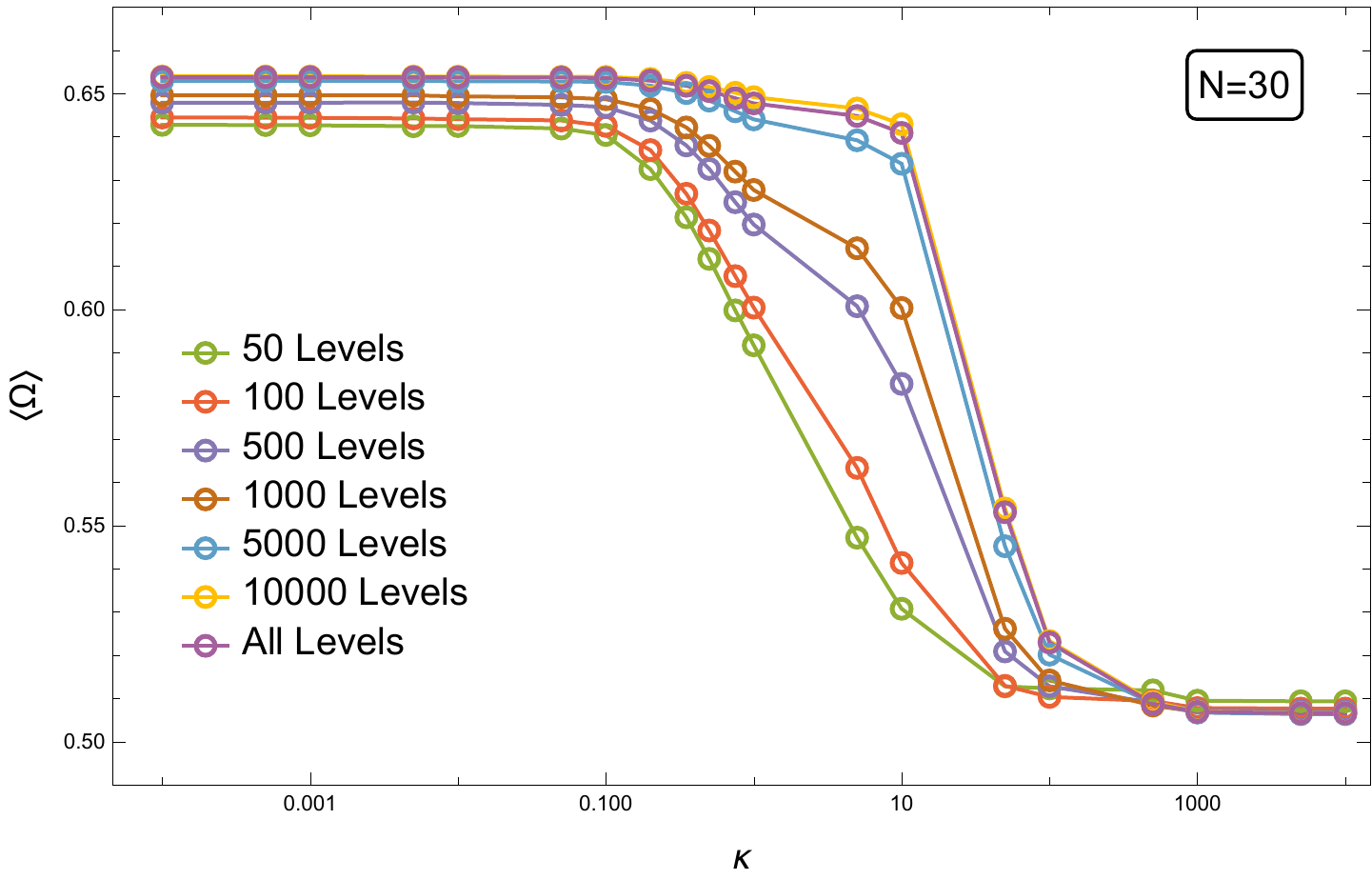}\label{fig: ir diversity av level}}
\caption{Left: The distributions of the IR diversity of the even parity sector~($D=2^{14}$) of $N=30$ SYK model with $\kappa=0, 1, 10$ and $100$ versus normalized energy $E$, respectively. Right: The mean value of the IR diversity of the subsectors of the spectra including $L$ levels starting from the ground state of the even parity sector~($D=2^{14}$) of $N=30$ SYK model versus $\kappa$.}
\label{fig: ir diversity dist and av level}
\end{figure}
\begin{figure}[t!]
\centering
\subfloat[$L=50$]{\includegraphics[width=8cm]{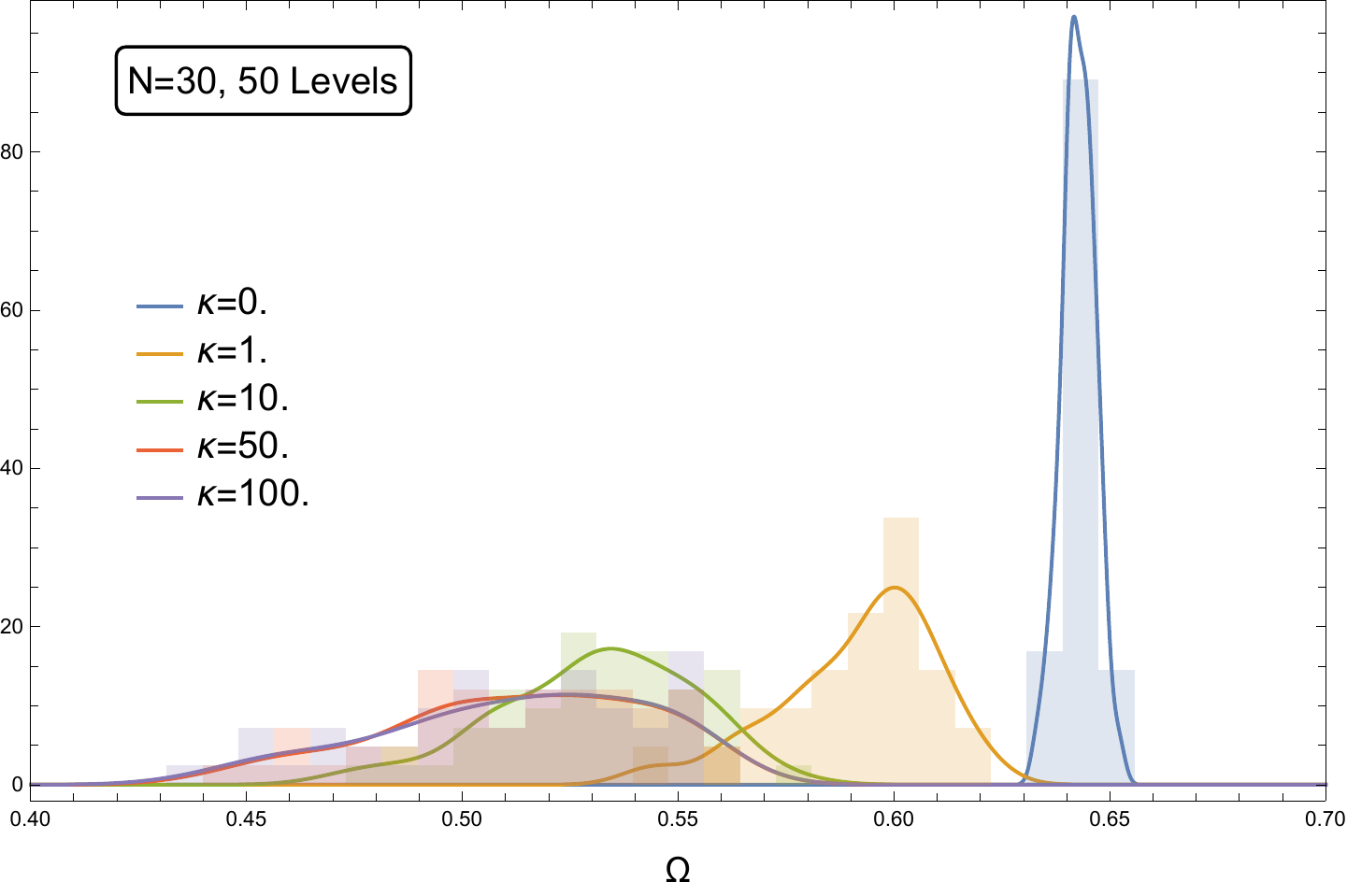}\label{fig: ir diversity dist 50}}\quad
\subfloat[$L=100$]{\includegraphics[width=8cm]{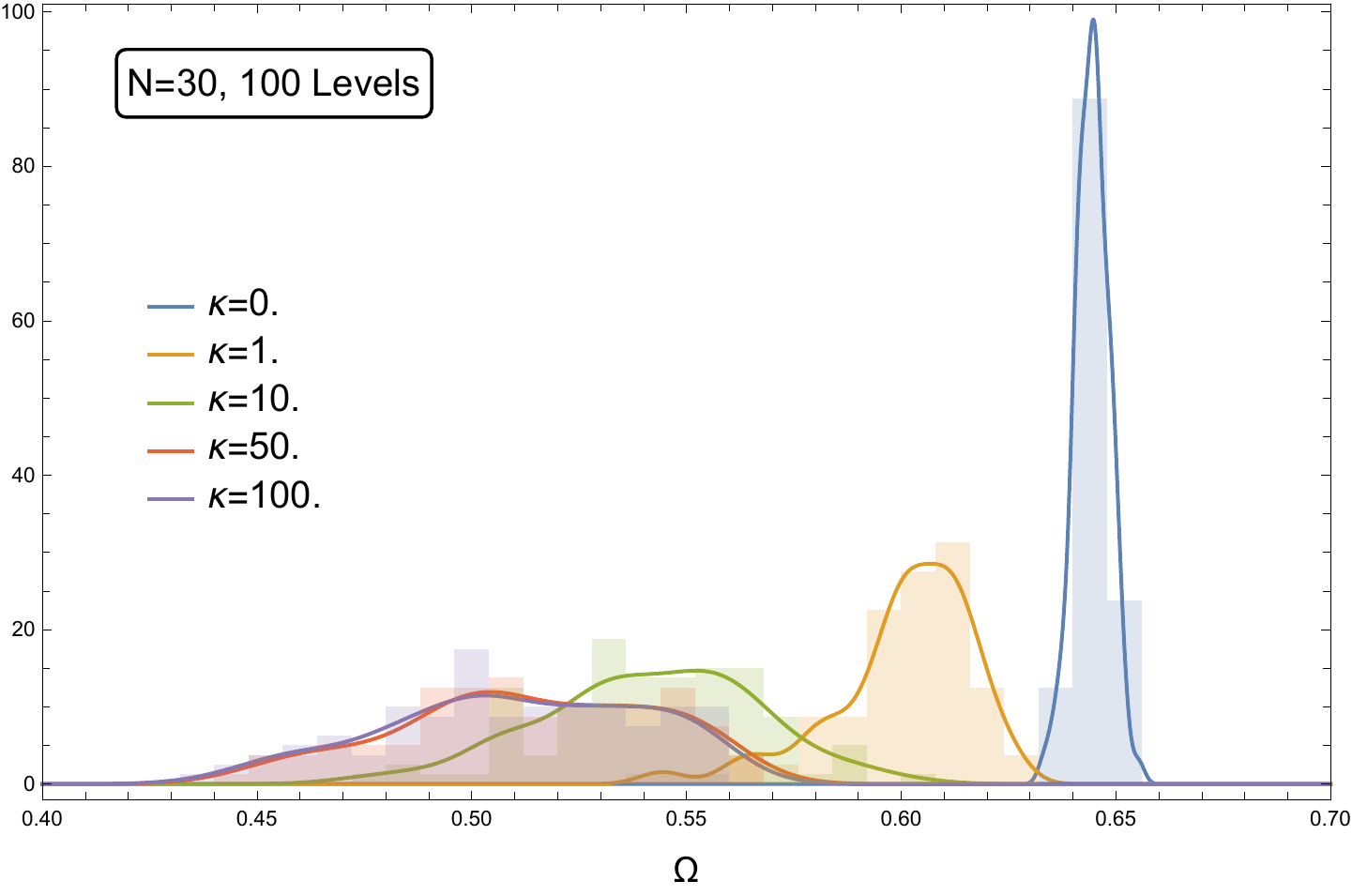}\label{fig: ir diversity dist 100}}\\
\subfloat[$L=5000$]{\includegraphics[width=8cm]{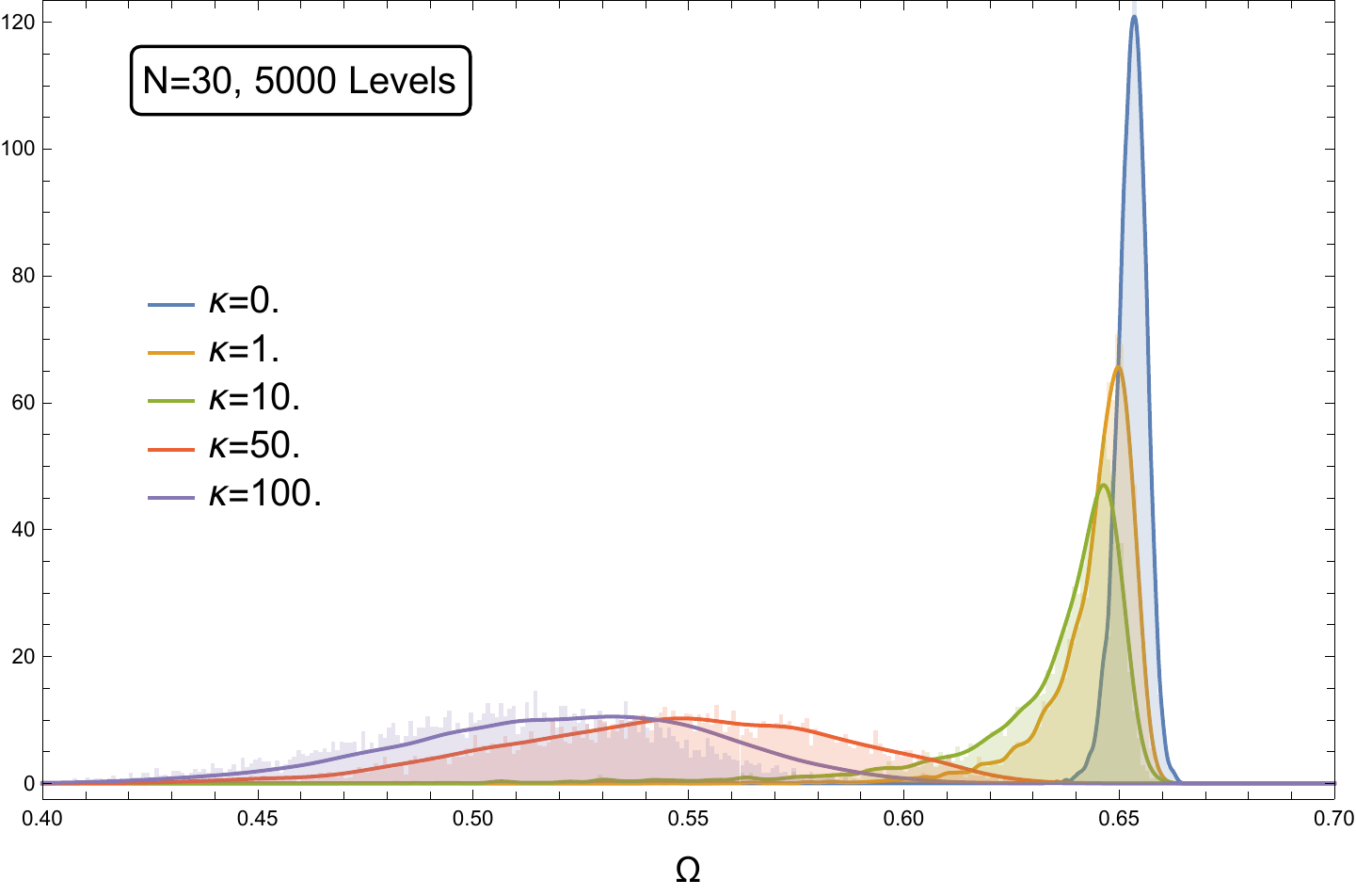}\label{fig: ir diversity dist 5000}}\quad
\subfloat[$L=10000$]{\includegraphics[width=8cm]{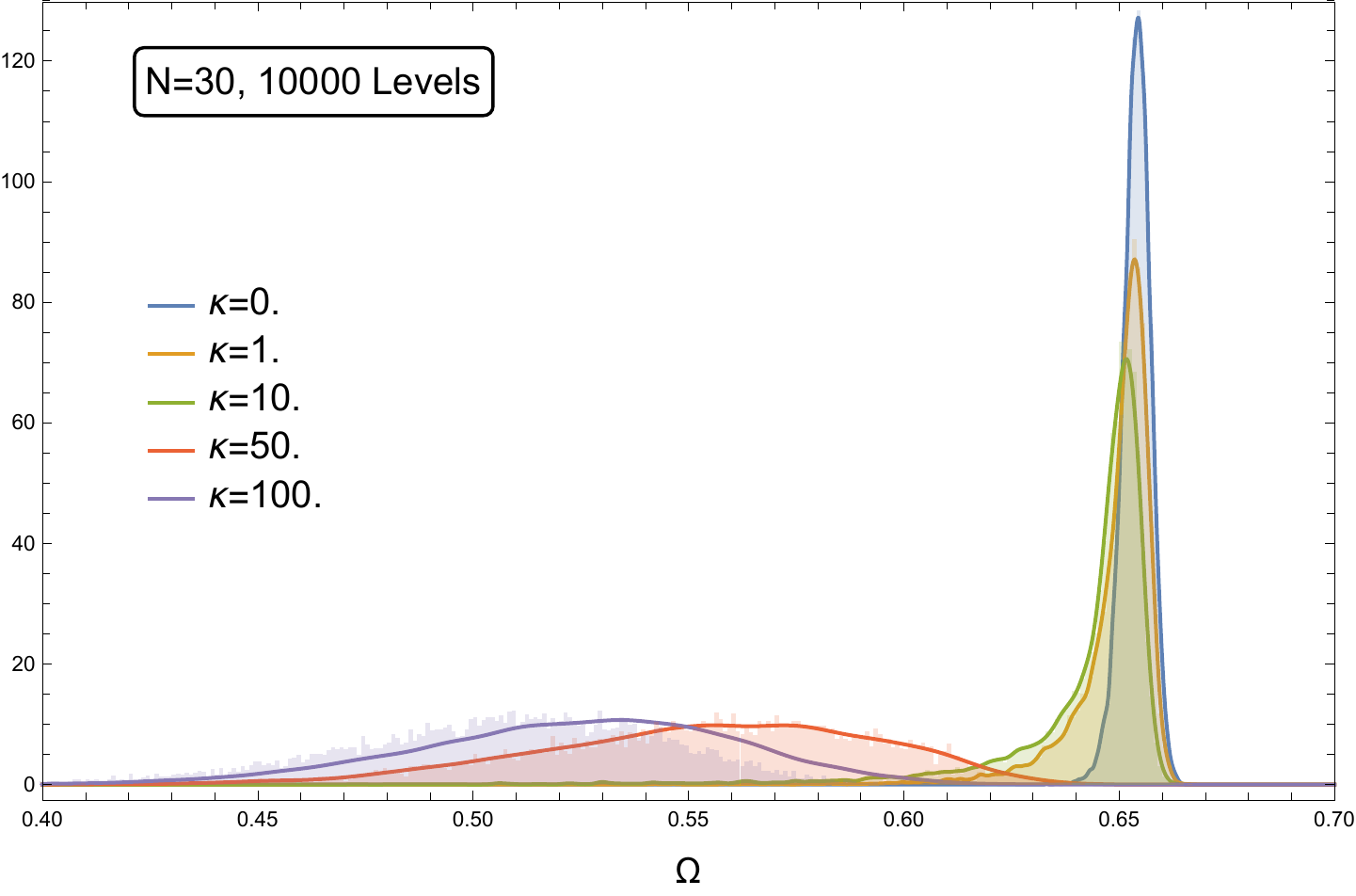}\label{fig: ir diversity dist 10000}}\\
\caption{The distribution of the IR diversity of the subsectors of the spectra including $L$ levels starting from the ground state of the even parity sector~($D=2^{14}$) of $N=30$ SYK model with $\kappa=0,1,10, 50, 100$ ($L=50,100,5000,10000$).}
\label{fig: ir diversity dist level}
\end{figure}

Now we present similar results for the IR diversity, as defined in~\eqref{def: ir diversity}, for both RMT (Figure~\ref{fig: random matrix ir diversity all}) and the even parity sector of $N=30$ mass-deformed SYK model (Figure~\ref{fig: ir diversity dist and av level}). Like the IPR in Section~\ref{sec: ipr}, we observed the delocalization/localization transition of the tails of the spectrum of $N=30$ SYK model at $\kappa\sim 1$ while it takes place at $\kappa\sim 10$ for whole spectrum (see Figure~\ref{fig: ir diversity dist and av level} and Figure~\ref{fig: ir diversity dist level}). 
%

%
%
%
Although the IPR and the IR entropy~(IR diversity) are independent observables, they seemingly show a qualitatively similar behavior. To study more quantitatively the difference between the IPR and IR entropy we compute another observable: the so-called {\it structural entropy}~\cite{Pipek:1992aa} defined by the difference between the first and the second R\'enyi entropy:
\begin{equation}
    S_n^{\text{\tiny str}}\,\equiv\, S^{\text{\tiny IR}}_n -\log \xi_n\ .
\end{equation}
The structural entropy is non-negative and bounded by the normalized IPR or the so-called {\it spatial filling factor}~\cite{Pipek:1992aa}, \ie
\begin{equation}
    0\,\leqq\, S_n^{\text{\tiny str}}\,\leqq \, -\log {\xi_n\over D}\ .\label{eq: inequality of str}
\end{equation}
Note that the structural entropy\footnote{Using Cauchy-Schwarz inequality one can easily show that the upper bound of \eqref{eq: inequality of str}, the logarithm of the normalizaed IPR, is also 0 if and only if the non-zero contribution of the basis to the corresponding eigenstate is uniform.} is $0$ if and only if the corresponding eigenstate is distributed uniformly over a subset of the given basis (\ie all non-zero $|\psi_{mn}|^2$ are equal). This can easily be seen by the Jensen's inequality with the definitions of IPR in~\eqref{def: ipr} and IR entropy in~\eqref{def: ir diversity}~\cite{Pipek:1992aa}. The structural entropy characterize the shape of the distribution of the probability ~$|\psi_{mn}|^2$ and has been applied in quantum chemistry and disordered system~\etc\ (see~\cite{Varga:2003aa} and reference therein).
\begin{figure}[h!]
\centering
\subfloat[]{\includegraphics[width=5cm]{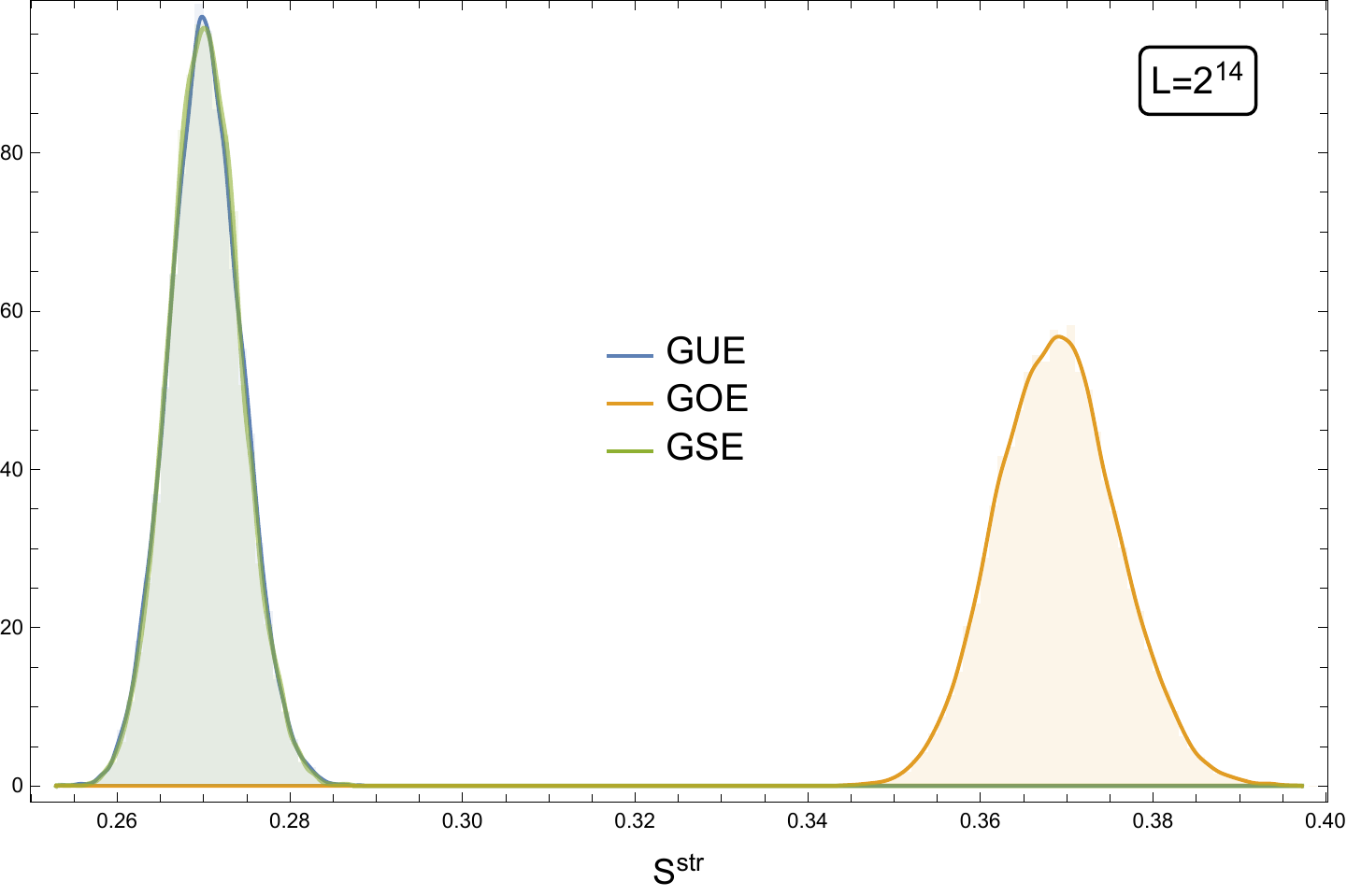}\label{fig: str rm}}\quad\subfloat[]{\includegraphics[width=5cm]{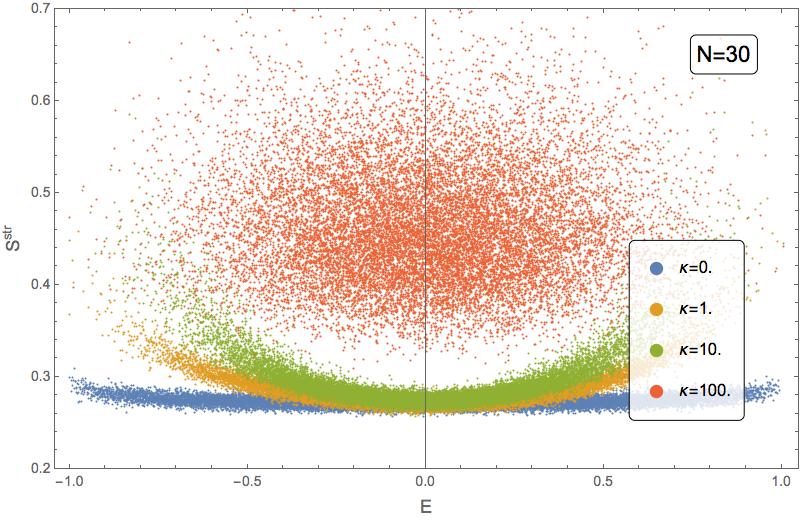}\label{fig: str dist all}}\quad
\subfloat[]{\includegraphics[width=5cm]{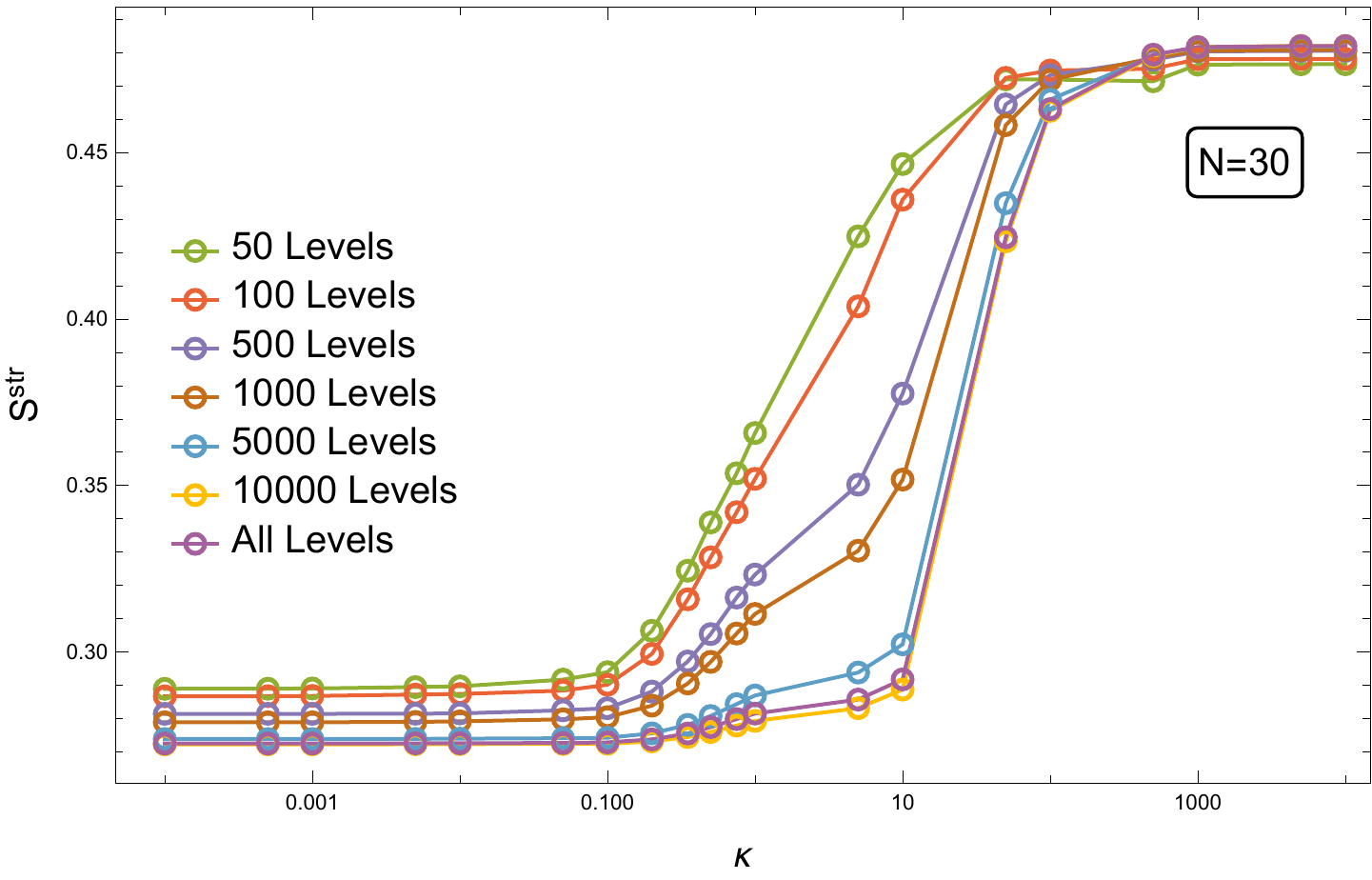}\label{fig: str av level}}
\caption{(a) The distribution of the structural entropy of the GUE, GOE and GSE of dimension $D=2^{14}$. The distribution of the GUE and GSE overlap. The structural entropy of the GUE, GOE and GSE follows the Gaussian distribution with mean $\approx 0.270, 0.369$ and $0.270$ and deviation $\approx 0.00424, 0.00696$ and $0.00416$, respectively. (b) The distributions of the structural entropy of the even parity sector~($D=2^{14}$) of $N=30$ SYK model with $\kappa=0, 1, 10$ and $100$ versus normalized energy $E$, respectively. (c) The mean value of the structural entropy of the subsectors of the spectra including $L$ levels starting from the ground state of the even parity sector~($D=2^{14}$) of $N=30$ SYK model versus $\kappa$.}
\label{fig: str dist and av level}
\end{figure}
We also present the structural entropy of the GUE, GOE and GSE in Figure~\ref{fig: str rm} and $N=30$ SYK model in Figure~\ref{fig: str dist all}. 

\begin{figure}[H]
\centering
\subfloat[$L=50$]{\includegraphics[width=7cm]{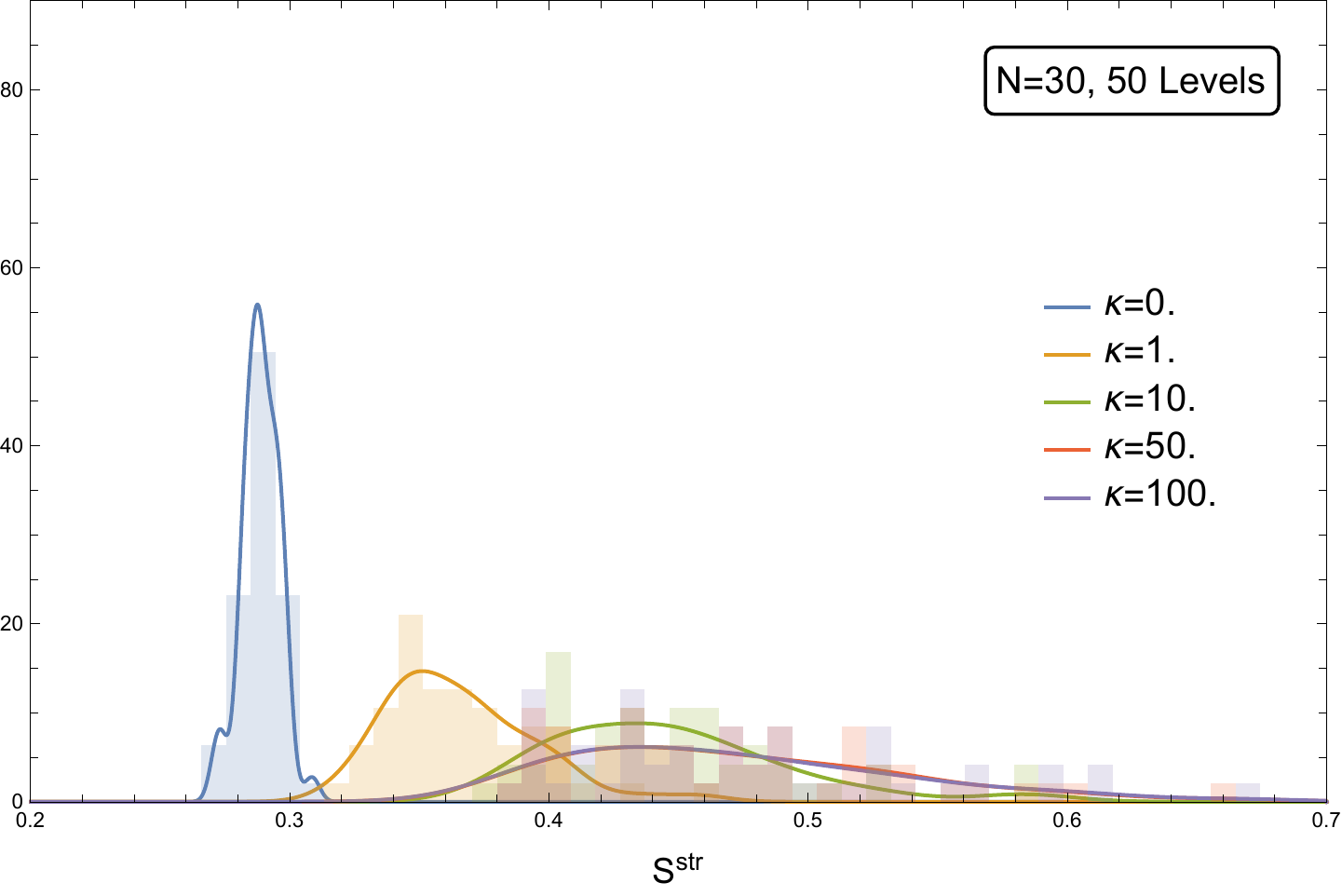}\label{fig: str dist 50}}\qquad
\subfloat[$L=100$]{\includegraphics[width=7cm]{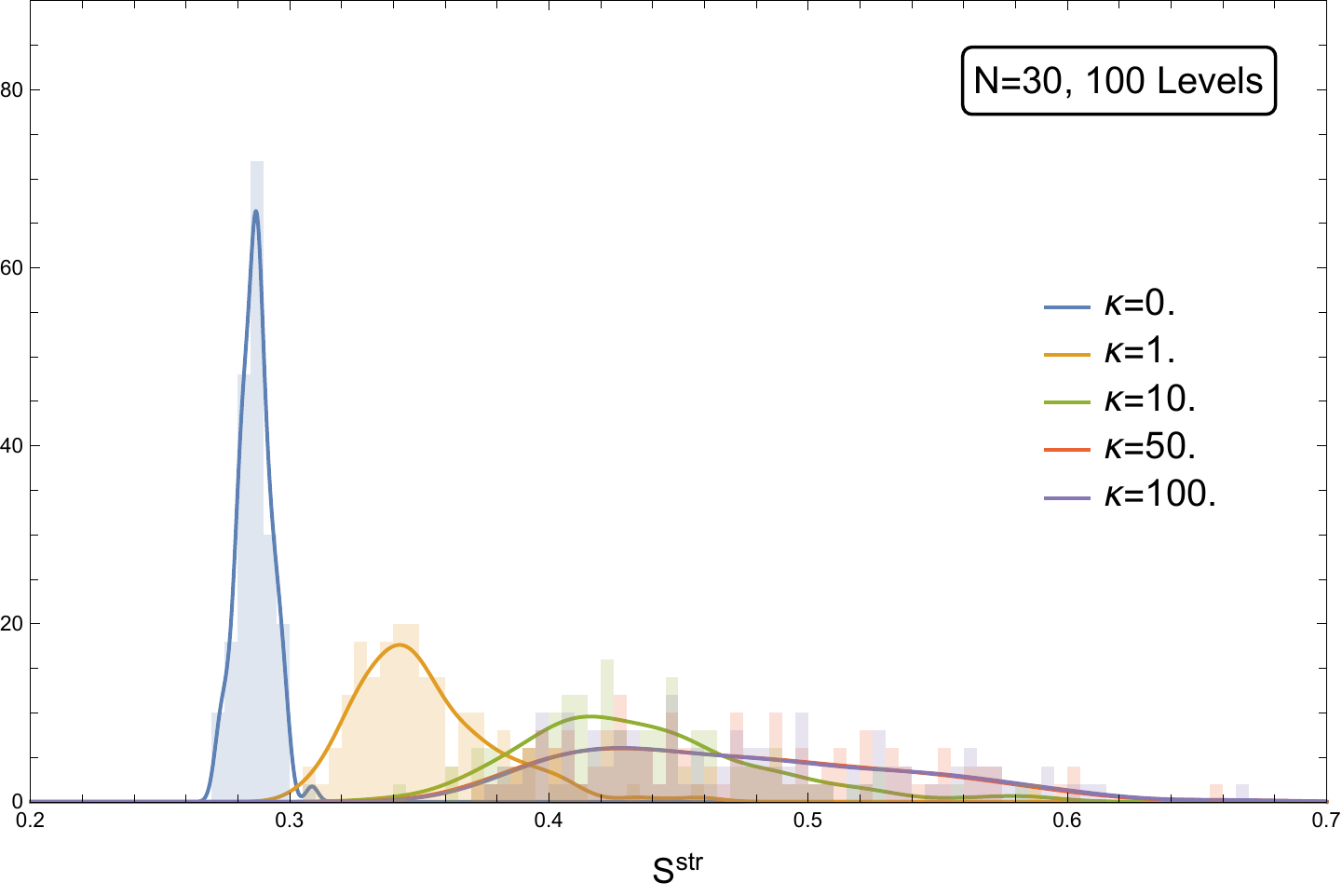}\label{fig: str dist 100}}\\
\subfloat[$L=5000$]{\includegraphics[width=7cm]{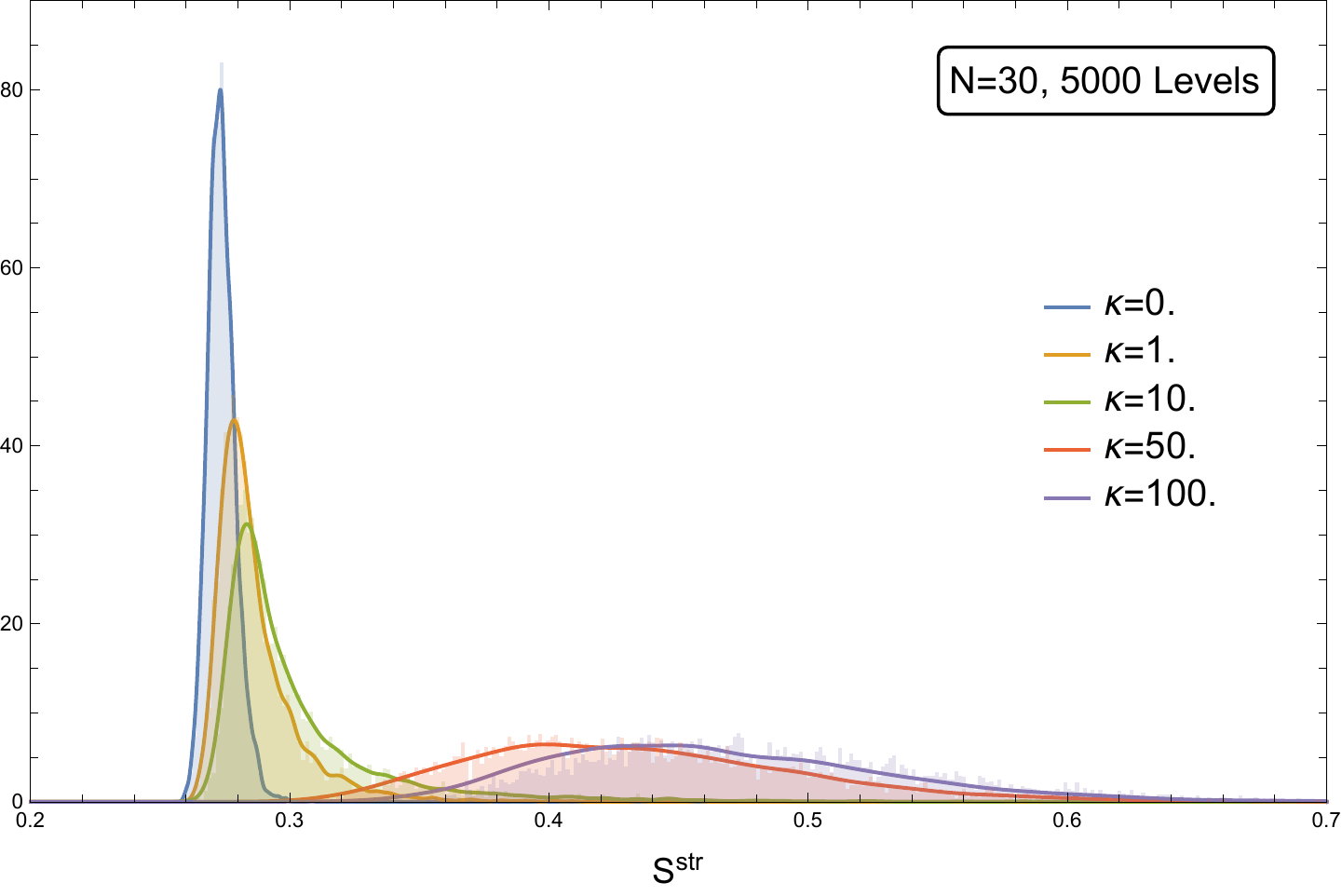}\label{fig: str dist 5000}}\qquad
\subfloat[$L=10000$]{\includegraphics[width=7cm]{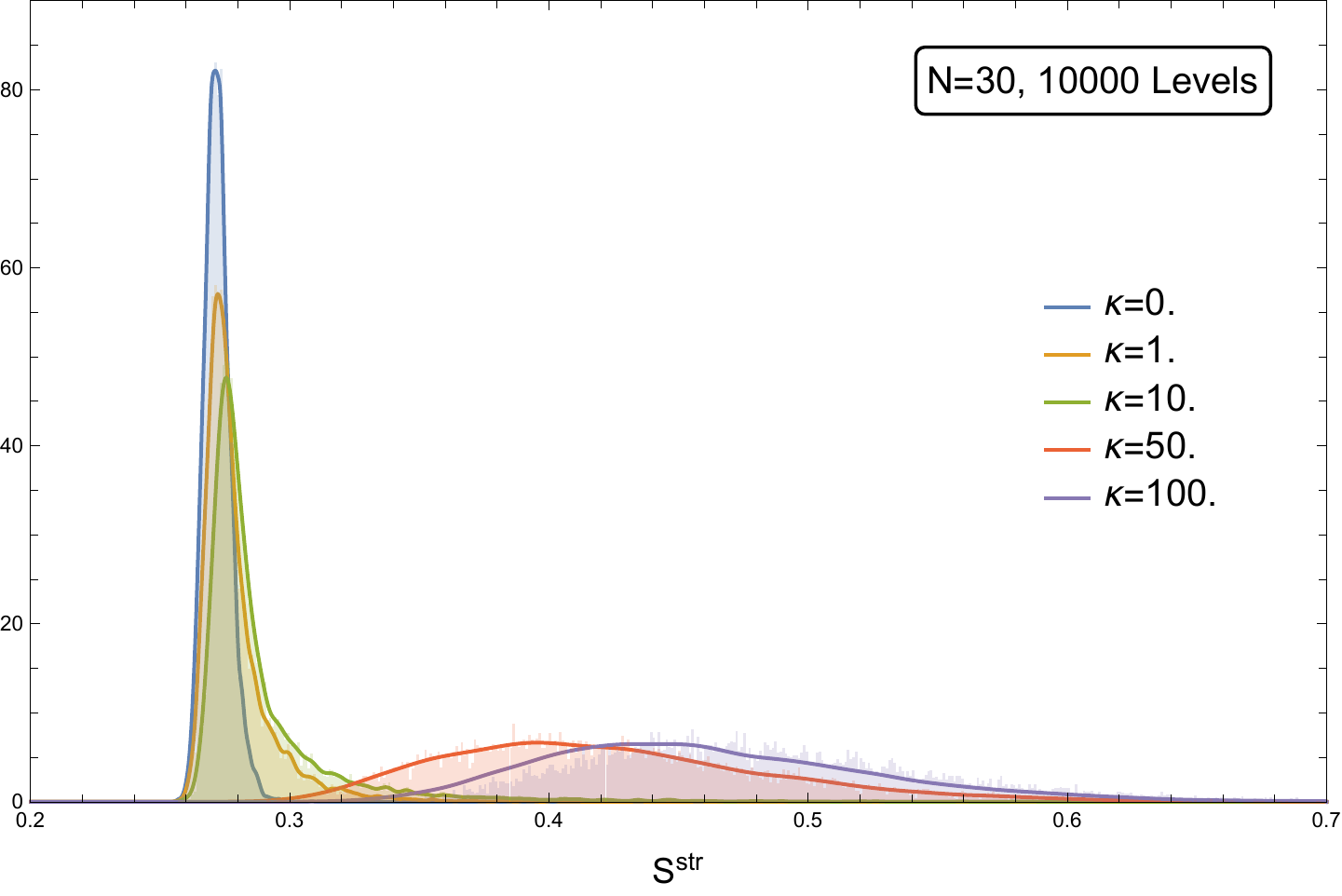}\label{fig: str dist 10000}}\\
\caption{The distribution of the structural entropy of the subsectors of the spectra including $L$ levels starting from the ground state of the even parity sector~($D=2^{14}$) of $N=30$ SYK model with $\kappa=0,1,10, 50, 100$ ($L=50,100,5000,10000$).}
\label{fig: ratio dist level}
\end{figure}

From the structural entropy, one can distinguish the integrable regime from the GOE class. Recall that the mean value of the IPR and the IR diversity of the integrable regime (\ie large $\kappa$) of the SYK model is close to those of the GOE although the shape of the distributions are different and hence, simply from the mean value of the IPRs, one cannot distinguish the integrable behavior from the GOE behavior. But, the averaged structural entropy of the integrable regime (\eg $\kappa=1000$) is $0.482$ (See Figure~\ref{fig: str av level}), which is far from the structural entropy $\approx 0.369$ of the GOE class (See Figure~\ref{fig: str rm}). The structural entropy also show the different chaotic/integrable transitions of the tails and the bulk of spectrum (See Figure~\ref{fig: ratio dist level}).

%
%
%
%
%
%

\clearpage

\section{Computing the Thouless time $t_\text{Th}$}
\label{sec:thouless_computation}

Having obtained a qualitative understanding on the behavior of the Thouless time as computed by the connected unfolded SFF in Section \ref{sec:numerics_preliminary}, and having confirmed these preliminary observations by the arguments in Section \ref{sec:chaos_energy}, we now want to characterize in a more quantitative way the Thouless time as a function of $\kappa$.
We restrict in this Section the analysis to the values $\kappa \leq 15$ and omit $\kappa\ge 25$ since, as we already observed, these are the values in which the connected unfolded SFF sees the transition from chaos to integrable behavior around $\kappa\sim 1$.
As we stressed this is also the range in which the chaotic to integrable transition is seen by studying the OTOCs, as  shown by \cite[Figure 3]{Garcia-Garcia:2017bkg}.

The results in Figure \ref{fig:beta0} and \ref{fig:counSFFN32} confirm that the Thouless time is an increasing function of $\kappa$. 
Interestingly, the connected unfolded SFF for $\kappa = 0.75$ is very similar to the connected unfolded SFF for $\kappa = 1$, when $N=30$.
This means that the Thouless time reaches the saturation for smaller values of $\kappa$: at least for $\kappa \sim 0.75$ when $\beta = 10^{-4}$ and $N = 30$.

Figure \ref{fig:beta0} also provides us a simple way to concretely characterize the Thouless time.
Indeed, it is evident that the time at which the ramp starts (which is the definition of the Thouless time) is also a local minimum of the connected unfolded SFF.
Hence, we can determine the Thouless time  by simply taking a zoom in the area close to the local minimum and then determining the minimum.
In concrete, we found the minimum for each connected unfolded SFF by taking points in a logarithmic scale, separated by  $\Delta\log t=(\log[10^2]-\log[10^{-4}])/(10^3-1)=0.0138$, and with a time window chosen, for $N=30$ and $\beta=10^{-4}$ for example, as follows:
\begin{alignat}{4}
&0.003\, &&<\,t\,<\,\,0.1\qquad&&\text{for}\;\kappa=0\ ,\nonumber \\
&0.003\,&&<\,t\,<\,\,0.1\qquad&&\text{for}\;\kappa=0.1\ ,\nonumber \\
&0.003\,&&<\,t\,<\,\,0.1\qquad&&\text{for}\;\kappa=0.2\ ,\nonumber \\
&0.02\,&&<\,t\,<\,\,0.1\qquad&&\text{for}\;\kappa=0.35\ ,\nonumber \\
&0.04\,&&<\,t\,<\,\,1\qquad&&\text{for}\;\kappa=0.5\ ,\nonumber \\
&0.3\,&&<\,t\,<\,\,3\qquad&&\text{for}\;\kappa=0.75\ ,\nonumber \\
&0.3\,&&<\,t\,<\,\,3\qquad&&\text{for}\;\kappa=1\ ,\nonumber \\
&0.3\,&&<\,t\,<\,\,3\qquad&&\text{for}\;\kappa=2\ ,\nonumber \\
&0.3\,&&<\,t\,<\,\,3\qquad&&\text{for}\;\kappa=5\ ,\nonumber \\
&0.3\,&&<\,t\,<\,\,10\qquad&&\text{for}\;\kappa=10\ ,\nonumber \\
&0.3\,&&<\,t\,<\,\,10\qquad&&\text{for}\;\kappa=15\ .
\end{alignat}

\begin{figure}[t!]
\centering
\fbox{\includegraphics[width=8cm]{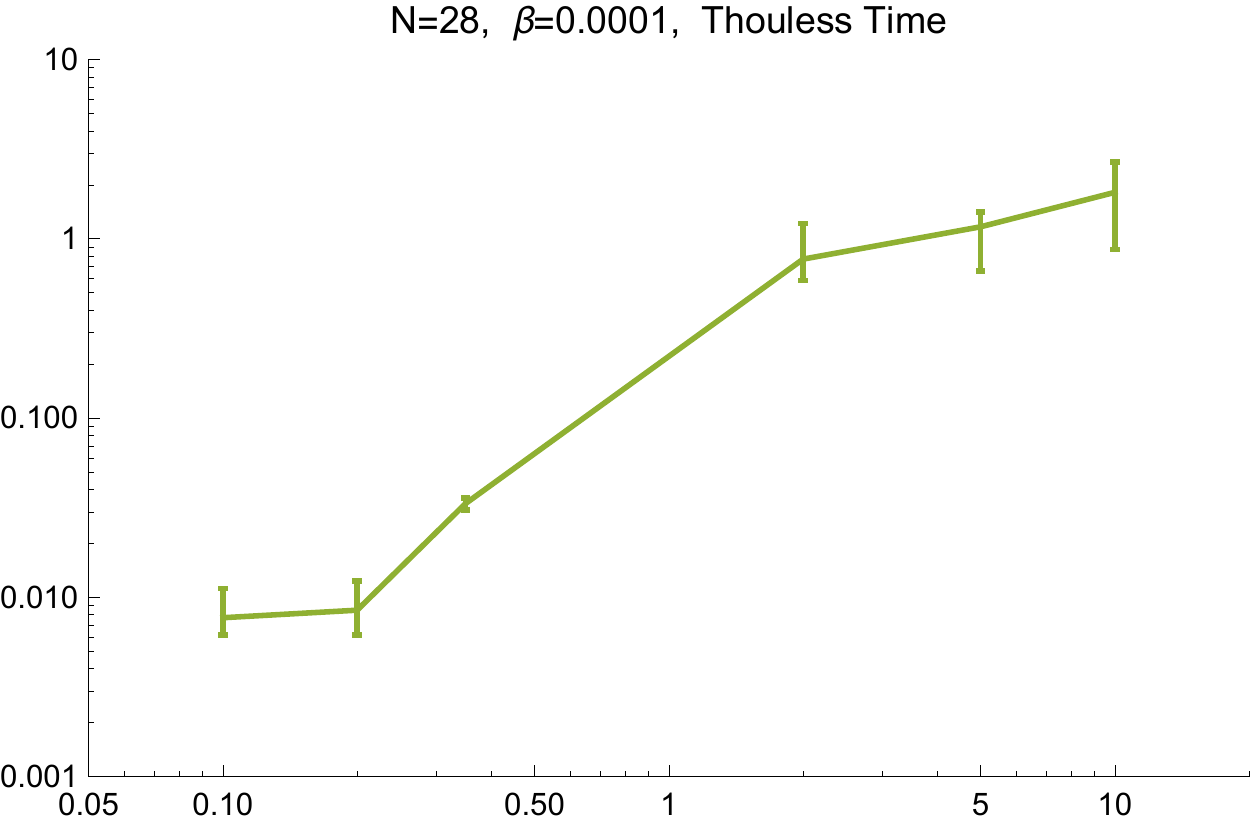}}\\
\fbox{\includegraphics[width=8cm]{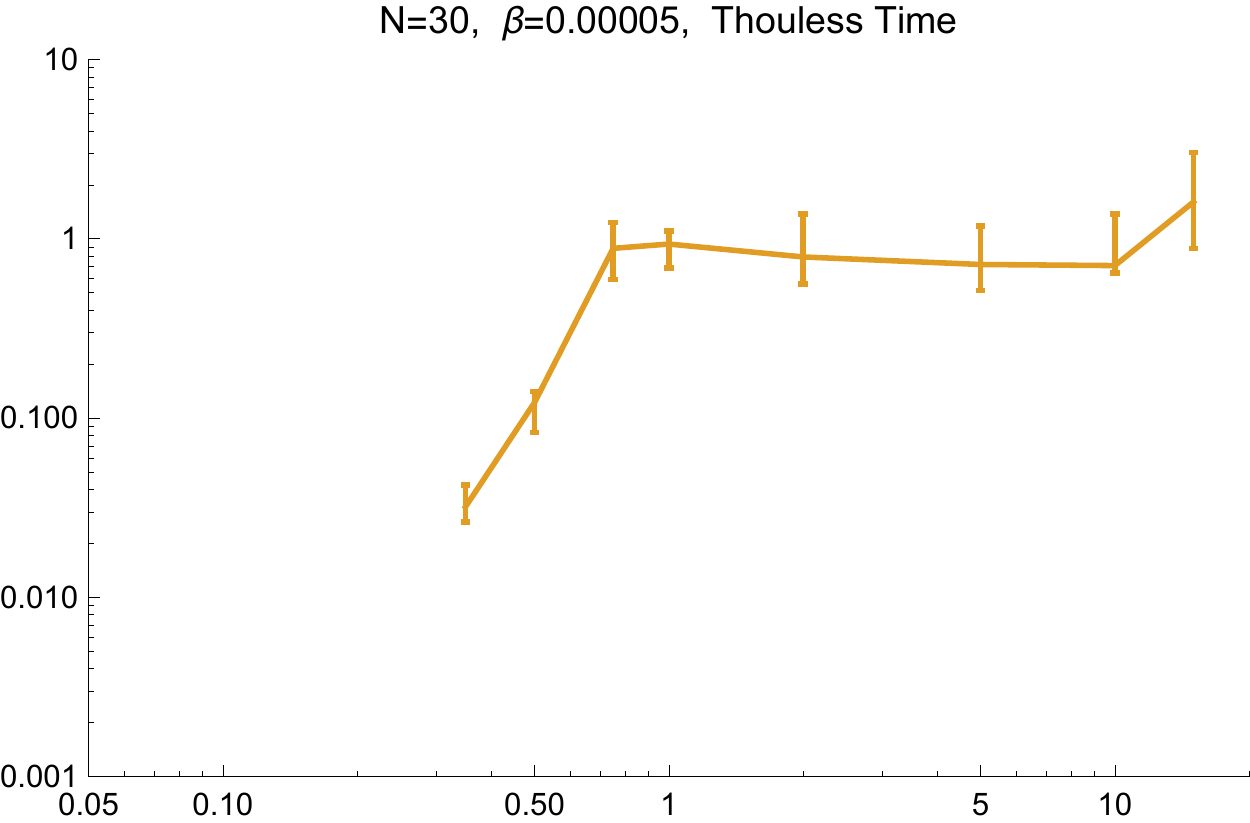}}\quad
\fbox{\includegraphics[width=8cm]{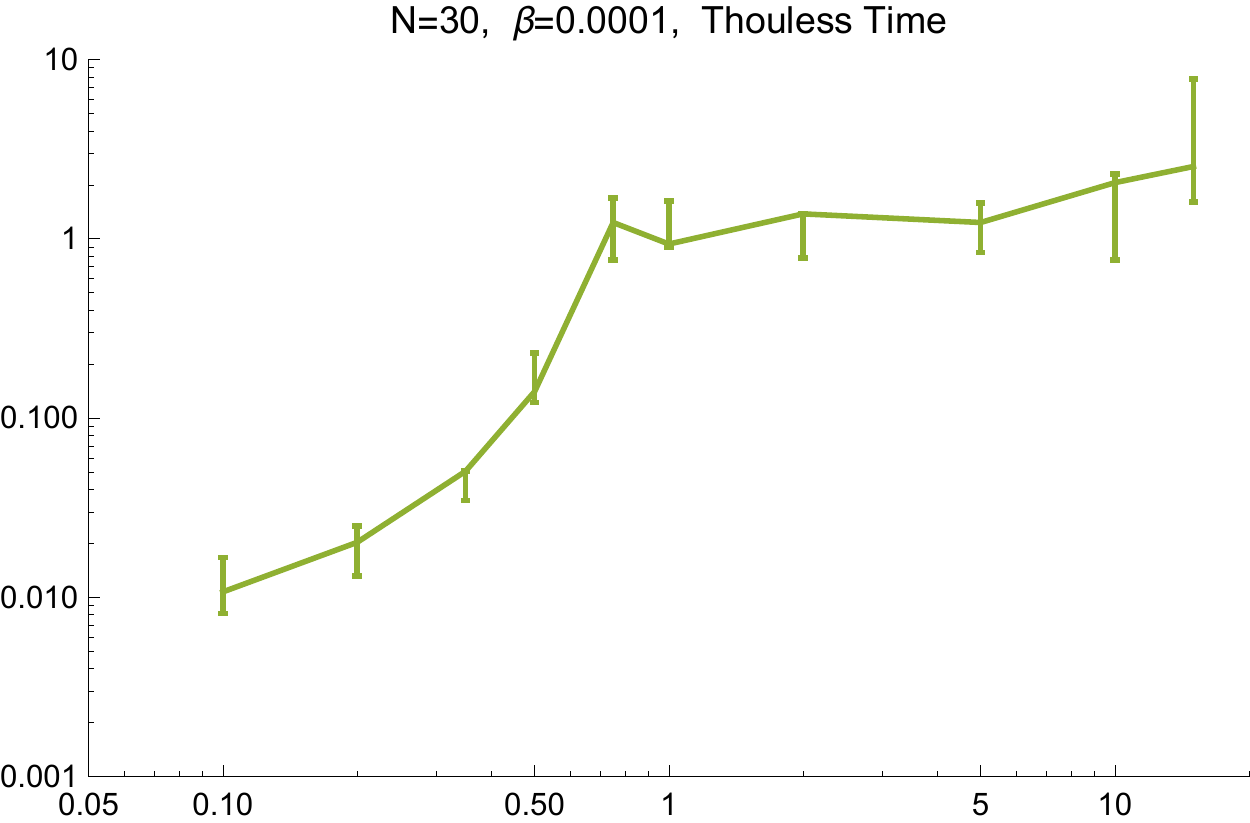}}\\
\fbox{\includegraphics[width=8cm]{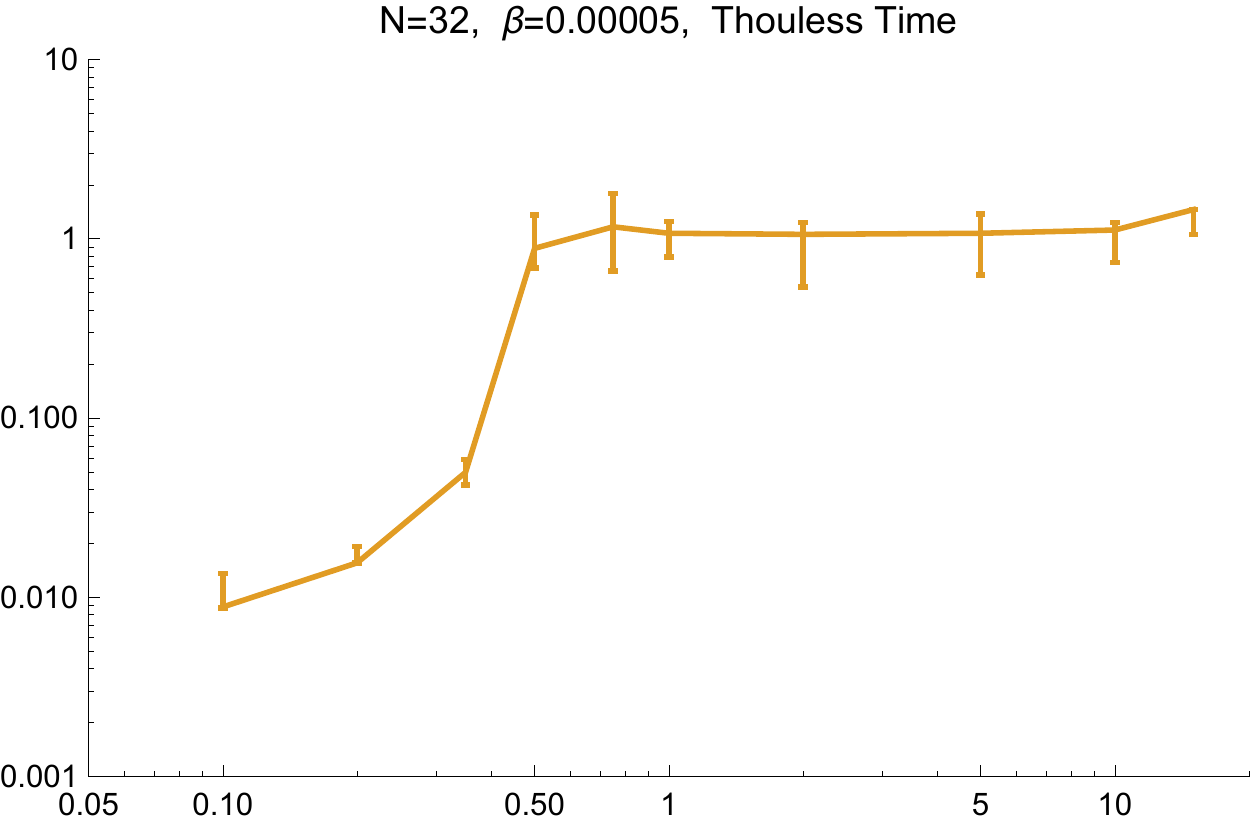}}\quad
\fbox{\includegraphics[width=8cm]{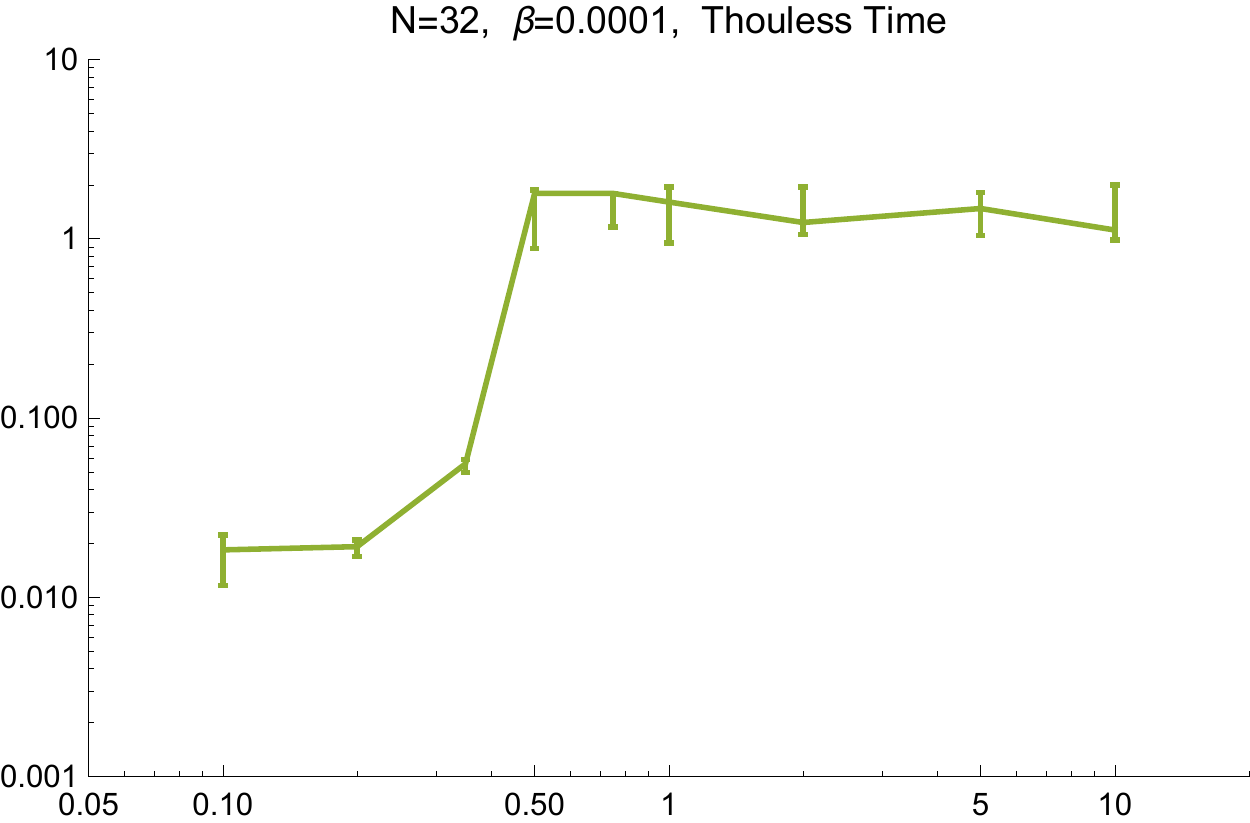}}
\caption{
Thouless time as a function of $\kappa$, for $N=28, \, 30, \, 32$ and $\beta=5\times 10^{-5}, \, 10^{-4}$.
}
\label{tTh_Ngamma30}
\end{figure}

We have also estimated the error by the maximal/minimal values of $t$ where the connected, unfolded SFF is valued  $1.1\times \text{(minima)}$, following a similar criterion to the one used in \cite{Cotler:2016fpe}. The results are shown in Figure \ref{tTh_Ngamma30}.
As we can see, the determination of the Thouless time looks accurate for low values of $\kappa$, except the case with $N=30$ and $\beta=5\times 10^{-5}$ where we could not determine $t_\text{TH}$ for $\kappa\le 0.75$ since the local minimum is hidden by the large oscillation.
For large values of $\kappa$ unfortunately the error bars are large, because, for $\kappa = 0.75 , \, 1$, the connected unfolded SFF around the dip shows an almost flat behavior and so it is hard to determine the true minimum.
It is possible that, by taking more samples, these error bars could be reduced a bit but the flatness of the connected unfolded SFF in that area prevents the error bars to become small even for very large ensemble realizations.
On the other hand, it is possible that increasing $N$ and going to the large $N$ limit, the flatness could be reduced.

Interestingly we see that, especially at $\beta = 10^{-4}$, the transition from the chaotic to integrable behavior becomes sharper and sharper when $N$ gets increased.
This is an expected feature which the determination of the Thouless time using the connected unfolded SFF seems to be able to capture.

\section{The Gaussian filtered SFF}
\label{sec:gaussian_filter}

In this Section, we want to compute the Gaussian filtered SFF, as defined in eq. (\ref{eq:gaussianfilter}), for the mass deformed SYK model, to have a comparison with the results we obtained for the connected unfolded SFF.
 
We remind that the Gaussian filtered SFF has been introduced in \cite{Gharibyan:2018jrp} as a variation of the full not-unfolded SFF, in which the contributions coming from the tails of the spectrum are {\it exponentially} suppressed with respect to the contributions coming from the bulk of the spectrum. 
It has been shown in \cite{Gharibyan:2018jrp} that this procedure allows to see the onset of RMT behavior at earlier times than the standard SFF.
It is not clear that the time at which the RMT behavior shows up in the Gaussian filtered SFF is the true time at which the RMT physics appear in the SYK dynamics, but at least this time provides an upper bound to the true time at which the RMT dynamics shows up.

As we already mentioned, the Gaussian filtered SFF, $Y (\alpha, \, t)$, depends on an additional parameter $\alpha$, which controls the dimension of the energy window which gets magnified by the Gaussian filtered SFF. 
The authors of \cite{Gharibyan:2018jrp} have found an optimum value of $\alpha$ for the SYK model corresponding to $\alpha = 2.9$ and, in the study of the Gaussian filtered SFF for the mass deformed SYK model, we have taken the same value of $\alpha$.

\begin{figure}[t!]
\begin{center}
\fbox{\includegraphics[width=8cm]{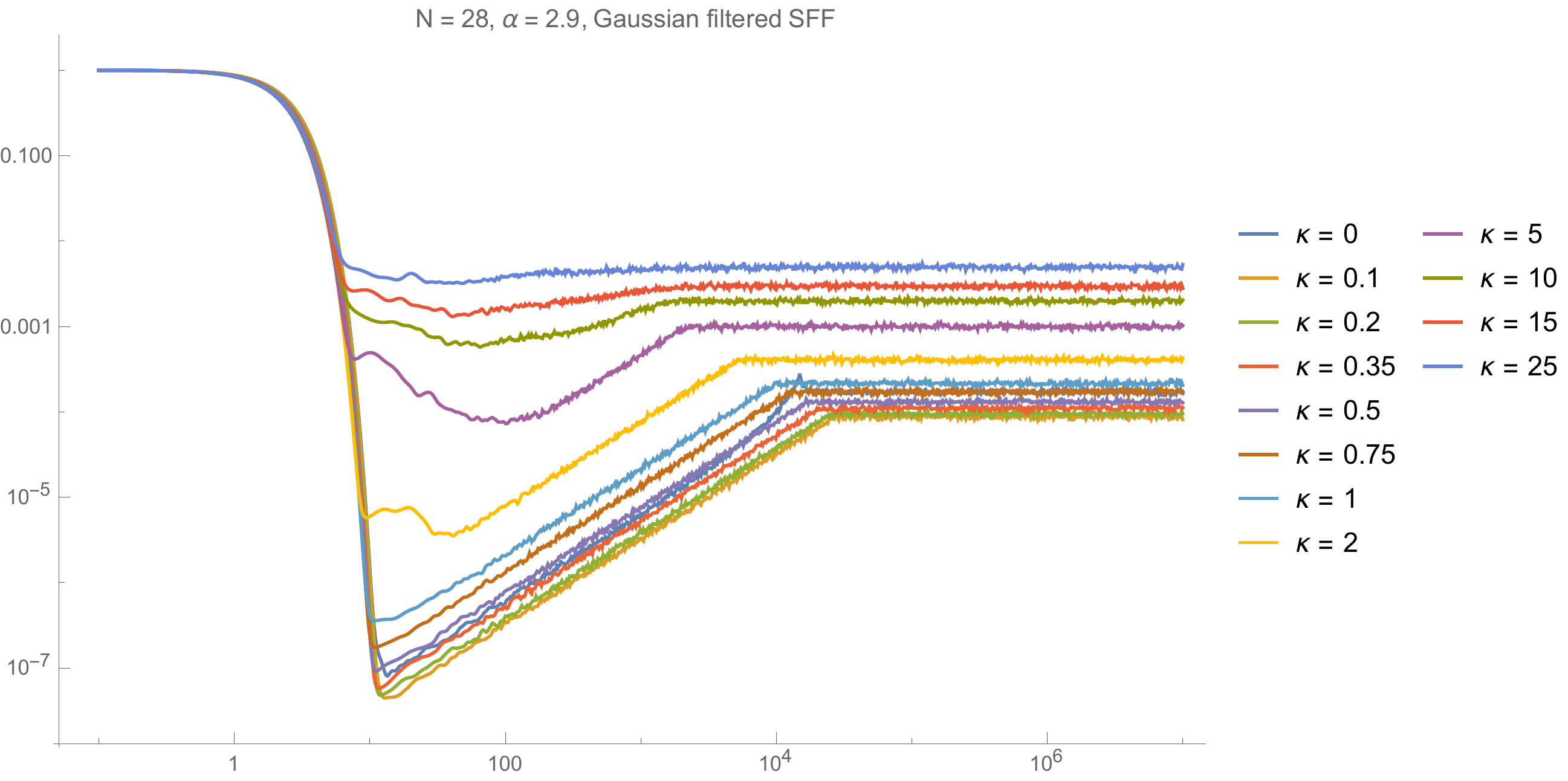}}\\
\fbox{\includegraphics[width=8cm]{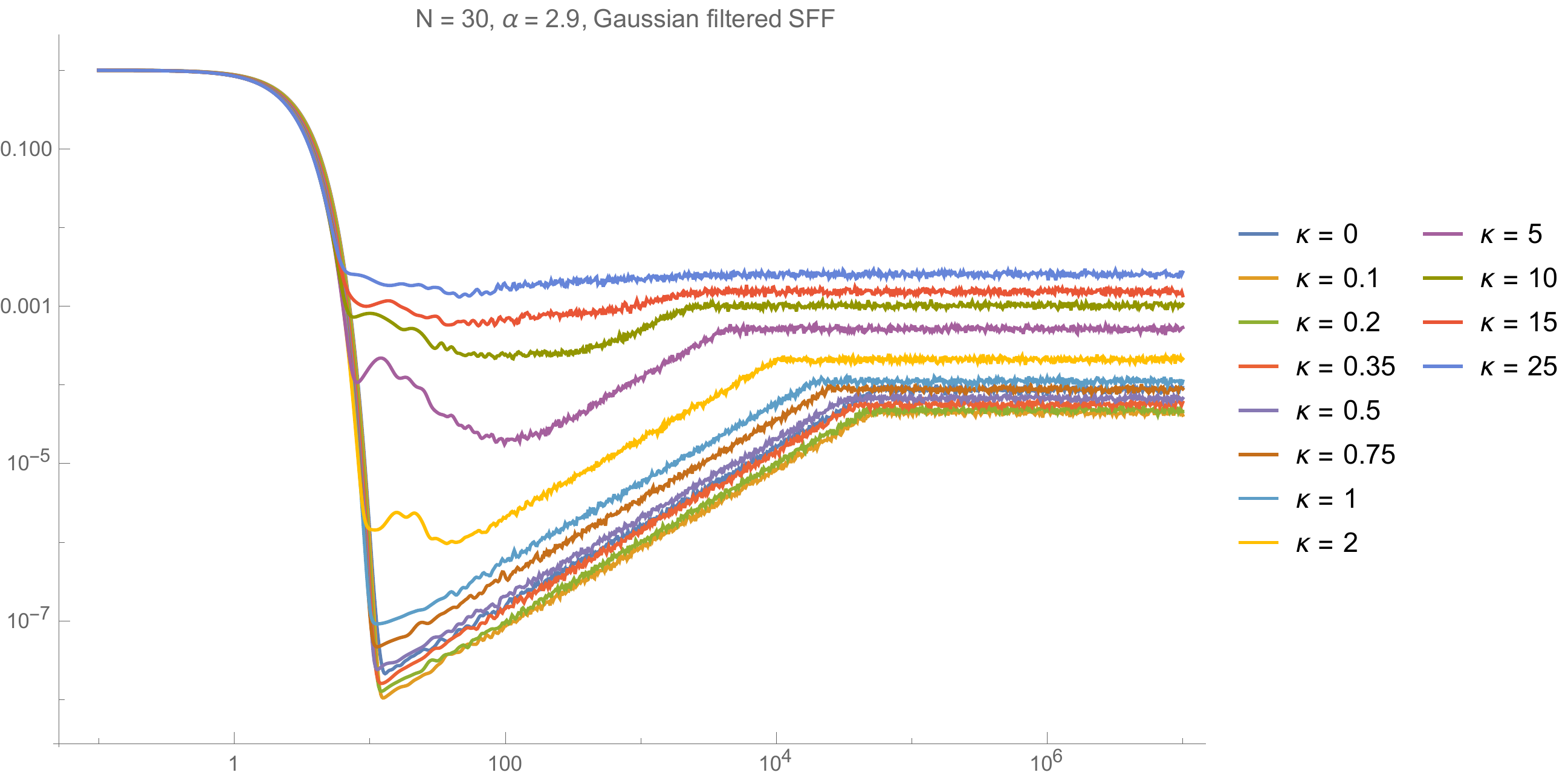}}\quad
\fbox{\includegraphics[width=8cm]{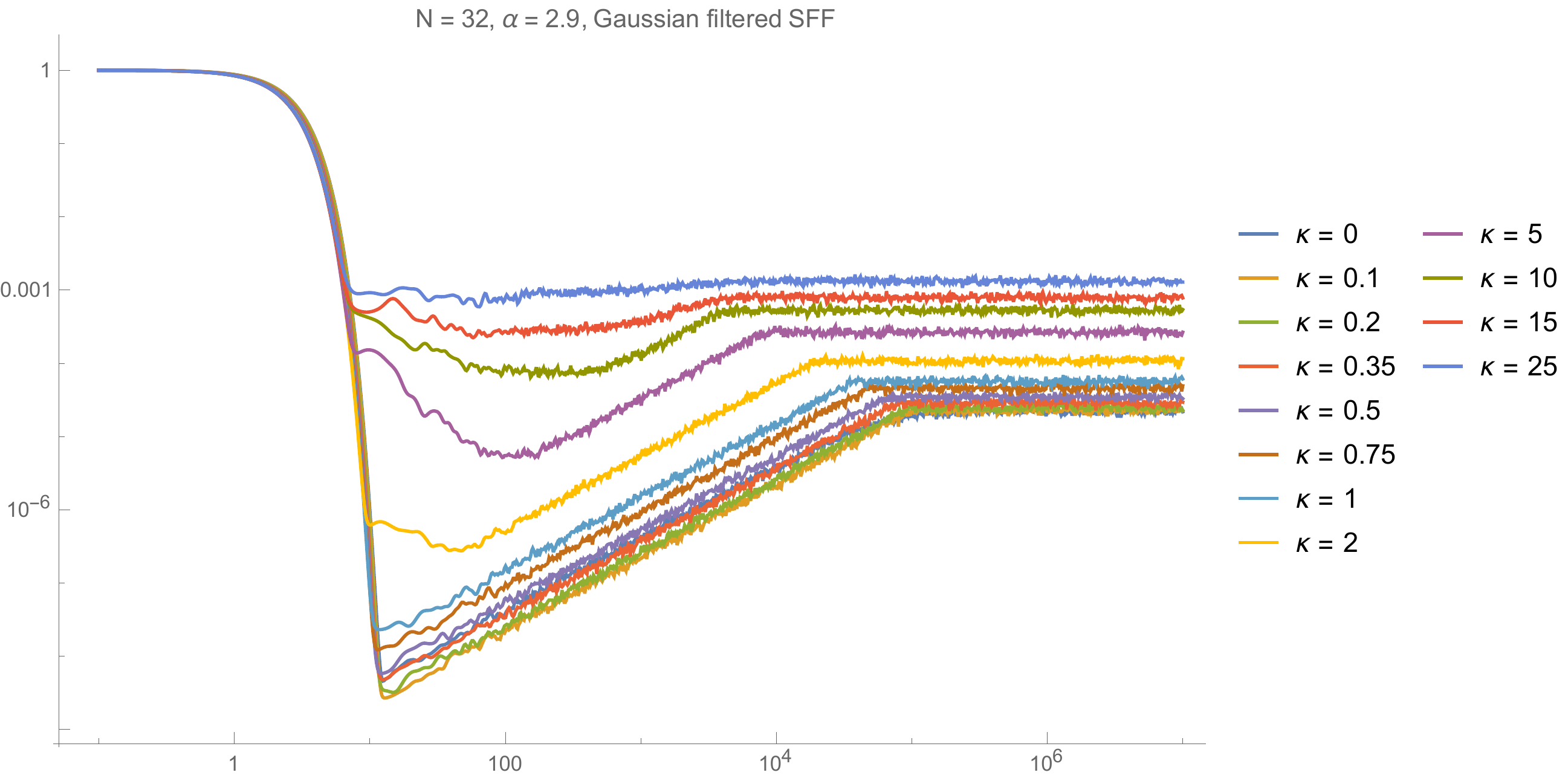}}
\end{center}
\caption{The Gaussian filtered SFF, for $N = 28, \, 30 , \, 32$, at different values of $\kappa$ and with $\alpha = 2.9$. The ensemble average is done over $M = 603$ ensemble realizations for $N = 28$, $M = 312$ ensemble realizations for $N = 30$ and $M = 150$ ensemble realizations for $N = 32$.}
\label{fig:gaussian_filter}
\end{figure}

In Figure \ref{fig:gaussian_filter} we have reported the Gaussian filtered SFF for the cases of $N = 28 , \, 30$ and $32$, averaged over $603, \, 312$ and $150$ ensemble realizations, respectively, at various values of $\kappa$.
It is immediate to recognize that the Gaussian filtered SFF is {\it not} seeing the transition from chaos to integrable dynamics at the low values of $\kappa$: indeed, from Figure \ref{fig:gaussian_filter}, we see that the plot is pretty insensitive to the value of $\kappa$ when we are in the regime $\kappa \leq 1$.

On the other hand, we see that the transition from chaos to integrable dynamics is well-described by the Gaussian filtered SFF at {\it higher} values of $\kappa$: starting from $\kappa = 2$, we see that the graph displays, after the initial Gaussian fall-off, an intermediate range in which the behavior is not very clear.
In total, we see that the RMT behavior, still present at $\kappa = 2$, shows up later, a signal that the chaotic features of the model are becoming weaker.

The RMT behavior continues to be reduced when we further increase $\kappa$ and we finally see the transition to the full integrable dynamics around $\kappa \sim 15$.
We recall that this value of $\kappa$ for the transition was already observed in \cite{Garcia-Garcia:2017bkg} from the RMT analysis, which therefore roughly agrees with the results of the Gaussian filtered SFF.
By contrast, we have seen in the previous Sections that the connected unfolded SFF is seeing a transition at lower values of $\kappa$, $\kappa \sim 1$, and that this value is in rough agreement with the results from the OTOCs.

Another interesting feature to observe, which agrees with the observations of \cite{Gharibyan:2018jrp}, is that the behavior of the Gaussian filtered SFF is almost unchanged when $N$ increases.

We can understand this discrepancy between the connected unfolded SFF and the Gaussian filtered SFF as a consequence of the way in which the integrable behavior enters in the system as a function of $\kappa$, that we have described in Section \ref{sec:chaos_energy}: we have seen that the integrable behavior affects first the states closed to the ground states and that later (for higher values of $\kappa$) affects the states in the bulk.
Since the Gaussian filtered SFF is, by construction, mostly controlled by the states in the bulk of the spectrum, it sees the integrable transition at high values of $\kappa$, while it is pretty insensitive to the changes in the states close to the ground states, which start to appear at low values of $\kappa$.
The connected unfolded SFF does exactly the opposite: it is mostly controlled by the states close to the ground state and so it sees the transition at lower values of $\kappa$.
This transition is also the transition that seems to control the behavior of the OTOCs.
On the other hand, it does not show any dependence on $\kappa$ when this is large.

Summarizing, we see that the connected unfolded SFF and the Gaussian filtered SFF --- at least for the particular case of the mass-deformed SYK model and as functions of $\kappa$ --- seem to be complementary: they are dominated by different portions of the spectrum and so they show different features of the transition from chaotic to integrable behavior.

\section{Discussion}
\label{sec: discussion}

In this paper, we have studied the onset of RMT dynamics in a mass-deformed version of the SYK model \cite{Garcia-Garcia:2017bkg}, where the usual SYK Hamiltonian with quartic interaction is perturbed by a quadratic term (a random mass-term). The strength of the quadratic deformation is controlled by a coupling constant $\kappa$.
As observed already in \cite{Garcia-Garcia:2017bkg}, when $\kappa = 0$ the model is equivalent to the standard SYK model  and so it is maximaly chaotic and shows RMT behavior. 
On the other hand, when $\kappa$ is very large, the model is dominated by the quadratic term and hence it is not chaotic.
It has been further shown in \cite{Garcia-Garcia:2017bkg} that this chaotic/integrable transition happens at certain finite value of $\kappa$.
However, the precise value of $\kappa$ at which the transition  takes place is dependent on the probe one is using to test chaos: using the OTOCs the transition becomes evident for small values of $\kappa$ ($\kappa \sim 1$), while the RMT analysis performed in \cite{Garcia-Garcia:2017bkg} (like the level spacing distribution using large portions of the spectrum) detects the chaotic/integrable transition for larger values of $\kappa$, around $\kappa \sim 15$.

With the aim of clarifying this point, we studied the chaotic/integrable transition as seen by other two probes of quantum chaos: the connected unfolded SFF and the newly introduced \cite{Gharibyan:2018jrp} Gaussian-filtered SFF.
These two observables have the property to probe different regions of the spectrum: the connected unfolded SFF, at sufficiently low values of the temperature, probe the tail of the spectrum --- i.e. the states closer to the ground state --- while the Gaussian-filtered SFF by construction is focused in the bulk of the spectrum.
We have checked that these two observables also see the transition to the integrable regime at different values of the coupling constant $\kappa$: the connected unfolded SFF sees a sharp transition of the Thouless time at small values of $\kappa$, which qualitatively agrees with the behavior seen by the OTOCs.
On the other hand, the Gaussian-filtered SFF sees the transition to the integrable regime at values of $\kappa$ in qualitative agreement with the RMT analysis of \cite{Garcia-Garcia:2017bkg}.

This observation suggested us that the chaotic/integrable transition does not take place homogeneously on the entire spectrum, but that states closer to the ground state (in the tail) move to the integrable behavior for lower values of $\kappa$ and that the connected unfolded SFF (and the OTOCs as well) is mostly controlled by these states. 
On the other hand, the states in the bulk of the spectrum are less influenced by the quadratic perturbation and so they move to the integrable regime at larger values of $\kappa$, and these are the values at which the Gaussian-filtered SFF sees the transtion to the integrable regime.
To confirm this picture, we studied the transition to the integrable regime for {\it portions} of the spectrum including only $L$ levels starting from the ground states.
We performed this confirming study both by analysing the level repulsion (and the $\tilde r$-statistics) and by studying the IPR distribution and the IR diversity (which depend also on the energy eigenstates) and found agreement between all these methods.

Given our results, we are tempted to conjecture that the connected unfolded SFF might be an interesting observable that deserves further studies and that it could be important in order to find a contact point between the early time definition of chaos (based on the OTOCs) and the late time definition of chaos.
For example, an intriguing question that one would like to address is the following, already formulated in \cite{Krishnan:2017lra}: is there a way to characterize maximally chaotic systems by looking at their spectrum?
It is possible that the answer to this question could be related to the study of the connected unfolded SFF.

Even though the connected unfolded SFF has some intriguing features that we described, we have to say that there are subtleties, already discussed in \cite{Gharibyan:2018jrp}, related to unfolding and that they should be clarified in order to make more robust the computations performed with the connected unfolded SFF.
In the unfolding procedure, we have adopted  the strategy that we first fit the number of states $n(E)=\sum\theta(E-E_i)$ of the spectrum of each ensemble realization by a fifth order polynomial in $E$, which we call $n_\text{av}(E)$, to define the unfolded eigenvalues $\tilde E_i=n_\text{av}(E_i)$.
Though the purpose of unfolding is to remove the global structure of the level distribution and to keep only the local fluctuations, hence $n_\text{av}(E)$ should almost coincide with $n(E)$ everywhere, in our current method we observe that the deviation is larger at the edges.
Indeed we also observed that if we perform the fitting on the edges and bulk separately, we obtain some different results. 
Hence, our results depend somehow on the way in which the unfolding is realized; even though the checks we performed, some of them are independent on the unfolding, suggest that the general picture should be robust.
However, since there is not an a-priori correct choice of fitting Ansatz and hence unfolding is not a well-defined procedure, so far we cannot fully trust our results.
In the random matrix model there is a canonical way to define $n_\text{av}(E)$: we should use the ``average density'' $\rho_\text{av}(E)\equiv \langle\sum_i\frac{1}{E-E_i}\rangle$ and define $n_\text{av}(E)=\int^E_{-\infty}dE'\rho_\text{av}(E')$.
Since the observables in the (mass-deformed) SYK model are also defined as ensemble averages, it could be another plausible choice to define $n_\text{av}$ in the same way.\footnote{
We thank Antonio Garc\'ia-Garc\'ia  and Fidel Schaposnik Massolo for useful discussions on this point.
}
Of course, it would be very nice if one could obtain an analytical good approximation of the large $N$ distribution $\rho_\text{av}(E)$.
For this purpose, the results of \cite{Garcia-Garcia:2017pzl}, \cite{Garcia-Garcia:2018kvh} might be very useful.
Once a robust unfolding procedure will be established, it will be very interesting to extrapolate the large $N$ limit of the connected unfolded SFF results. 
To this purpose, the methods of \cite{Gur-Ari_progress} could be very useful.

%
\section*{Acknowledgement}
%

We would like to thank Subhro Bhattacharjee, Antonio Garc\'ia-Garc\'ia, Guy Gur-Ari, Masanori Hanada, Sungjay Lee, Loganayagam R. and Fidel Schaposnik Massolo for extensive discussions and/or correspondence.
We thank Korea Institute for Advanced Study for providing computing resources (KIAS Center for Advanced Computation Abacus System) for this work. 
A part of calculations were performed using the computing facilities at ICTS; in particular, JY acknowledge the use of the high-performance computing clusters Mario and Pacman. JY thanks the Korea Institute for Advanced Study~(KIAS) for the hospitality and support during the initial stages of this work, within the program ``IPMU-KIAS-Kyunghee Univ. joint workshop''. This research was supported in part by the International Centre for Theoretical Sciences (ICTS) for participating in the program - Kavli Asian Winter School (KAWS) on Strings, Particles and Cosmology 2018 (Code: ICTS/Prog-KAWS2018/01). JY also thanks the Mandelstam Institute for Theoretical Physics for the hospitality and partial support within the program ``The Second Mandelstam Theoretical Physics School and Workshop''. JY gratefully acknowledge support from International Centre for Theoretical Sciences (ICTS), Tata Institute of Fundamental Research, Bangalore. 
DR thanks the Jiao Tong University, Shanghai, and Antonio Garc\'ia-Garc\'ia for the warm hospitality during the final stages of this work.

\appendix

\section{The $\tilde{r}$-parameter statistics}
\label{app:r-statistics}

In this Appendix, we compute the distribution of the ratio of the nearest-neighborhood spacing, the so-called $\tilde{r}$-parameter statistics~\cite{Oganesyan:aa,Atas:2012aa}. 
We will compute it for the full spectrum of the mass-deformed SYK model (at various values of $\kappa$) in Appendix \ref{app:r-statistics full spectrum} and for subsectors of the mass-deformed SYK spectrum in Appendix \ref{app: r-statistics of subsector} (again, at different values of $\kappa$).
In agreement with the results of Section \ref{sec:chaos_energy}, we will see that the chaotic/integrable transition affect the different portions of the spectrum inhomogeneously. 
However, we will notice that the results of Appendix \ref{app: r-statistics of subsector} seem to be dependent on the unfolding, in the sense that the separation between the transitions as seen by the low-lying modes and the  states in the bulk of the spectrum is more evident when we consider the unfolded spectrum than when we consider the not unfolded spectrum (even though they are still visible also in this case).
This is an unexpected feature, since it is known that the results of the $\tilde{r}$-parameter statistics should be independent on the unfolding (though this statement is usually applied to the $\tilde{r}$-parameter statistics computed over the full spectrum).
We do not know whether this behavior is an artifact due to the fact that our unfolding procedure is not optimal and, also for this reason, we decided to study the IPR and IR diagnostics (in Sections \ref{sec: ipr} and \ref{sec: ir diversity}) to be more confident about the robustness of our findings.

\subsection{The $\tilde{r}$-parameter of full spectrum}
\label{app:r-statistics full spectrum}

The $\tilde{r}$-parameter statistics of SYK(-like) models has been discussed in~\cite{Chaudhuri:2017vrv,Ho:2017nyc,Garcia-Garcia:2018pwt}. Given the nearest-neighbor energy spacing $s_n\equiv E_{n+1} -E_n$ of an ordered energy spectrum $\{E_n\}$, the $\tilde{r}$-parameter is defined by
\begin{equation}
    \tilde{r}_n\, \equiv\, {\min(s_n,s_{n-1})\over \max(s_n,s_{n-1}) }\,=\,\min \left(r_n,{1\over r_n}\right)\ ,
\end{equation}
where we defined $r_n\equiv {s_n\over s_{n-1}}$. Note that, by definition, the $r$-parameter lies in $[0,1]$.\footnote{One may also consider the statistics of $r_n$ instead of $\tilde{r}_n$. However, the mean value of this ratio diverges for the Poisson distribution~\cite{Atas:2012aa}, which is not suitable for our case.} The Wigner-like surmise for $3\times 3$ matrices gives the distribution of the ratio of consecutive level spacing:
\begin{equation}
    p(r)\,=\,{1\over Z_\beta} {(r+r^2)^\beta\over (1+r+r^2)^{1+{3\over 2}\beta}}\ ,
\end{equation}
where $\beta=1,2$ or $4$ for GOE, GUE or GSE, respectively, and $Z_\beta$ is a normalization constant. From the distribution of $r$'s, one can evaluate the mean value of $\tilde{r}$-parameter by
\begin{equation}
	\langle \tilde{r} \rangle_W\,=\,  \int_0^1 r\; p(r) dr\,+\,\int_1^\infty r^{-1}\; p(r) dr \ .
\end{equation}
The mean values of $\tilde{r}$ for Poisson, GOE, GUE and GSE are summarized in Table~\ref{tab: r parameter}~\cite{Atas:2012aa}.
\begin{table}[t!]
\centering
{\centering
{\renewcommand{\arraystretch}{1.5}
\begin{tabular}{| >{\centering\arraybackslash}p{10mm} | >{\centering\arraybackslash}p{25mm}| >{\centering\arraybackslash}p{25mm}| >{\centering\arraybackslash}p{25mm}| >{\centering\arraybackslash}p{25mm}|}
\hline
& Poisson & GOE & GUE & GSE\\
\hline
\multirow{ 2}{*}{$\langle \tilde{r}\rangle$ }  & $2\log 2-1$ & $4-2\sqrt{3}$ & $2{\sqrt{3}\over \pi}-{1\over 2}$ & ${32\over 15}{\sqrt{3}\over \pi}-{1\over 2}$ \\
 &  $\approx 0.38629$ & $\approx 0.53590$ & $\approx 0.60266$ & $\approx 0.67617$ \\
\hline
\end{tabular}}
}
\caption{The mean value of $\tilde{r}$-parameter for Poisson, GOE, GUE and GSE.}
\label{tab: r parameter}
\end{table}
\begin{table}[t!]
\centering
{\centering
{\renewcommand{\arraystretch}{1.5}
\begin{tabular}{| c| c| c| c| c|}
\hline
\ $q=0 \; (\text{mod}\;\;4)$ & $N=0 \; (\text{mod}\;\;8)$ & $N=2 \; (\text{mod}\;\;8)$ & $N=4 \; (\text{mod}\;\;8)$ & $N=6 \; (\text{mod}\;\;8)$\\
\hline
Ensemble & GOE &  GUE  & GSE & GUE \\
\hline
\hline
$q=2 \; (\text{mod}\;\;4)$ & $N=0 \; (\text{mod}\;\;8)$ & $N=2 \; (\text{mod}\;\;8)$ & $N=4 \; (\text{mod}\;\;8)$ & $N=6 \; (\text{mod}\;\;8)$\\
\hline
Ensemble & BdG(D) &  GUE  & BdG(C) & GUE \\
\hline
\end{tabular}}
}
\caption{Random Matrix Classification of SYK Models: $q=0,2  \; (\text{mod}\;\;4)$ ($q\geqq 4$)}
\label{tab: ensemble 1}
\end{table}
The random matrix classification of SYK models has been studied in~\cite{You:2016aa,Garcia-Garcia:2016mno,Cotler:2016fpe,Li:2017aa,Kanazawa:2017dpd}, which is summarized in Table~\ref{tab: ensemble 1}. The quartic interaction in our hamiltonian~\eqref{eq:hamiltonian} belongs to GOE, GUE or GSE class depending on $N$ (See Table~\ref{tab: ensemble 1}). On the other hand, the quadratic interaction plays a role of (random) mass term and can be analytically solved. Hence, it does not show the random matrix universality. Therefore, when the quadratic interaction dominates the quartic one in~\eqref{eq:hamiltonian} (\ie for large $\kappa$), one can expect that the level spacing exhibits Poisson statistics. On the other hand, for $\kappa=0$, it will shows the corresponding random matrix behavior, and there would be a transition between the random matrix class and the Poisson statistics as $\kappa$ increases. However, since the quadratic interaction breaks the symmetry of the quartic one for $N\equiv 0,\, 4 \, (\text{mod}\, 8)$, the perturbation by the quadratic interaction could lead to additional transition to other matrix class, for sufficiently small\footnote{Note that this is not the chaotic/integrable transition which appear at larger value of $\kappa$.} $\kappa$. This transition between different classes can also be captured by the averaged $\tilde{r}$-parameters. Hence, we now discuss $\tilde{r}$-parameters as a function of $\kappa$. 

As already remarked, the $\tilde{r}$-parameter analysis would not require the unfolding procedure. However, since we have unfolded the spectrum for the connected SFF, we also consider the unfolded spectrum in addition to the spectrum without unfolding. Furthermore, we discuss the difference between the $\tilde{r}$-parameters of the unfolded spectrum and the spectrum without unfolding.

\begin{figure}[t!]
\centering
\subfloat[No Unfolding]{\includegraphics[width=8cm]{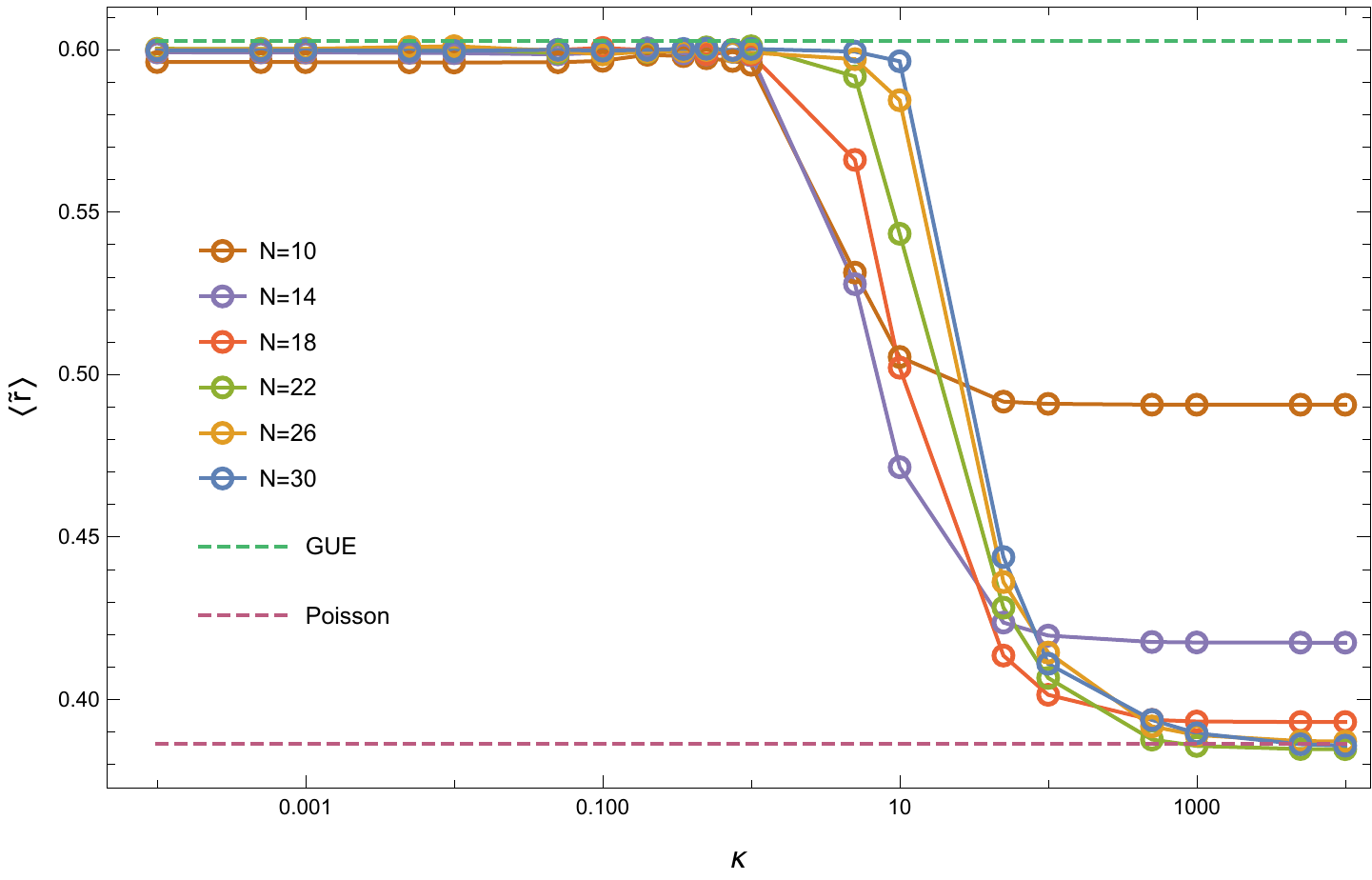}\label{fig: rpara gue 10 30}}\quad
\subfloat[Unfolded]{\includegraphics[width=8cm]{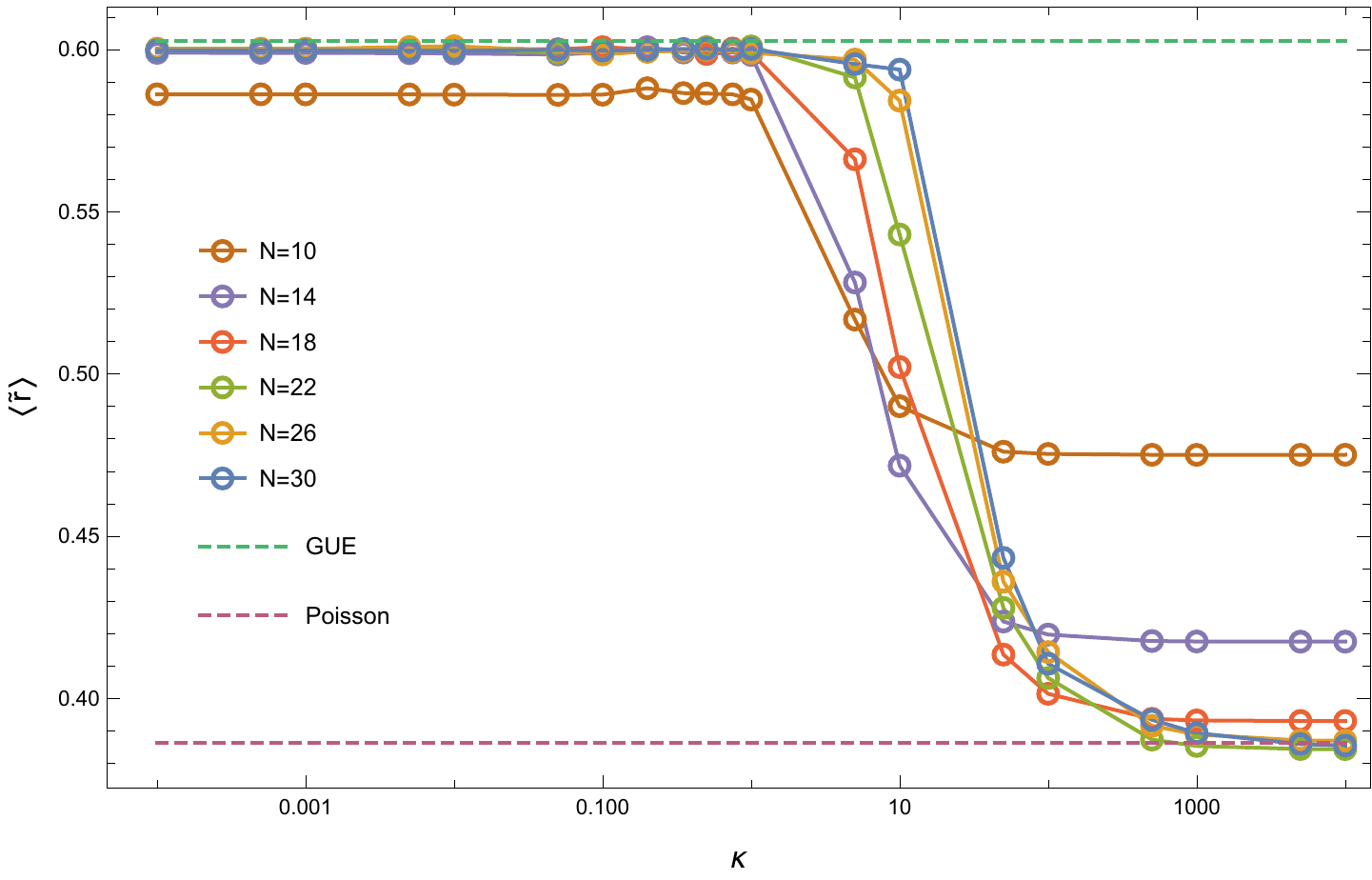}\label{fig: urpara gue 10 30}}
\caption{The mean value of the $\tilde{r}$-parameter for $N=10,14,18,22,26,30 \equiv 2, 6\;(\text{mod} 8)$ of (a) the spectrum without unfolding (b) the unfolded spectrum. The average is performed over $M = 5000,3000,1000,300,100,100$ ensembles, respectively. The transition from GUE to Poisson is observed.}
\label{fig: all rpara gue 10 30}
\end{figure}
%
%
%
%
%
%
%
%
%
%

First, let us consider $N\equiv 2, 6 \;(\text{mod} 8)$ where level statistics is expected to be GUE for $\kappa=0$. In this case, the quadratic interaction does not break the symmetry of the quartic one. Indeed, we observe only the chaotic/integrable transition from GUE to Poisson\footnote{A similar transition from GUE to Poission depending on a cutoff of the interaction was studied in~\cite{Garcia-Garcia:2018pwt}.} in Figure~\ref{fig: all rpara gue 10 30} which becomes sharper and clearer as $N$ increases. Note that the averaged $\tilde{r}$-parameter for small $N$ (\eg $N=10$ and $N=14$) is far from the Poisson statistics. This  is a consequence of the fact that the level spacing for the small $N$ does not have enough statistics.

%
%
%
%
%

\begin{figure}[t!]
\centering
\subfloat[No Unfolding]{\includegraphics[width=8cm]{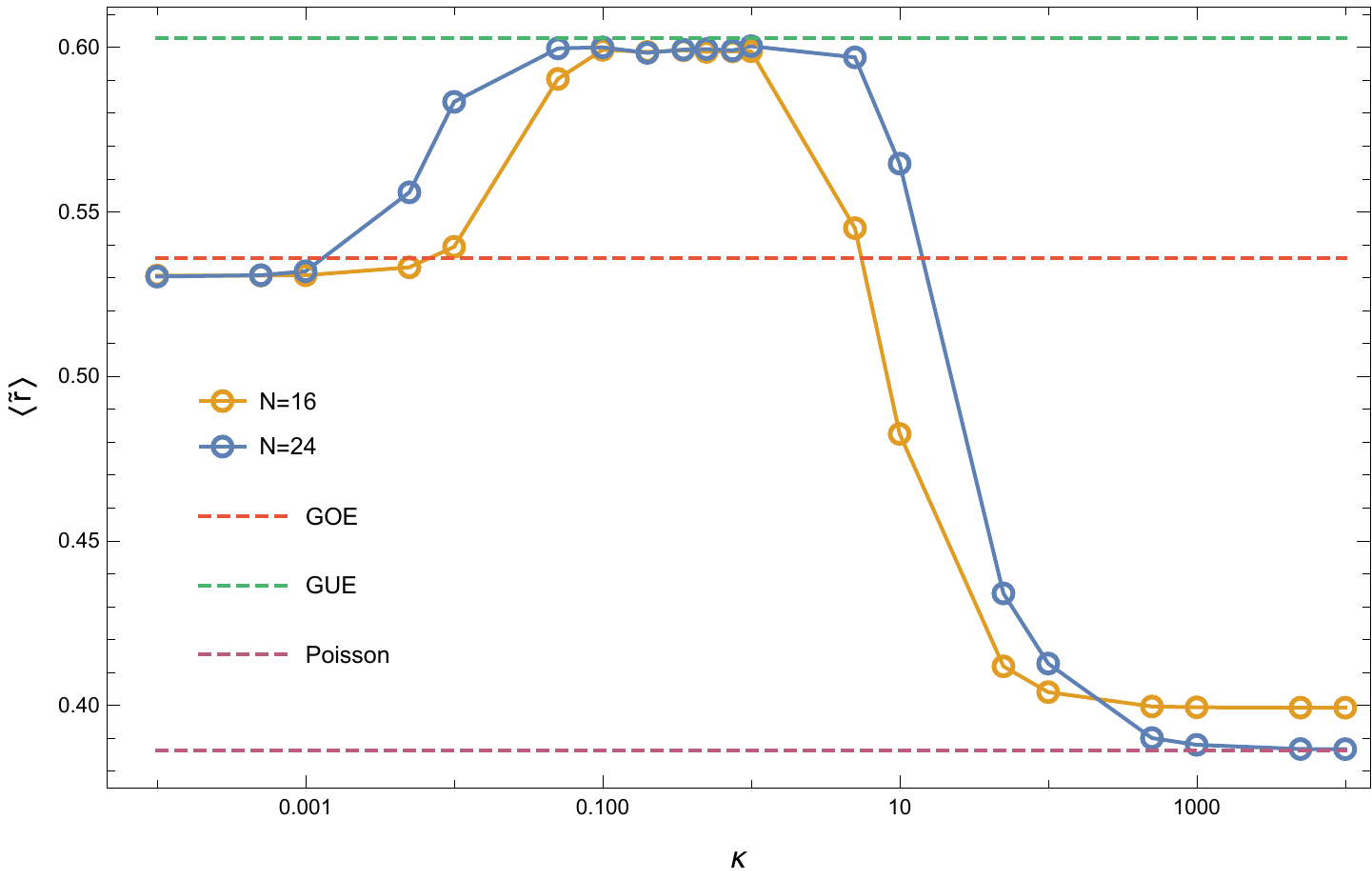}\label{fig: rpara goe 16 24}}\quad
\subfloat[Unfolded]{\includegraphics[width=8cm]{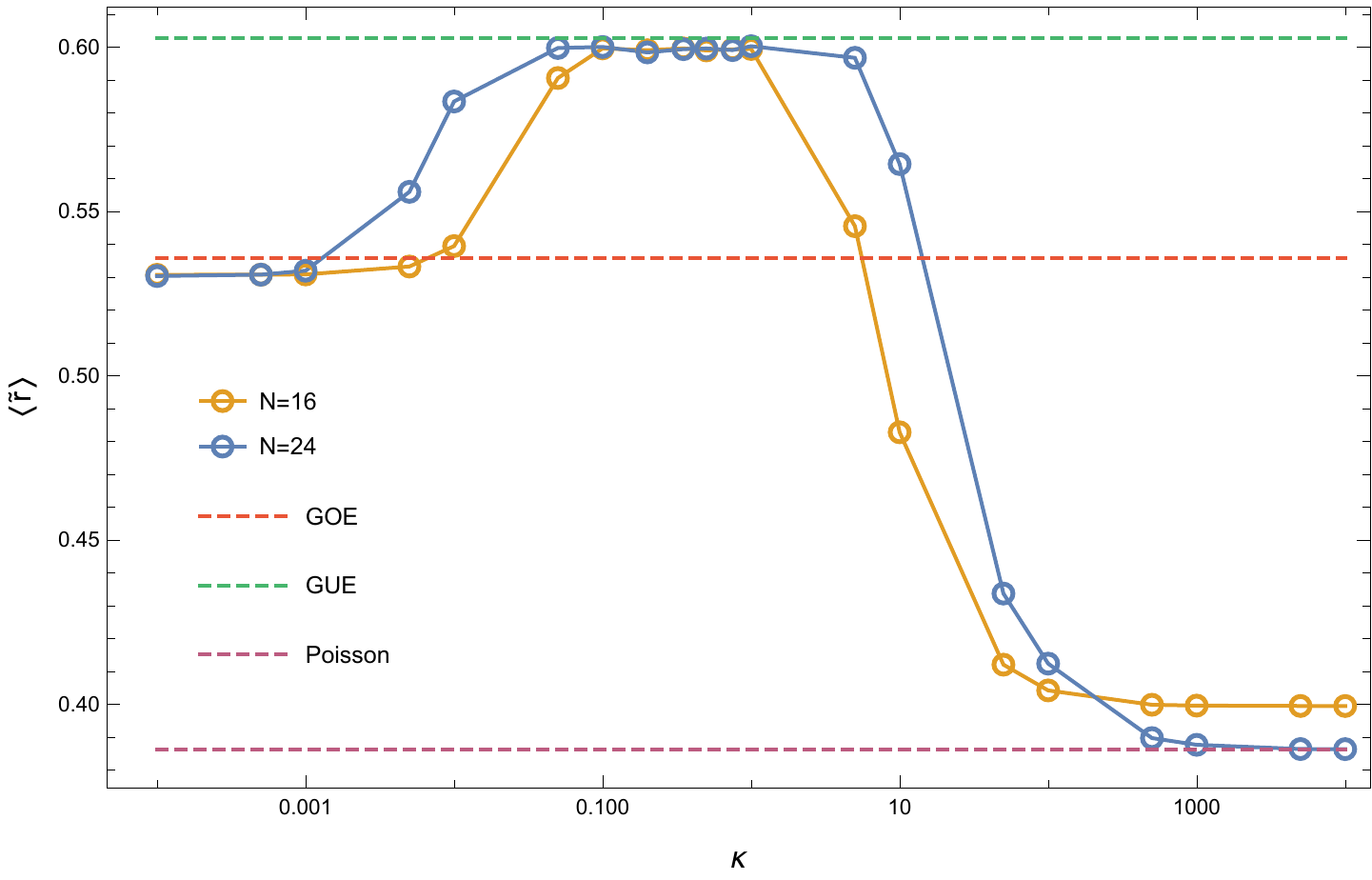}\label{fig: urpara goe 16 24}}
\caption{The mean value of the $\tilde{r}$-parameter for $N=16,24 \equiv 0\;\;(\text{mod} 8)$ of (a) the spectrum without unfolding (b) the unfolded spectrum. The average is performed over $M = 2000,100$ ensembles, respectively. We observe transitions from GOE to GUE and from GUE to Poisson.}
\label{fig: all rpara goe 16 24}
\end{figure}
For $N\equiv 0\;(\text{mod} 8) $, the quartic interaction belongs to the GOE class. However, the quadratic one breaks the symmetry of the GOE, which leads to a transition from GOE to GUE at small $\kappa\sim 10^{-2}$ (see Figure~\ref{fig: all rpara goe 16 24}). It would be interesting to study the intermediate transition from GOE to GUE in larger $N$. For instance, one can check whether in large $N$ the intermediate transition takes place at non-zero value of $\kappa$ or not. But, the limitation of numerical computation makes it difficult to perform the large $N$ interpolation of this transition. As $\kappa$ increases further, one can observe the second transition from GUE to Poisson at $\kappa\sim 10$ due to the dominating quadratic interaction  (see Figure~\ref{fig: all rpara goe 16 24}).

%

%
%
%
%
%

%

%
\begin{figure}[t!]
\centering
\subfloat[No Unfolding]{\includegraphics[width=8cm]{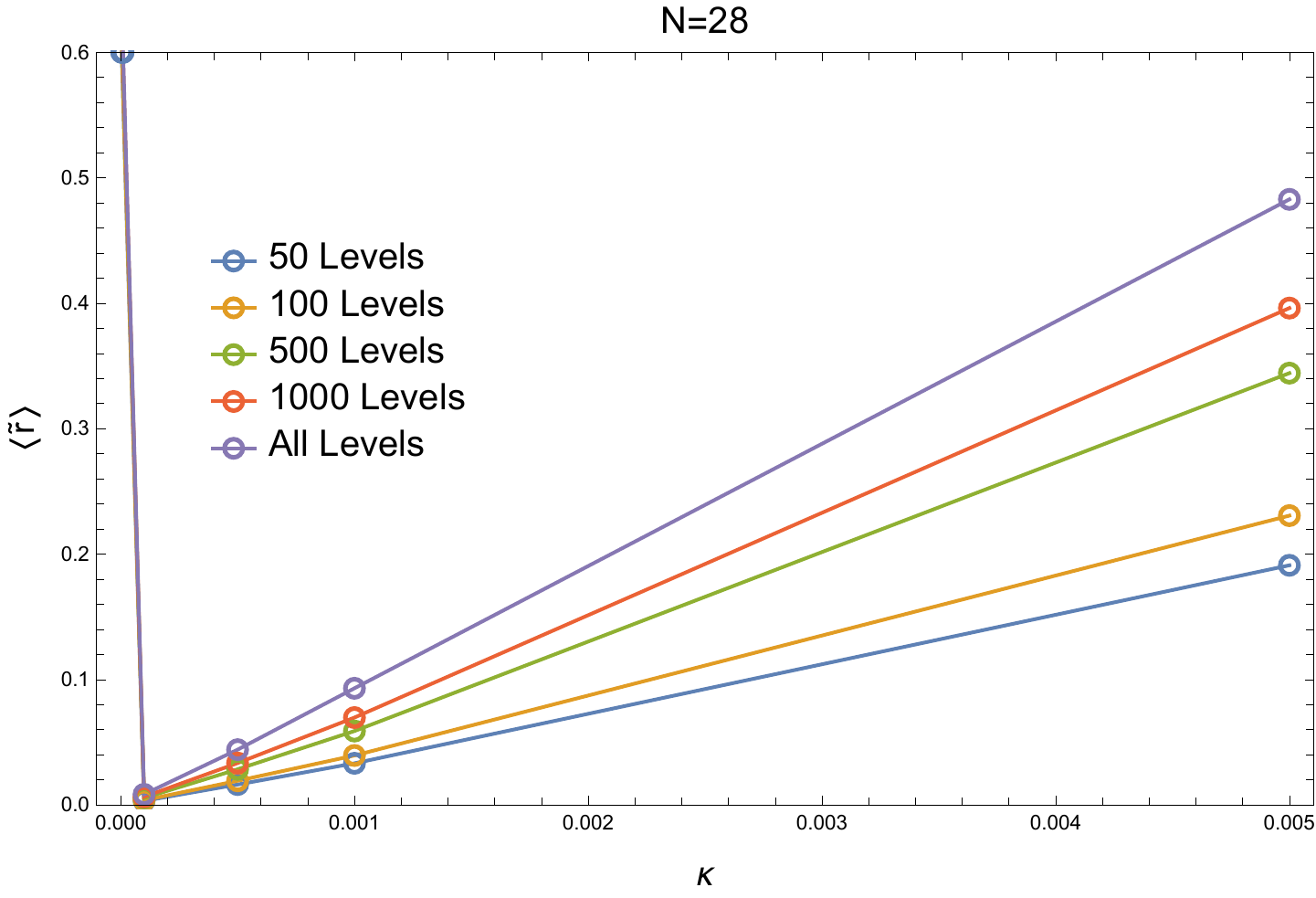}\label{fig: rpara 28 small k}}\quad
\subfloat[Unfolded]{\includegraphics[width=8cm]{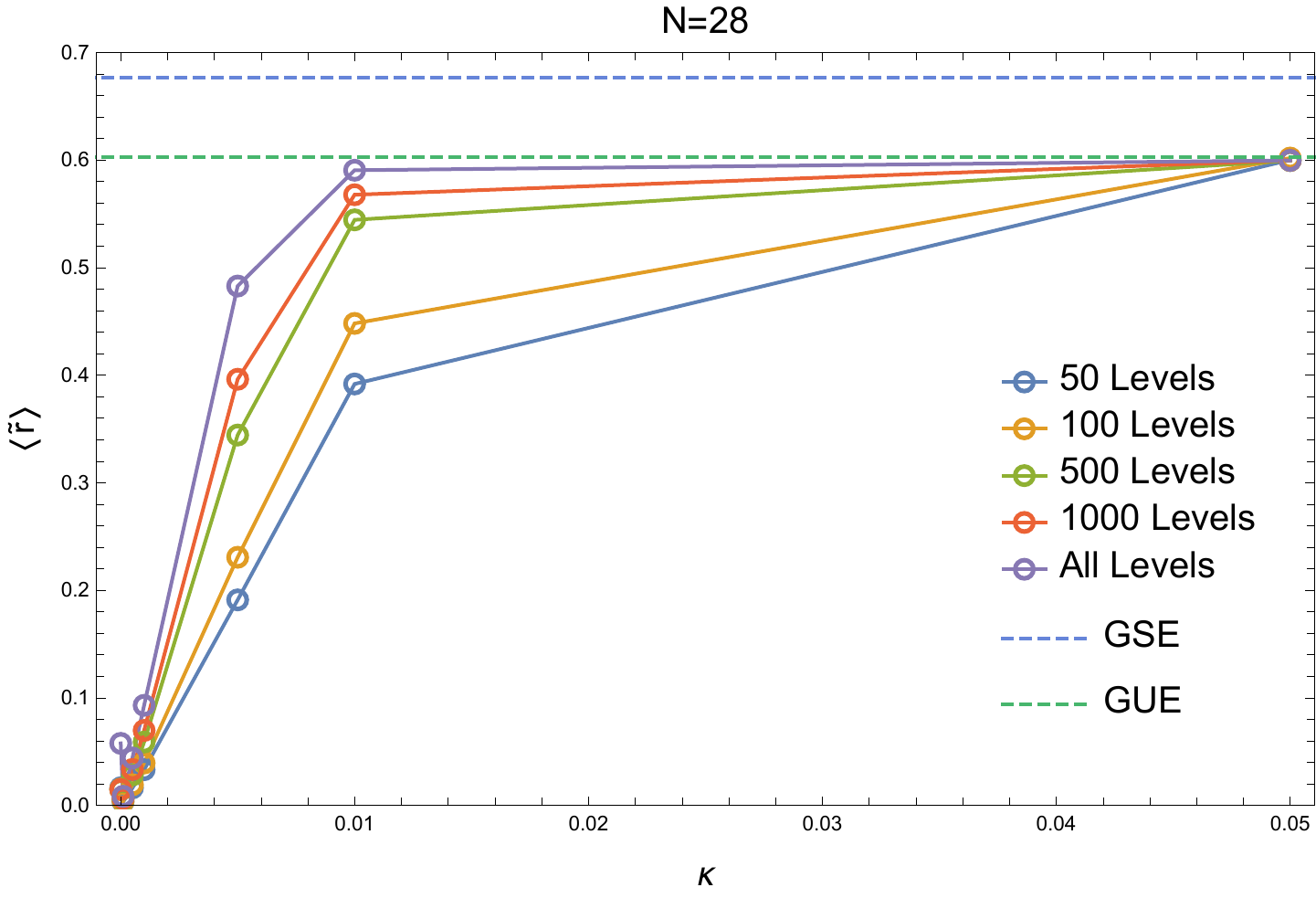}\label{fig: urpara 28 small k}}
\caption{The mean value of the $\tilde{r}$-parameter for $N=28$ (a) with unfolding and (b) without unfolding. The average is performed over $M = 100$ ensembles, respectively. The $\tilde{r}$-parameter are computed subsectors of the spectra including $M$ levels starting from the ground state ($M=50,100,500,1000,$ All).}
\label{fig: all rpara 28 small k}
\end{figure}

\begin{figure}[t!]
\centering
\subfloat[No Unfolding]{\includegraphics[width=8cm]{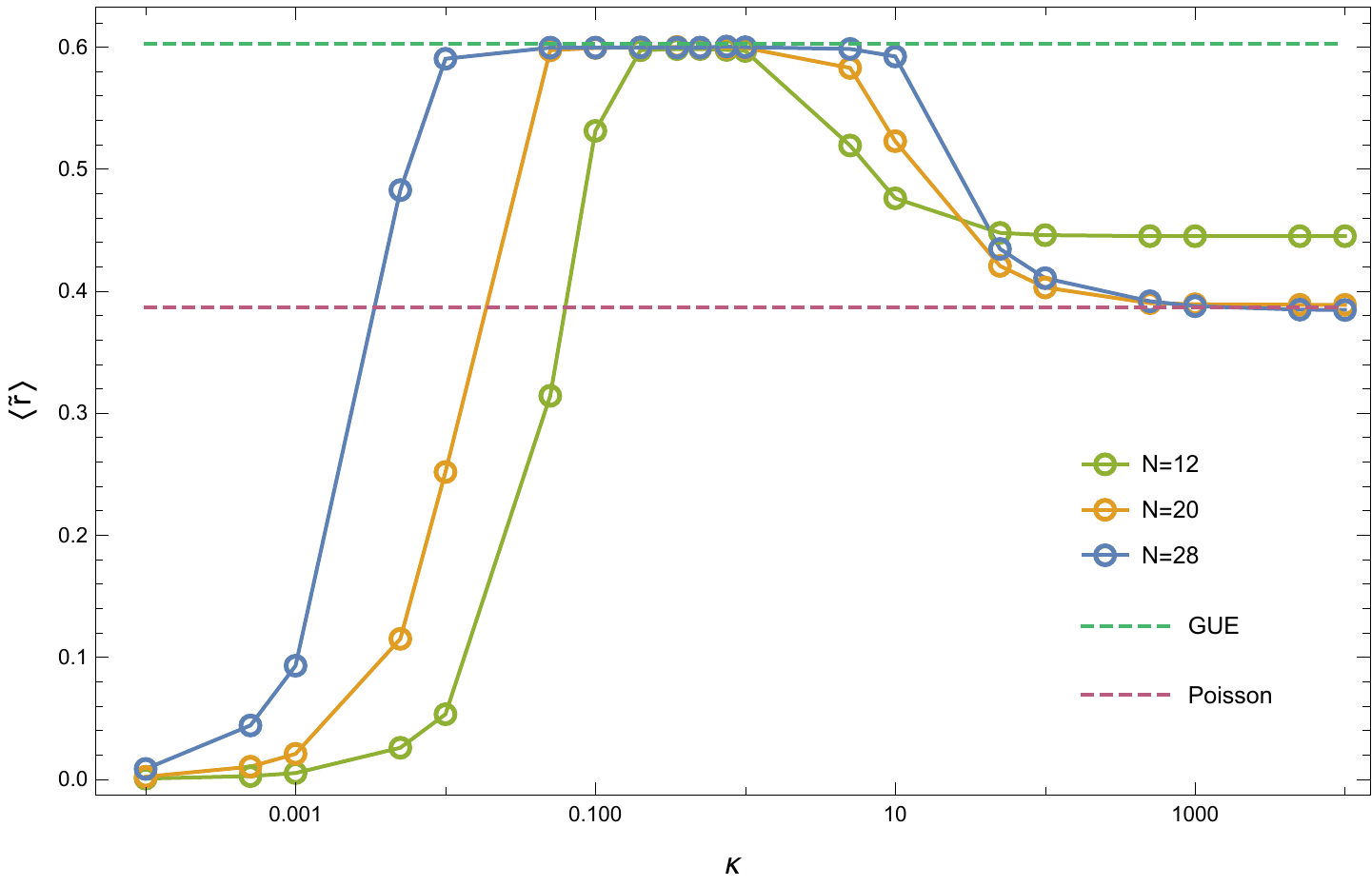}\label{fig: rpara gse 12 28}}\quad
\subfloat[Unfolded]{\includegraphics[width=8cm]{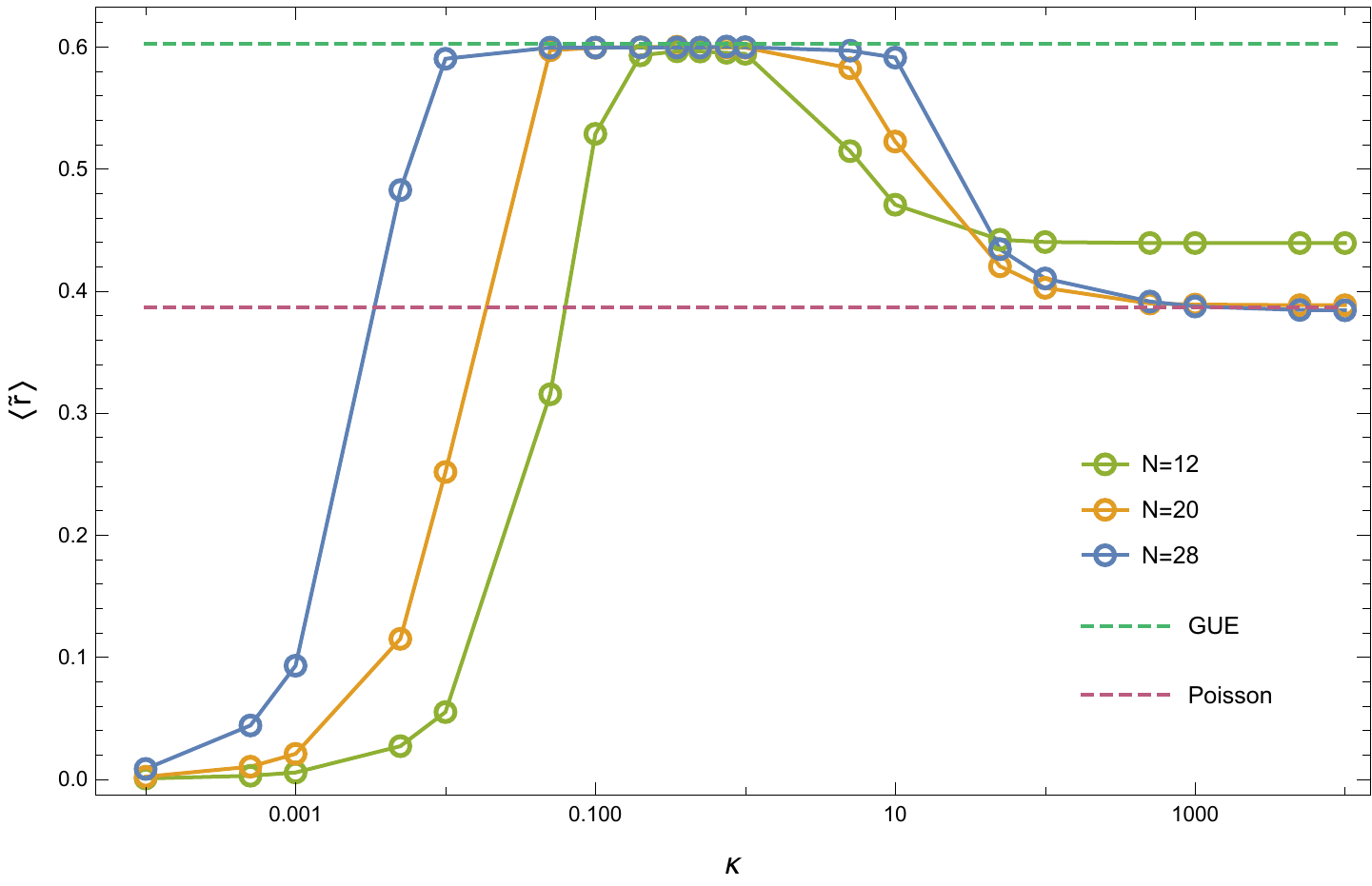}\label{fig: urpara gse 12 28}}
\caption{The mean value of the $\tilde{r}$-parameter for $N=12,20,28 \equiv 4\;(\text{mod} 8)$ of (a) the spectrum without unfolding (b) the unfolded spectrum. The average is performed over $M = 4000,500,100$ ensembles, respectively. We observe a transition from GUE to Poisson.}
\label{fig: all rpara gse 12 28}
\end{figure}

Note that the $\tilde{r}$-parameters of the spectrum without unfolding and the unfolded spectrum for $N\equiv 0,2,6$ ($\text{mod} 8$) do not show crucial difference (See Figure~\ref{fig: all rpara gue 10 30} and \ref{fig: all rpara goe 16 24}). However, one can see a huge discrepancy in the mean value of the unfolded $\tilde{r}$-parameter for $N\equiv 4\;(\text{mod} 8)$ in Figure~\ref{fig: all rpara 28 small k}. Recall that At $\kappa=0$ the quartic interaction is supposed to belong to the GSE class in which we have $\langle\tilde{r}\rangle \approx0.67617$. However, the $\tilde{r}$-parameter of the unfolded spectrum at $\kappa=0$ is $0.65429,0.450978$ and $0.0576467$ for $N=12,20$ and $28$, respectively (See Figure~\ref{fig: urpara gse 12 28}), which are far from the GSE class. On the other hand, the $\tilde{r}$-parameter of spectrum without unfolding at $\kappa=0$ is close to the expected value. \textit{i.e.\,} $0.67102, 0.670404$ and $0.666468$ for $N=12,20$, respectively (See Figure~\ref{fig: rpara 28 small k}).

As soon as the small quadratic interaction appears, it breaks the symmetry of the quartic Hamiltonian, which makes the $\tilde{r}$-parameter close to zero for both cases (see Figure~\ref{fig: all rpara 28 small k} and \ref{fig: all rpara gse 12 28}). As $\kappa$ increases, the $\tilde{r}$-parameter increases until it reaches the $\tilde{r}$-parameter of the GUE class. Then, for example, the $\tilde{r}$-parameter for $N=28$ exhibits the GUE universality from $\kappa\sim 10^{-2}$ to $\kappa\sim 10$. Moreover, as $\kappa$ is increased further, we can also observe the transition from GUE to Poisson around $\kappa\sim 10$. Note that the $\tilde{r}$-parameter of the spectrum without unfolding and the unfolded spectrum exhibit a qualitatively similar behavior for non-zero $\kappa$.

\begin{figure}[t!]
\centering
\subfloat[No Unfolding]{\includegraphics[width=8cm]{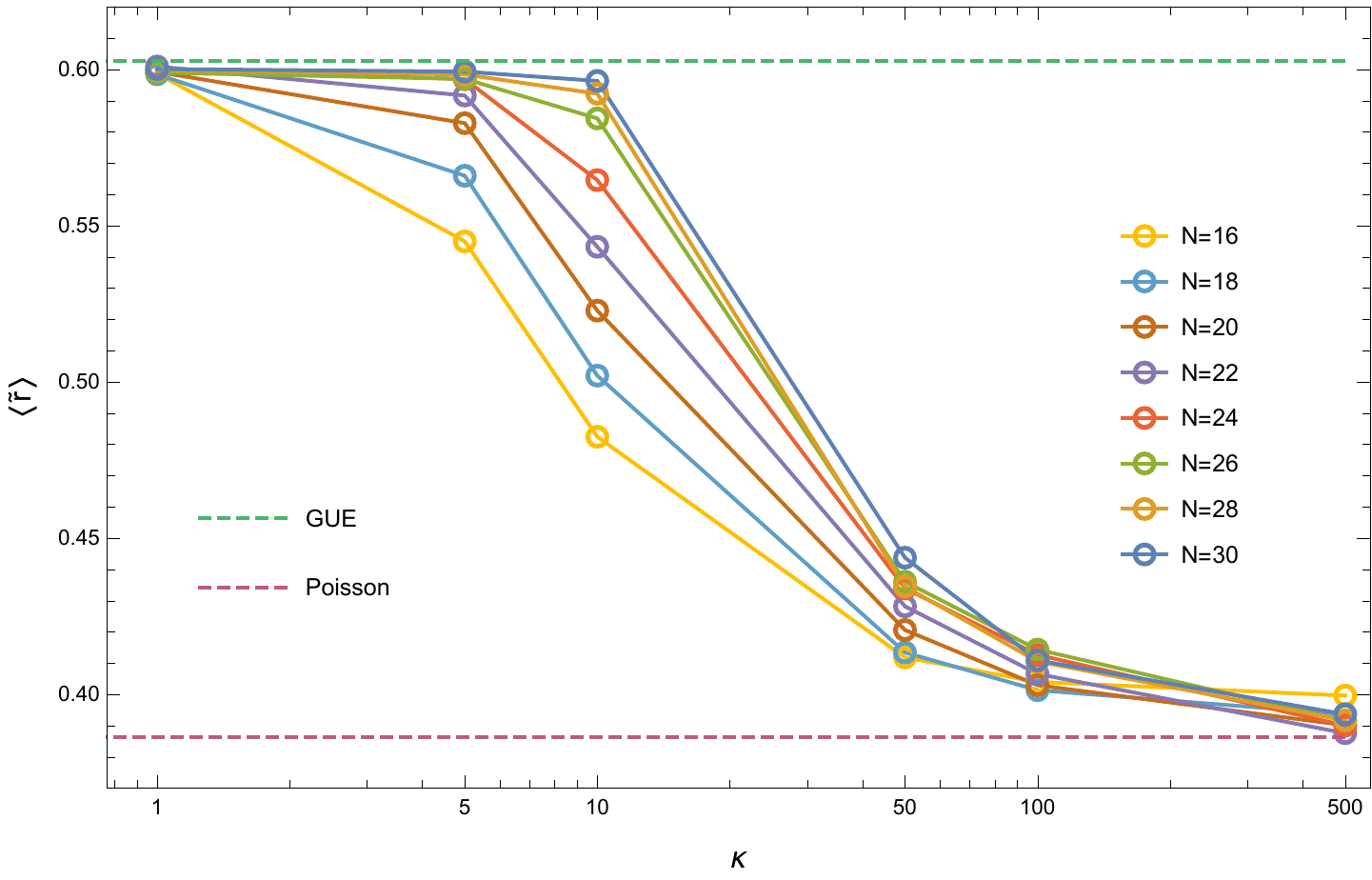}\label{fig: rpara 16 30 middle}}\quad
\subfloat[Unfolded]{\includegraphics[width=8cm]{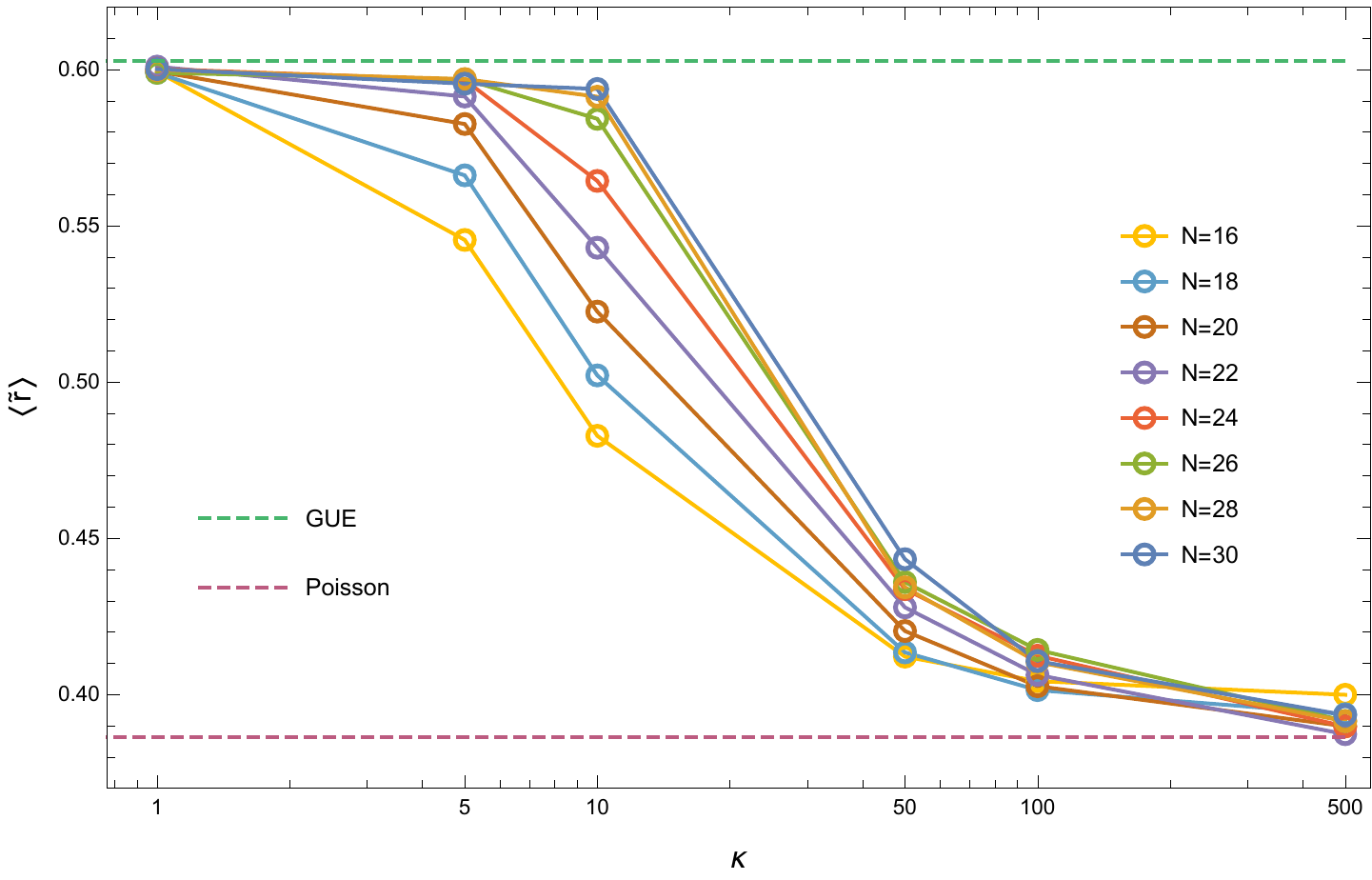}\label{fig: urpara 16 30 middle}}
\caption{The mean value of the $\tilde{r}$-parameter for $N=16,18,20,22,24,26,28,30$ ($1\leqq \kappa\leqq 500$) of (a) the spectrum without unfolding (b) the unfolded spectrum. The average is performed over $M = 2000,1000,500,300,100,100,100,100$ ensembles, respectively.}
\label{fig: all rpara 16 30 middle}
\end{figure}
Although the $\tilde{r}$-parameters for small $\kappa<0.1$ show distinct behavior depending on $N$, they all approach to the GUE $\tilde{r}$-parameter. In addition, as $\kappa$ increases further, the transition from GUE to Poisson appears at $\kappa\sim 10$ for every sufficiently large $N$ (See Figure~\ref{fig: all rpara 16 30 middle}). In this chaotic/integrable transition, one cannot find any qualitative difference between the spectrum without unfolding and the unfolded spectrum.

\subsection{The $\tilde{r}$-parameter of subsectors of the spectrum}
\label{app: r-statistics of subsector}

In the previous Section, we focus our attention on the $\tilde{r}$-parameter for whole spectrum where the chaotic/integrable transition seems to occur around $\kappa\sim 10$. This result is consistent with the Gaussian filtered SFF, introduced in \cite{Gharibyan:2018jrp} and studied for this particular model in Section \ref{sec:gaussian_filter}. However, as we discussed in Section \ref{sec:numerics_preliminary}, and as we confirmed with the analysis of Section \ref{sec:chaos_energy}, the low-lying spectrum seems to be responsible for the chaotic/integrable transition in the connected unfolded SFF. Hence, we now study the $\tilde{r}$-parameter of the low-lying spectrum without unfolding as well as  the low-lying unfolded spectrum.\footnote{As already mentioned in Section.\ref{sec:level_repulsion} we unfold the whole spectrum first, and then take the low-lying unfolded levels.}

\begin{figure}[t!]
\centering
\subfloat[No Unfolding]{\includegraphics[width=8cm]{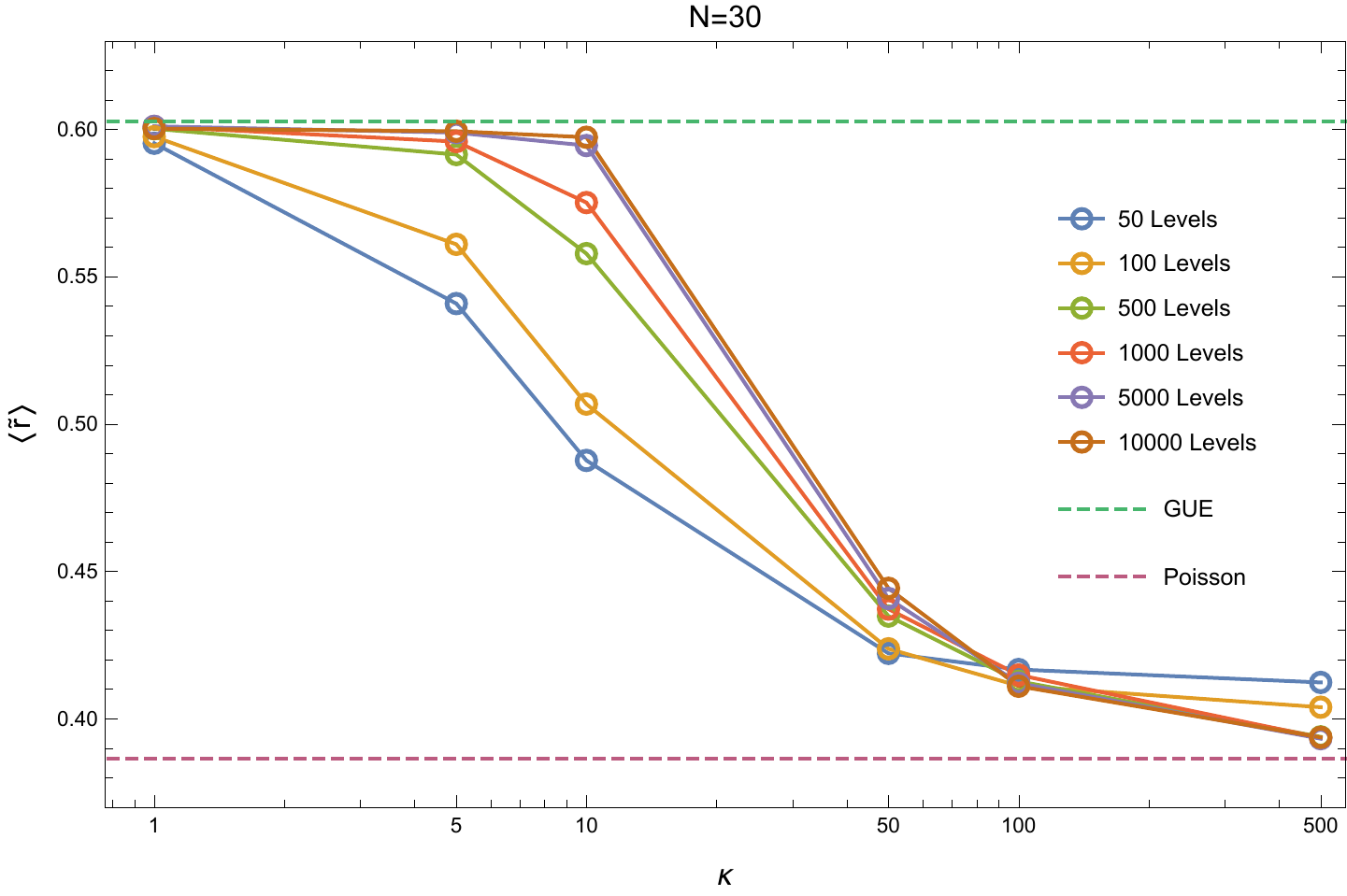}\label{fig: rpara middle 30 level}}\quad
\subfloat[Unfolded]{\includegraphics[width=8cm]{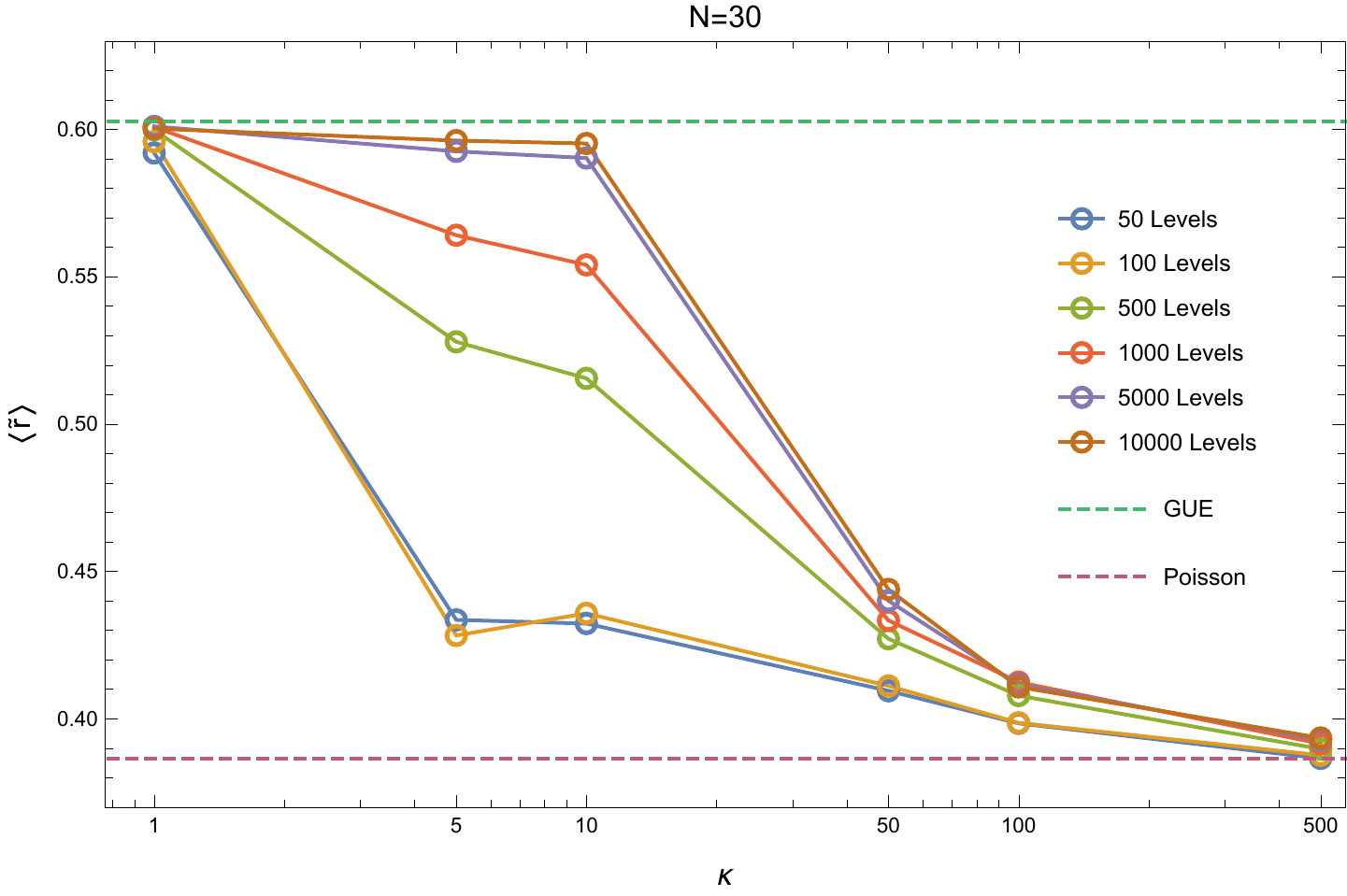}\label{fig: urpara middle 30 level}}
\caption{The mean value of the $\tilde{r}$-parameter for $N=30$ ($1\leqq \kappa\leqq 500$) of (a) the spectrum without unfolding (b) the unfolded spectrum. The average is performed over $M = 100$ ensembles, respectively. The $\tilde{r}$-parameter are computed subsectors of the spectra including $M$ levels starting from the ground state ($M=50,100,500,1000,5000,10000$).}
\label{fig: all rpara middle 30 level}
\end{figure}
For the case of $N=30$, we observe that the low-lying levels~($L=50,100$) begin to transit to the value of Poisson at smaller value of $\kappa<10$ than the whole spectrum (See Figure~\ref{fig: all rpara middle 30 level}). Moreover, as we include more levels, the value of $\kappa$ at which the transition occurs becomes larger. And, for $L=5000, 10000$ low-lying levels, the chaotic/integrable transition is close to that of whole spectrum ($\kappa\sim 10$). Note that during the transition, the $\tilde{r}$-parameter of the low-lying spectrum is affected by the unfolding procedure in contrast to the bulk of the spectrum. Namely, the transition of the low-lying unfolded spectrum seems to be sharper and takes place at lower value of $\kappa\sim 1$ than that of the low-lying spectrum without unfolding. 

%
%
%
%
\begin{figure}[t!]
\centering
\subfloat[No Unfolding]{\includegraphics[width=8cm]{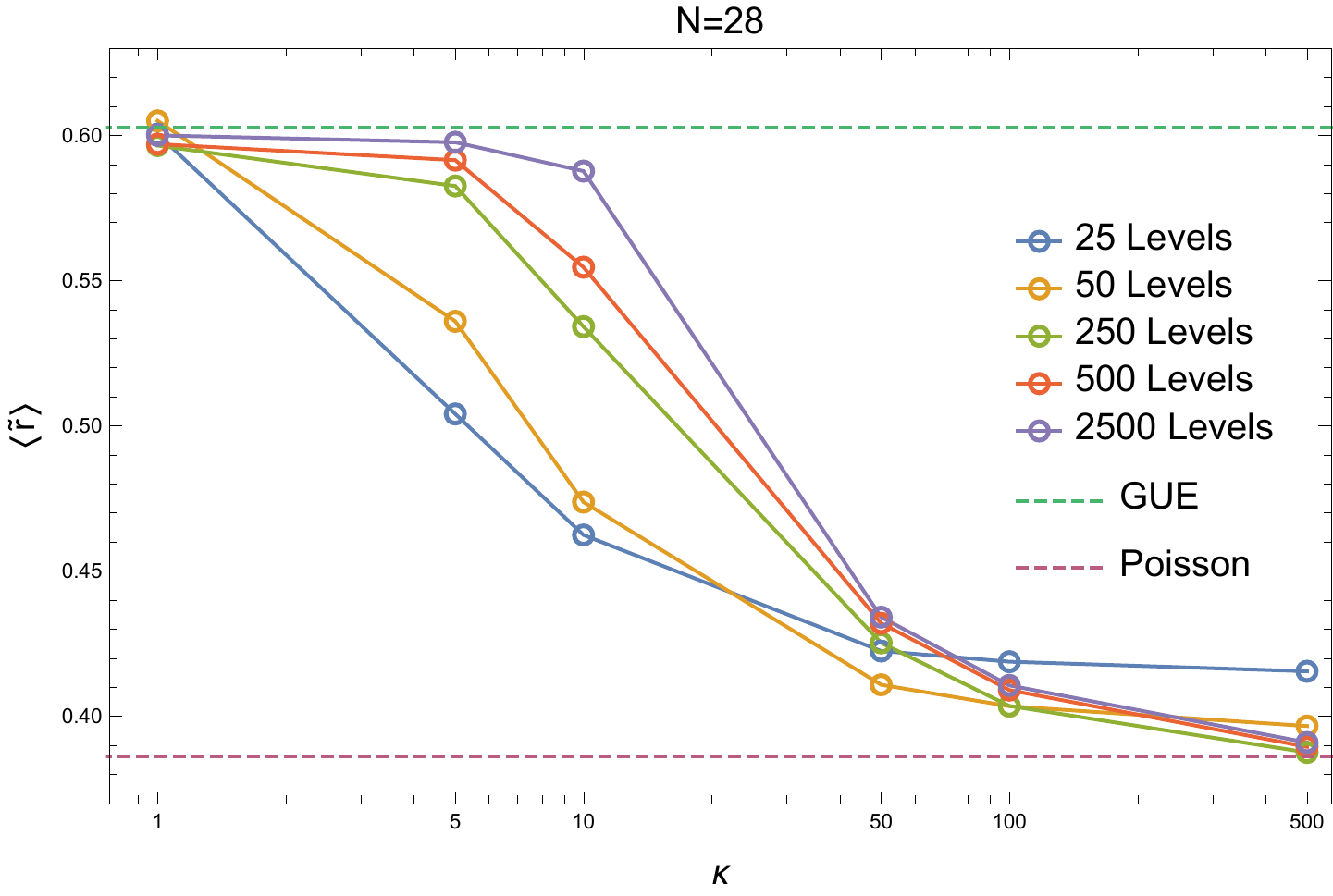}\label{fig: rpara middle 282 level}}\quad
\subfloat[Unfolded]{\includegraphics[width=8cm]{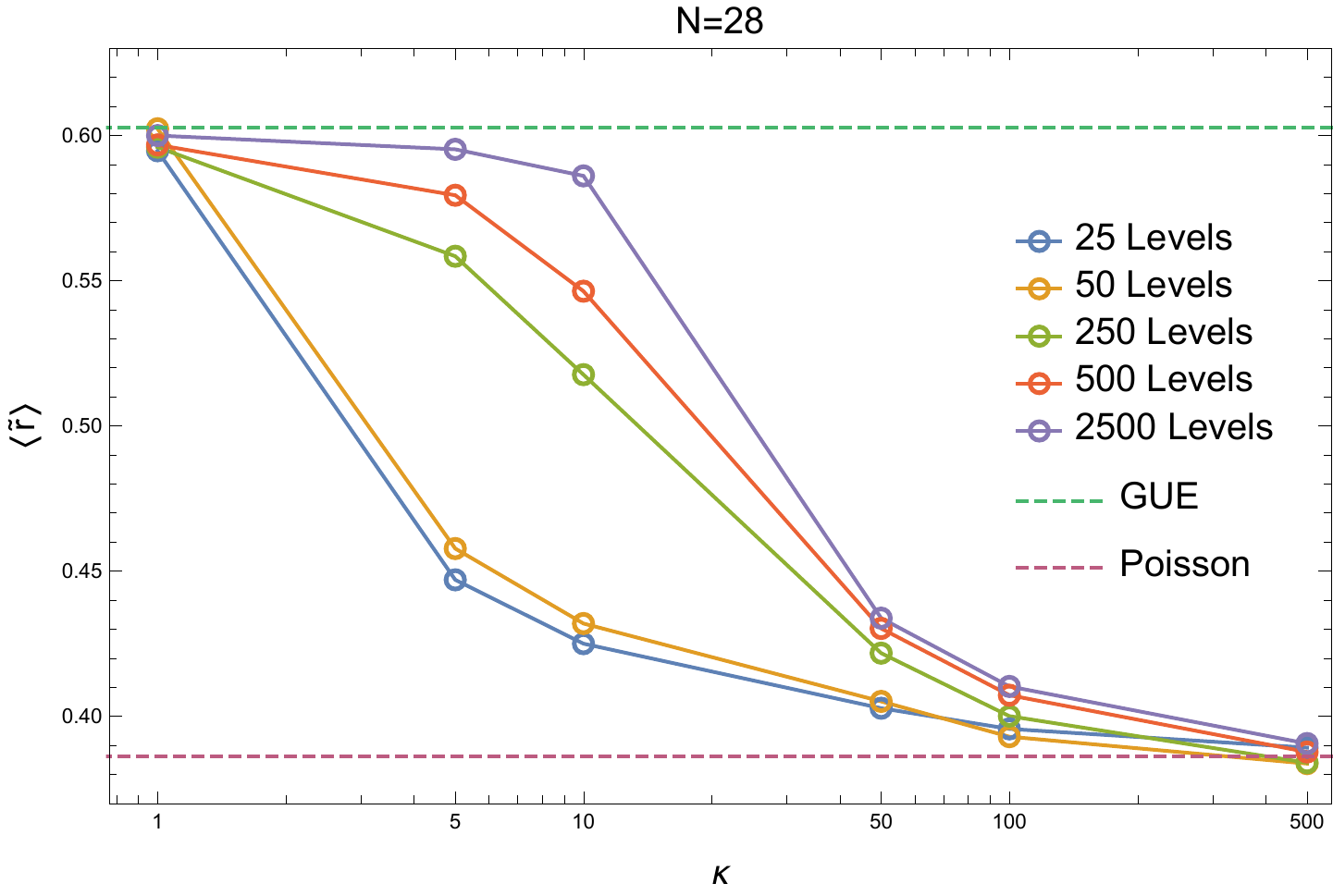}\label{fig: rpara middle 282 level2}}\\
\subfloat[No Unfolding]{\includegraphics[width=8cm]{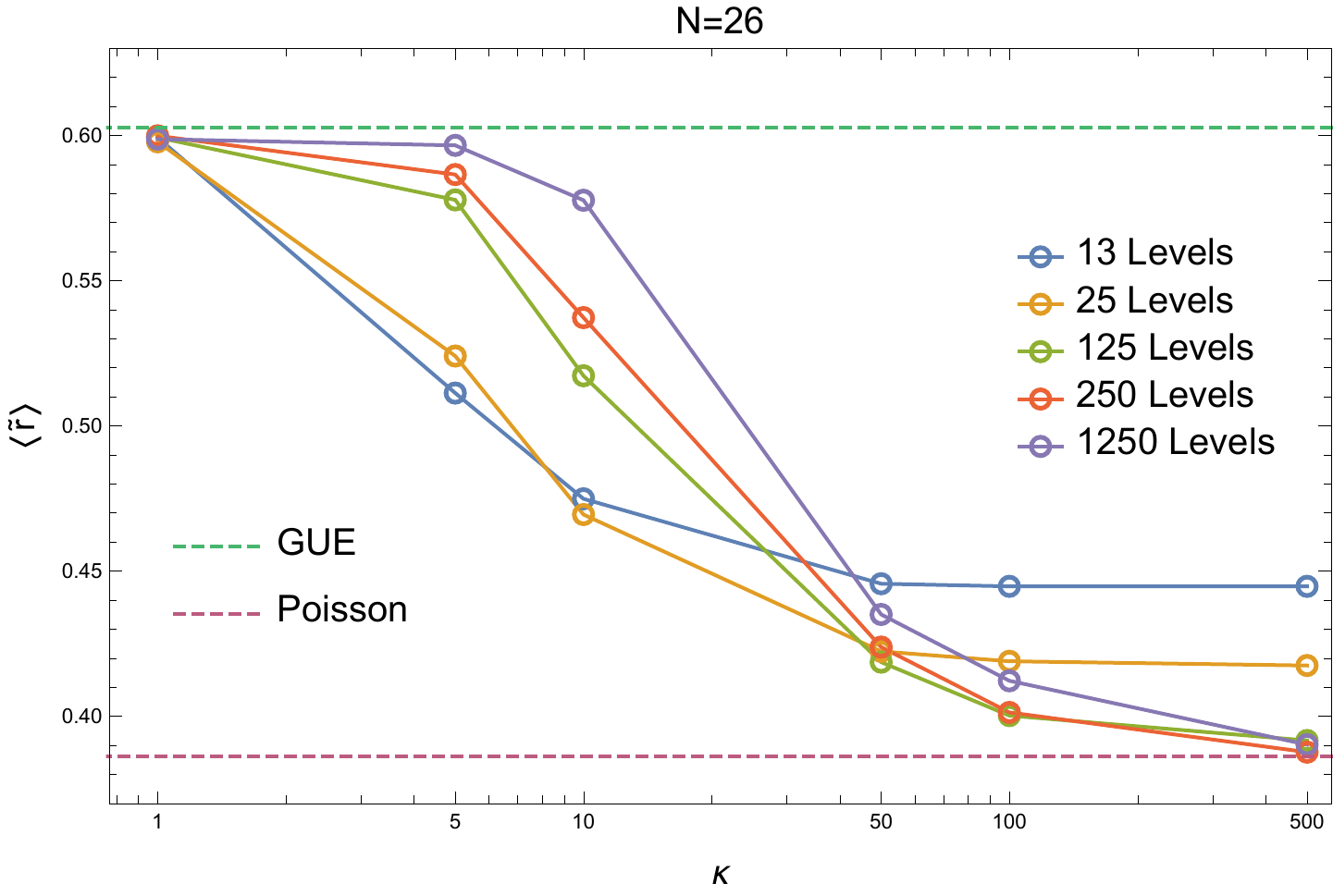}\label{fig: rpara middle 262 level}}\quad
\subfloat[Unfolded]{\includegraphics[width=8cm]{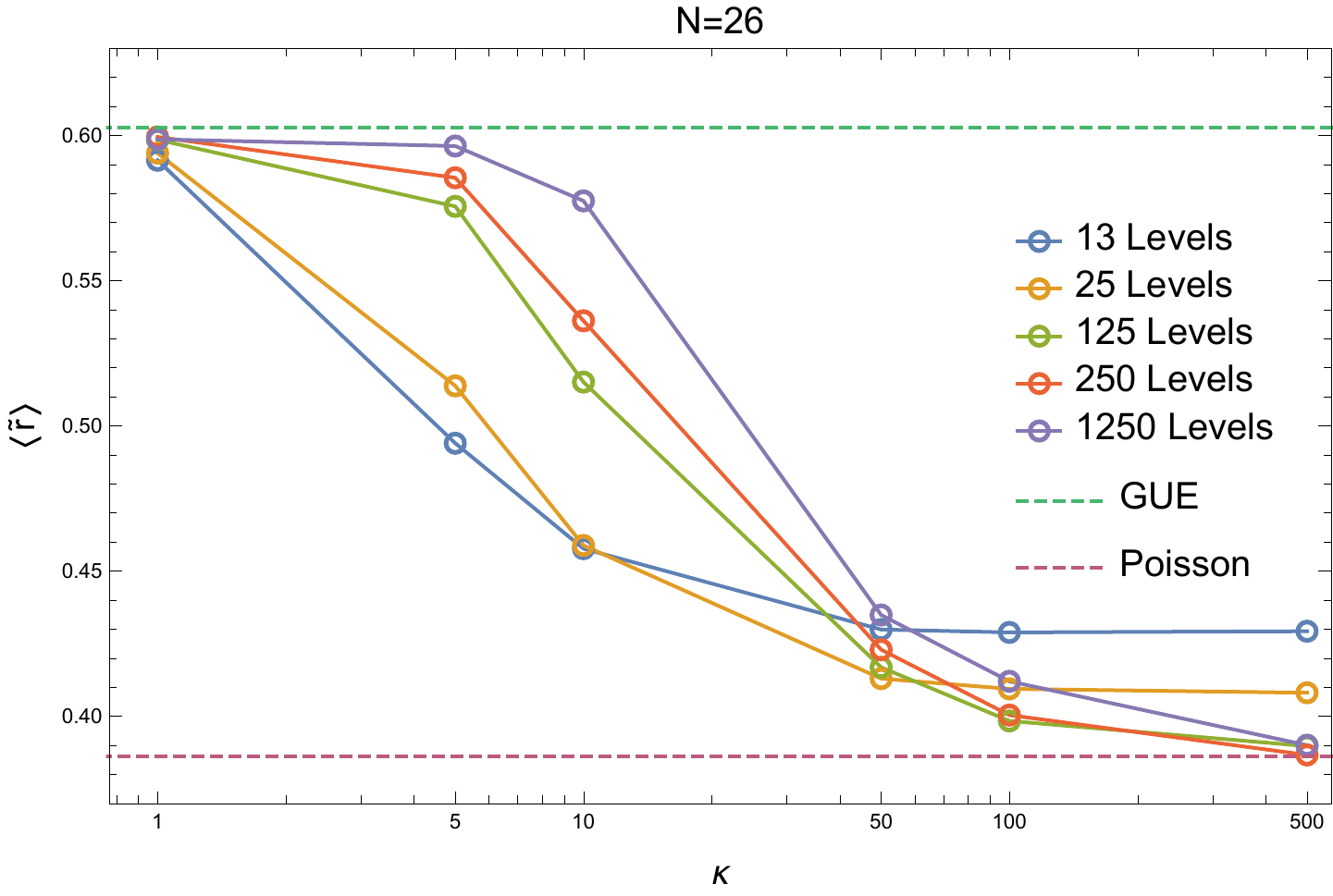}\label{fig: rpara middle 262 level2}}

\caption{The mean value of the $\tilde{r}$-parameter of (a) $N=28$ spectrum without unfolding, (b) $N=28$ unfolded spectrum, (c) $N=26$ spectrum without unfolding and (d) $N=26$ unfolded spectrum ($1\leqq \kappa\leqq 500$). The average is performed over $M = 100$ ensembles, respectively. The $\tilde{r}$-parameter are computed subsectors of the spectra including $M$ levels starting from the ground state ($M=50,100,500,1000$).}
\label{fig: all rpara middle 26 28 level2}
\end{figure}
In Figure~\ref{fig: all rpara middle 26 28 level2}, we also evaluate the transition of $\tilde{r}$-parameter of the same portions of the spectrum for $N=26, 28$ as in $N=30$ case. For example, we take $L=25,50, 250, 500, 2500$ levels for $N=28$ case. Unlike $N=30$ case, the low-lying levels for $N=26, 28$ do not have enough statistics to show the transition to Poisson. Also, the unfolding procedure lower the $\tilde{r}$-parameter of the low-lying levels during the transition while the bulk of spectrum is not affected. Furthermore, the effect of the unfolding on the low-lying spectrum becomes more drastic for larger value of $N$ (See Figure~\ref{fig: all rpara middle 30 level} and Figure~\ref{fig: all rpara middle 26 28 level2}). 

As we already mentioned in the beginning of this Appendix, this difference in the behavior of the low-lying modes with respects to the bulk of the spectrum, which becomes sharper when $N$ gets large and by using the unfolded spectrum, could be a hint of some interesting large $N$ behaviors of the model.
It is possible that this behavior is somehow magnified artificially by the unfolding procedure but given the results of Sections \ref{sec: ipr} and \ref{sec: ir diversity} we are optimistic about the robustness of our findings.

%
%

\bibliographystyle{JHEP}
\bibliography{ChaosandRMT}

\end{document}